\documentclass[longauth]{aa}  

\usepackage{graphicx}
\usepackage{txfonts}
\usepackage[]{hyperref}
\usepackage{longtable}
\usepackage{xspace,xcolor}
\usepackage{newtxtext,newtxmath}
\usepackage{natbib}
\usepackage{outlines}
\usepackage{savesym}
\savesymbol{tablenum}
\usepackage{siunitx}
\usepackage{comment}
\usepackage{graphicx}
\usepackage{float}
\usepackage{orcidlink}

\usepackage{amsmath}	
\usepackage{amssymb}	
\usepackage{multicol}        
\usepackage{bm}		
\usepackage{pdflscape}	

\usepackage[T1]{fontenc}
\usepackage{ae,aecompl}

\usepackage{newtxtext,newtxmath}
\begin{document} 
   \title{Normal or transitional? The evolution and properties of two\\ type Ia supernovae in the Virgo cluster}
   \titlerunning{Virgo cluster SNe~Ia: SN 2020ue and SN 2020nlb}

   \author{L.~Izzo\inst{1,2}\orcidlink{0000-0001-9695-8472},
          C.~Gall\inst{1}\orcidlink{0000-0002-7272-5129},
          N.~Khetan\inst{1,3},
          N.~Earl\inst{4},\orcidlink{0000-0003-1714-7415},
          J.~Hjorth\inst{1}\orcidlink{0000-0002-4571-2306},
          W.~B.~Hoogendam\inst{5}\orcidlink{0000-0003-3953-9532},
          Y.~Q.~Ni\inst{6,7,8}\orcidlink{0000-0003-3656-5268},
          A.~Sedgewick\inst{1}\orcidlink{0000-0002-9158-750X},
          S.~M.~Ward\inst{9},
          Y.~Zenati\inst{10,11,12,13}\orcidlink{0000-0002-0632-8897},
          K.~Auchettl\inst{14,15},
          S.~Bhattacharjee\inst{16}\orcidlink{0000-0002-7350-7043},
          S.~Benetti\inst{17},
          M.~Branchesi\inst{18},
          E.~Cappellaro\inst{17},
          A.~Catapano\inst{19},
          K.~C.~Chambers\inst{5}\orcidlink{0000-0001-6965-7789},
          D.~A.~Coulter\inst{10,11}\orcidlink{0000-0003-4263-2228},
          K.~W.~Davis\inst{13}\orcidlink{0000-0002-5680-4660},
          M.~Della Valle\inst{2}\orcidlink{0000-0003-3142-5020},
          S.~Dhawan\inst{20},
          T.~de~Boer\inst{4}\orcidlink{0000-0001-5486-2747},
          G.~Dimitriadis\inst{21}\orcidlink{0000-0001-9494-179X},
          R.~J.~Foley\inst{13},
          M.~Fulton\inst{22},
          H.~Gao\inst{5}\orcidlink{0000-0003-1015-5367},
          W.~J.~Hon\inst{15},
          M.~E.~Huber\inst{4}\orcidlink{0000-0003-1059-9603}
          D.~O.~Jones\inst{5}\orcidlink{0000-0002-6230-0151},
          C.~D.~Kilpatrick\inst{23}\orcidlink{0000-0002-5740-7747},
          C.~C.~Lin\inst{5}\orcidlink{0000-0002-7272-5129},
          T.~B.~Lowe\inst{5},
          E.~A.~Magnier\inst{5}\orcidlink{0000-0002-7965-2815},
          K.~S.~Mandel\inst{9}\orcidlink{0000-0001-9846-4417},
          R.~Margutti\inst{24},
          G.~Narayan\inst{4}\orcidlink{0000-0001-6022-0484}
          P.~Ochner\inst{17,25},
          Y.~C.~Pan\inst{16}\orcidlink{0000-0001-8415-6720},
          A.~Reguitti\inst{26,17}\orcidlink{0000-0003-4254-2724},
          C.~Rojas-Bravo\inst{14}\orcidlink{0000-0002-7559-315X},
          M.~Siebert\inst{11}\orcidlink{0000-0003-2445-3891},
          S.~J.~Smartt\inst{22,27}\orcidlink{0000-0002-8229-1731},
          K.~W.~Smith\inst{18,20}\orcidlink{0000-0001-9535-3199},
          S.~Srivastav\inst{18}\orcidlink{0000-0003-4524-6883},
          J.~J.~Swift\inst{28}\orcidlink{0000-0002-9486-818X},
          K.~Taggart\inst{5},
          G.~Terreran\inst{29}\orcidlink{0000-0003-0794-5982},
          S.~Thorp\inst{9,30}\orcidlink{0009-0005-6323-0457},
          L.~Tomasella\inst{17}\orcidlink{0000-0002-3697-2616},
          R.~J.~Wainscoat\inst{5}\orcidlink{0000-0002-1341-0952},
          }

   \authorrunning{Izzo et al.}

   \institute{DARK, Niels Bohr Institute, University of Copenhagen, Jagtvej 155A, 2200 Copenhagen, Denmark
        \email{luca.izzo@nbi.ku.dk,luca.izzo@inaf.it}
    \and
        INAF, Osservatorio Astronomico di Capodimonte, Salita Moiariello 16, I-80121 Naples, Italy
    \and
        School of Mathematics and Physics, University of Queensland, Brisbane, QLD 4072, Australia
    \and
        Department of Astronomy, University of Illinois at Urbana-Champaign, 1002 W. Green St., IL 61801, USA
    \and
        Institute for Astronomy, University of Hawaii, 2680 Woodlawn Drive, Honolulu, HI 96822, USA
    \and
        Kavli Institute for Theoretical Physics, University of California, Santa Barbara, 552 University Road, Goleta, CA 93106-4030, USA
    \and
        Las Cumbres Observatory, 6740 Cortona Drive, Suite 102, Goleta, CA 93117, USA
    \and
        David A. Dunlap Department of Astronomy and Astrophysics, University of Toronto, 50 St. George Street, Toronto, ON M5S 3H4, Canada
    \and
        Institute of Astronomy and Kavli Institute for Cosmology, Madingley Road, Cambridge, CB3 0HA, UK
    \and 
        Physics and Astronomy Department, Johns Hopkins University, Baltimore, MD 21218, USA
    \and 
        Space Telescope Science Institute, Baltimore, MD 21218, USA
    \and
        Astrophysics Research Center of the Open University (ARCO), The Open University of Israel, Ra’anana 4353701, Israel
    \and
        Department of Natural Sciences, The Open University of Israel, Ra'anana 4353701, Israel
    \and
        Department of Astronomy and Astrophysics, University of California, Santa Cruz, CA 95064, USA
    \and
        School of Physics, The University of Melbourne, VIC 3010, Australia
    \and
        Graduate Institute of Astronomy, National Central University, 300 Zhongda Road, Zhongli, Taoyuan 32001, Taiwan
    \and
        INAF – Osservatorio Astronomico di Padova, Vicolo dell'Osservatorio 5, I-35122 Padova, Italy
    \and
        Gran Sasso Science Institute (GSSI), I-67100 L’Aquila, Italy 
    \and 
        Osservatorio Astronomico S. di Giacomo, AstroCampania, via S. di Giacomo, I-80051 Agerola, Italy
    \and
        School of Physics and Astronomy, Birmingham University, B15 2TT, UK
    \and
        Department of Physics, Lancaster University, Lancaster, LA1, UK
    \and
        Department of Physics, University of Oxford, Keble Road, Oxford, OX1 3RH, UK
    \and 
        Center for Interdisciplinary Exploration and Research in Astrophysics (CIERA) and Department of Physics and Astronomy, Northwestern University, Evanston, IL 60208, USA
    \and
        Department of Astronomy, University of California, Berkeley, CA 94720-3411, USA
    \and
        Dipartimento di Fisica e Astronomia, Universit\'a degli Studi di Padova, Via F. Marzolo 8, I-35131 Padova, Italy
    \and  
        INAF – Osservatorio Astronomico di Brera, Via E. Bianchi 46, I-23807 Merate (LC), Italy
    \and
        Astrophysics Research Centre, School of Mathematics and Physics. Queen's University Belfast, BT7 1NN, UK
    \and 
        The Thacher School, 5025 Thacher Rd., Ojai, CA 93023, USA
    \and
        Adler Planetarium, 1300 South DuSable Lake Shore Drive, Chicago, IL, 60605, USA
    \and
        The Oskar Klein Centre, Department of Physics, Stockholm University, AlbaNova University Centre, SE 106 91 Stockholm, Sweden
    }


\date{}


\abstract
{Type Ia supernovae (SNe~Ia) are among the most precise cosmological distance indicators used to study the expansion history of the Universe.
The vast increase of SN~Ia data due to large-scale astrophysical surveys has led to the discovery of a wide variety of SN~Ia sub-classes, such as transitional and fast-declining SNe~Ia. However, their distinct photometric and spectroscopic properties differentiate them from the population of normal SNe~Ia such that their use as cosmological tools remains challenged.  
Here, we present a high-cadenced photometric and spectroscopic dataset of two SNe~Ia, SNe~2020ue and 2020nlb, which were discovered in the nearby Virgo cluster of galaxies. Our study shows that SN 2020nlb is a normal SN Ia whose unusually red color is intrinsic, arising from a lower photospheric temperature rather than interstellar reddening, providing clear evidence that color diversity among normal SNe Ia can have a physical origin.
In contrast, SN 2020ue has photometric properties, 
such as color evolution and light-curve decay rate, similar to those of transitional SNe, spectroscopically it is more aligned with normal SNe~Ia. This is evident from 
spectroscopic indicators such as the pseudo-equivalent width of \ion{Si}{II} lines. 
Thus, such SNe~Ia that are photometrically at the edge of the standard normal SNe~Ia range may be missed in cosmological SNe~Ia samples. 
Our results highlight that spectroscopic analysis of SNe Ia around peak brightness is crucial for identifying intrinsic color variations and constructing a more complete and physically homogeneous SN Ia sample for precision cosmology. }

\keywords{
Supernovae: individual: SN 2020ue -- Supernovae: individual:SN 2020nlb -- Cosmology: distance scale
}
                    

\maketitle
\section{Introduction}
\label{s:intro}

Type Ia supernovae (SNe~Ia) result from the thermonuclear detonation of carbon-oxygen (CO) white dwarfs (WDs), completely disrupting them \citep{Nugent2011,Bloom2012,Maguire2017,Branch2017}. Early spectra are characterized by the absence of hydrogen and helium, with prominent silicon and other iron-group and intermediate-mass elements \citep{Filippenko1997}. The strength of these spectral signatures is likely connected to the progenitor system and explosion mechanism, both of which remain uncertain. Various progenitor scenarios and explosion mechanisms have been proposed due to the complex physics of SN~Ia explosions \citep{2018PhR...736....1L,ZhangWei_Ropke_Han23,RuiterSeitenzahl24}. Leading scenarios include the single-degenerate (SD) \citep{Whelan1973} and double-degenerate (DD) systems \citep{Iben1984}.

The SD scenario involves a Chandrasekhar-limit WD accreting matter from a non-degenerate companion star (a red giant or low-mass main sequence star) until it reaches critical mass and undergoes carbon detonation. Conversely, the DD scenario is initiated by the dynamical merging of two WDs, triggering a thermonuclear runaway due to a rapid temperature increase or an accretion-induced collapse \citep[see, e.g.,][and references therein]{RuiterSeitenzahl24}. The classical model explaining CO WD explosions involves the delayed-detonation mechanism \citep{Khokhlov1991,Yamaoka1992,Iwamoto1999}. This process begins with a subsonic deflagration, transitioning into a supersonic detonation as the burning rate rapidly increases. The detonation wave propagates through the entire WD, enabling the burning of exterior material. This mechanism effectively explains the distribution of elements determined via high-cadence spectroscopic analysis and spectral synthesis modeling techniques \citep{Stehle2005,Mazzali2008,Tanaka2011,Mazzali2015}. This includes the confinement of Fe-peak elements, particularly the $^{56}$Ni radioactive isotope, within the innermost layers.

The progenitor system(s) and explosion mechanism(s) of SNe~Ia  influence the photometric evolution, including lightcurves. The synthesis of $^{56}$Ni, which decays into $^{56}$Co \citep{Colgate1969}, and the diffusion of radiation within the SN ejecta \citep{BranchWheeler2017} drive the increase in brightness for 15-20 days \citep{Woosley2007, Firth2015, Hillebrandt+13, Perets+19_Ia}. As a result, the amount of $^{56}$Ni synthesized correlates with the peak brightness of SN~Ia \citep{Arnett1982}, and regulates the heating budget. High-energy photon diffusion from the ejecta's core to its envelope results in thermalization and optical detection as the ejecta becomes optically thin. Moreover, temperature affects opacities in the SN ejecta, influencing diffusion time scales \citep{Hoeflich2013}, which dictate the light curve decline rate. This provides a theoretical basis for the empirical SN~Ia luminosity-decay relation \citep{Pskovskii1977,Phillips1993}. Brighter SN~Ia peak magnitudes exhibit longer rise and decay times due to larger $^{56}$Ni synthesis, while fainter SNe~Ia with faster rising and decaying light-curves have produced smaller amounts.
 
`Normal' SNe~Ia display a secondary NIR maximum 20--40 days post-optical peak brightness \citep{Hamuy1993,Kasen2006,Jack2012,Phillips2012,Dhawan2015}, which was first noted for SN1980N and SN1981B \citep{Elias1985}. This is likely due to significant iron-group element synthesis during the explosion, absorbing UV radiation and re-emitting it in the near infrared (NIR) via fluorescence \citep{Kasen2006}. Subsequently, optical and NIR light curves enter a shallow, steady decay powered by the $^{56}$Co to $^{56}$Fe decay chain, characterized by a longer half-life (77.12 days) time than the $^{56}$Ni to $^{56}$Co chain  \citep[6.10 days,][]{Nadyozhin1994}. 
Around 10-15\% of SNe~Ia are sub-luminous, fast-declining events \citep{Li2011,Desai2024}, with a gradual rather than sharp division between sub-luminous and normal SNe~Ia \citep{Graur2024}. Classical examples include SN~1991bg-like events \citep{Filippenko1992b,Leibundgut1993}, characterized by faint peak luminosities, rapid $B$-band decline rates ($\Delta m_{15} > 1.5$ mag), and redder colors ($B-V \sim 0.4$ mag) than normal SNe~Ia \citep{Hoeflich2017}. Lacking secondary NIR maxima, these events also exhibit prominent \ion{Ti}{II} features in early spectra, indicating low photospheric temperatures \citep{Leibundgut1993, Nugent1995}. Intriguingly, such SNe~Ia are predominantly found in early-type galaxies.

{\it Transitional} SNe~Ia constitute another sub-class with mixed photometric and spectroscopic characteristics \citep{Taubenberger2008, Gonzalez2011, Hsiao2015, Ashall2016, Gall2018}. They exhibit fast lightcurve decline rates ($\Delta m_{15} \gtrsim 1.5$ mag), brighter peak magnitudes than SN~1991bg-like events, and a secondary NIR maximum. Spectroscopically, they lack \ion{Ti}{II} at similar epochs to SN~1991bg-like events but display enhanced higher ionization features such as \ion{Fe}{III}, as well as an intense, but lower ionization \ion{Si}{II} 5972~\AA\ transition. 
Progenitor systems and explosion mechanisms for transitional SNe~Ia remain ambiguous, with both SD and DD scenarios plausible. Within the delayed detonation model, sub-luminous SNe~Ia's total $^{56}$Ni production and distribution is likely dominated by the deflagration phase \citep{Hoeflich2017, Gall2018, Ashall2018}. The WD progenitor's central density influences the $^{56}$Ni produced during this phase; high densities can result in the overproduction of stable $^{58}$Ni at the expense of $^{56}$Ni \citep{Seitenzahl2017}. Lower $^{56}$Ni production translates to reduced ejecta heating and cooler photospheric temperatures. For low-luminosity SNe~Ia sub-classes, central density variations can significantly impact peak luminosities and spectral elemental signatures, while effects are minimal for normal SNe~Ia \citep{Ashall2018}. Alternatively, recent multi-dimensional simulations of sub-Chandrasekhar mass explosion models with double-detonation triggered in either SD or DD scenarios \citep{Nomoto1984, Woosley1986, Fink2007,Pakmor+11,Fisher_Jumper2015,DanM+15,Perets+19_Ia,Boos2021} demonstrate a wide range of $^{56}$Ni masses. This can account for the observed variations in light curve properties and spectral signatures of sub-luminous and normal SNe~Ia \citep{ShenK2021}.

Low-luminosity SNe~Ia sub-classes (fast-declining and transitional) challenge progenitor and explosion models and their cosmological use as distance indicators \citep{2010AJ....140.2036S, Gall2018}. This is mainly because cosmological models rely on parameters derived from SN~Ia photometric data, i.e., lightcurves. However, it remains unclear if transitional SNe~Ia are at the faint/red end of normal SNe~Ia or the blue/bright end of truly sub-luminous SNe~Ia. Thus, lightcurve parameters of low-luminosity SNe~Ia sub-classes obtained using lightcurve fitter codes \citep{2005A&A...443..781G, 2007A&A...466...11G, Burns2011, Burns2014} primarily employing normal SN~Ia templates can result in distance uncertainties up to 14\% \citep{Gall2018}. Including fast-declining SN~Ia templates can improve distance estimates \citep{Hoogendam2022}. Moreover, fast-declining SNe~Ia may follow a different luminosity- light curve shape relation than normal SNe~Ia, enabling their use as standalone distance indicators \citep{Graur2024}. Incorporating specific spectroscopic information into the SN~Ia standardization process can also help reduce uncertainties \citep{2021ApJ...912...70B, 2021ApJ...912...71B,2023JCAP...11..046M}.

In this work, we present two nearby SNe~Ia, SN 2020ue and SN 2020nlb, in early-type Virgo cluster galaxies. Distances inferred from light-curve fitting agree well with independent host galaxy measurements. SN 2020ue exhibits distinct photometric properties, such as stretch-color parameter and a $B - V$ color evolution, consistent with transitional SNe~Ia. The Lira law, describing the $B - V$ evolution of Type Ia supernovae, features an initial blue color, rapid reddening around 10 days post-peak brightness, and subsequent linear color decline useful for estimating host galaxy extinction \citep{1996MsT..3L,Phillips1999, Forster2013}. Conversely, SN 2020nlb displays normal SN~Ia characteristics. Despite photometric differences, detailed spectroscopic analysis reveals common spectral feature patterns, suggesting that incorporating spectroscopic information into photometric relations could enhance discrimination between normal and transitional events.

In Sec.~\ref{s:obs}, we describe the photometric and spectroscopic datasets of both SN~2020nlb and SN~2020ue. The inferred results and insights from the photometric analysis and modeling are presented in Sec.~\ref{s:photana}, while the spectroscopic analysis, including spectral modeling, from early epochs to the nebular phase, is described in Sec.~\ref{s:specana}. In Sec.~\ref{s:discu}, we discuss the use of spectroscopic information to identify normal vs transitional/fast decliner events, while in the last section, Sec.~\ref{s:concl}, we draw our conclusions. 

\section{Observations}
\label{s:obs}

\subsection{Discovery, classification and host galaxy}\label{sec:2_1}

\begin{figure*}
    \centering
    \includegraphics[width=0.90\linewidth]{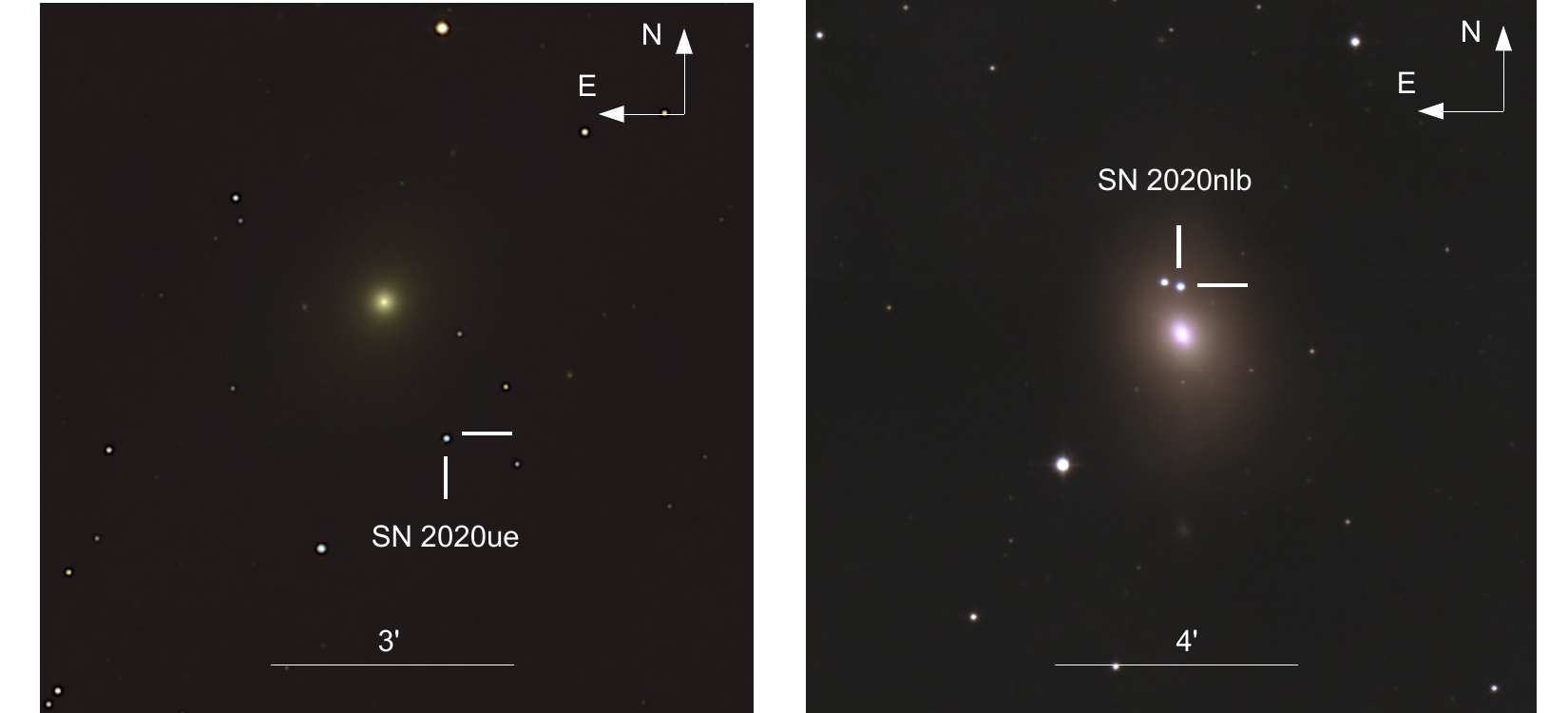}
    \caption{A trichromy image made using $BVr$ single filter images obtained with the Asiago Schmidt telescope a few days after the SN discoveries. The positions of SN 2020ue (left panel) and SN 2020nlb (right panel) in their respective host galaxies are marked using white indicators. Courtesy of Giovanni Benetti (DFA, University of Padova).}
    \label{fig:images}
\end{figure*}

SN 2020ue was discovered by \citet{Itagaki2020} with a brightness of C (Clear) = 15 mag in the galaxy NGC 4636 on January 12, 2020 (MJD = 58860.72) at the coordinates RA=12:42:46.780, $\delta$=+02:39:34.13. The transient was classified as an SN~Ia by \citet{Kawabata2020} a few hours later, on MJD = 58860.86. The last non-detection by the Asteroid Terrestrial-impact Last Alert System \citep[ATLAS][]{Tonry2018} was on January 4, 2020 (MJD = 58852.62). The host galaxy of SN 2020ue, NGC 4636, is the southernmost early-type member of the Virgo cluster \citep{deVaucouleurs1961}. This galaxy has been the subject of multi-wavelength studies in the last years, given its relatively high concentration of dark matter \citep{Schuberth2006} and the evidence of extended interstellar dust \citep{Temi2007}, suggesting a relatively recent merger with some faint galaxy companions. There are several distance measurements in the literature for NGC 4636, obtained with different methods. Within the {\it CosmicFlows-3} compilation of distances and peculiar velocities \citep{Kourkchi2020}, the distance to NGC 4636 was estimated as a weighted average between the surface brightness fluctuations (SBF) methodology and the Fundamental Plane (FP) values to be $\mu = 30.90 \pm 0.24$ mag. 

SN 2020nlb was discovered by ATLAS on June 25, 2020 (MJD = 59025.25) as a new object of magnitude $o = 17.44 \pm 0.08$ mag \citep[orange filter,][]{Townsend2020} at the coordinates RA=12:25:24.190, $\delta$=+18:12:12.80. The host galaxy is NGC 4382, also known as M85. The last non-detection by ATLAS was two days prior to discovery, at MJD = 59023.28. The spectroscopic confirmation as SN~Ia in NGC 4382 was reported by the Nordic optical telescope Unbiased Transient Survey 2 (NUTS2) collaboration \citep{Fiore2020}. 
The galaxy has been classified as an S0 type galaxy \citep{Trentham2002}, and is located in the E cloud of the Virgo cluster \citep{deVaucouleurs1961}, opposite to the location of NGC 4636. Similar to NGC 4636, the distance modulus for NGC 4382 was estimated by {\it CosmicFlows-3} to be $\mu = 31.00 \pm 0.24$ mag \citep{Tully2016}. 
This SN has already been studied by \citet{Sand2021} and \citet{Williams24}.

\subsection{Photometry}

\begin{figure*}
    \centering
   \includegraphics[width=0.47\linewidth]{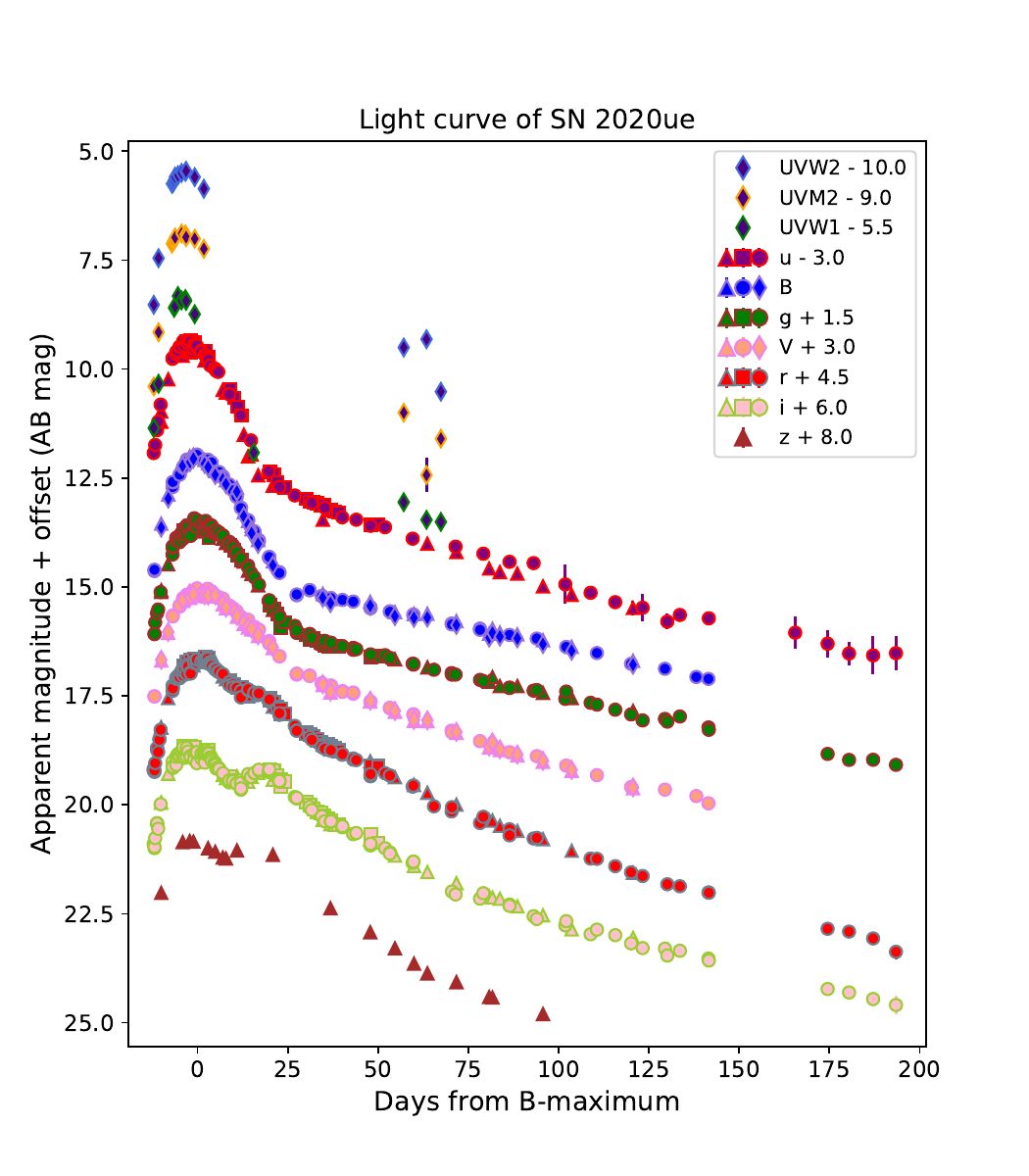}
    \includegraphics[width=0.47\linewidth]{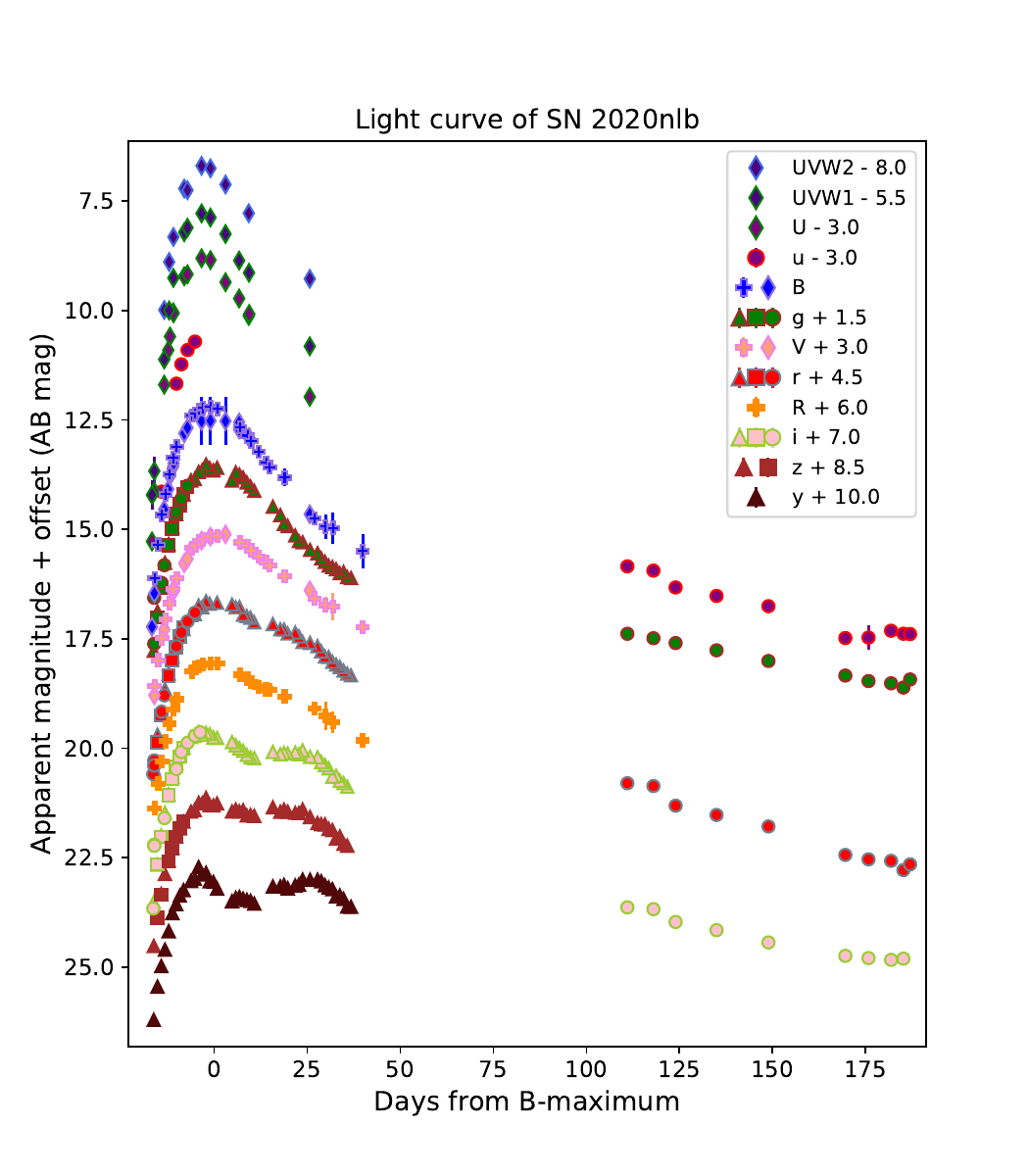}
    \caption{Light curve evolution of SN 2020ue (left panel) and SN 2020nlb (right panel). Data from different facilities are shown with different symbols, with colors corresponding to different filters: {\it Swift}-UVOT data are shown with diamond markers, Copernico-AFOSC and NOT-ALFOSC (for SN 2020ue) and Pan-STARRS (for SN 2020nlb) data with triangles, Swope and Thacher data with squares, LCO data with circles, OASDG data with plus markers. Both light curves have not been corrected for Galactic reddening.}
    \label{fig:LC2020uenlb}
\end{figure*}

Our photometric dataset includes data covering the UV and optical evolution of both SNe~Ia from discovery to around 150 days post-peak brightness, obtained with various observational facilities. Optical observations were conducted at a nearly 1-day cadence near peak brightness, decreasing to a lower cadence during the slow-declining $^{56}$Co decay phase. However, a 70-day data gap exists for SN 2020nlb, starting approximately 40 days after peak brightness due to Sun proximity. Optical photometry in $griz$ bands was obtained for both SN~2020nlb and SN~2020ue with the Panoramic Survey Telescope and Rapid Response System (Pan-STARRS) used in normal survey modes via the Young Supernova Experiment (YSE, \citealt{Jones2021, Aleo2023}; data accessed via \texttt{YSE-PZ} \citealt{YSEPZ1, Coulter2023}) and the Pan-STARRS search for kilonovae \citep{McBrien2021}. The Pan-STARRS1 (PS1) telescope is a 1.8-meter facility installed on top of the Haleakala on Maui, Hawaii, US \citep{Chambers2016}. It observes a 7.06 square degree field with a {\it grizy} SDSS-like photometric filter system. PS1 data are processed in real-time at the University of Hawaii where transient sources are selected \citep{Magnier2020}, and later on, filtered and classified by the Transient Science Server at Queen’s University Belfast \citep{Smith2020}. PS1 data for both SNe presented here are shown in Fig. \ref{fig:LC2020uenlb}.

 Additional optical and ultraviolet photometry of SN~2020ue and SN~2020nlb was obtained from multiple facilities. Optical observations were carried out with the Alhambra Faint Object Spectrograph and Camera (ALFOSC) at the Nordic Optical Telescope (NOT), the Asiago Faint Object Spectrograph and Camera (AFOSC) at the Schmidt 67/92 cm telescopes at Asiago, the Swope 1 m at Las Campanas, the Thacher 0.7 m, and the Las Cumbres Observatory (LCO) 1 m network, using various combinations of $uBVgri(z)$ filters. The data were reduced using dedicated employing PSF or forced photometry calibrated to the Sloan or Pan-STARRS systems. Ultraviolet observations were obtained with the Neil Gehrels \textit{Swift}/UVOT and reduced with \textsc{heasoft} standard procedures. Additional $BVR$ photometry for SN 2020nlb was collected with the Osservatorio Astronomico "Salvatore Di Giacomo" (OASDG) 0.5 m telescope and processed with an \texttt{Astropy}-based pipeline. A detailed description of the data reduction and analysis of each single dataset is provided in the Appendix A. The combined datasets is available online\footnote{\url{https://github.com/lucagrb/SN2020nlb_SN2020ue}} provide well-sampled light curves from peak brightness through late epochs.

\subsection{Spectroscopy}

\begin{figure*}
    \centering
    \includegraphics[width=0.49\linewidth]{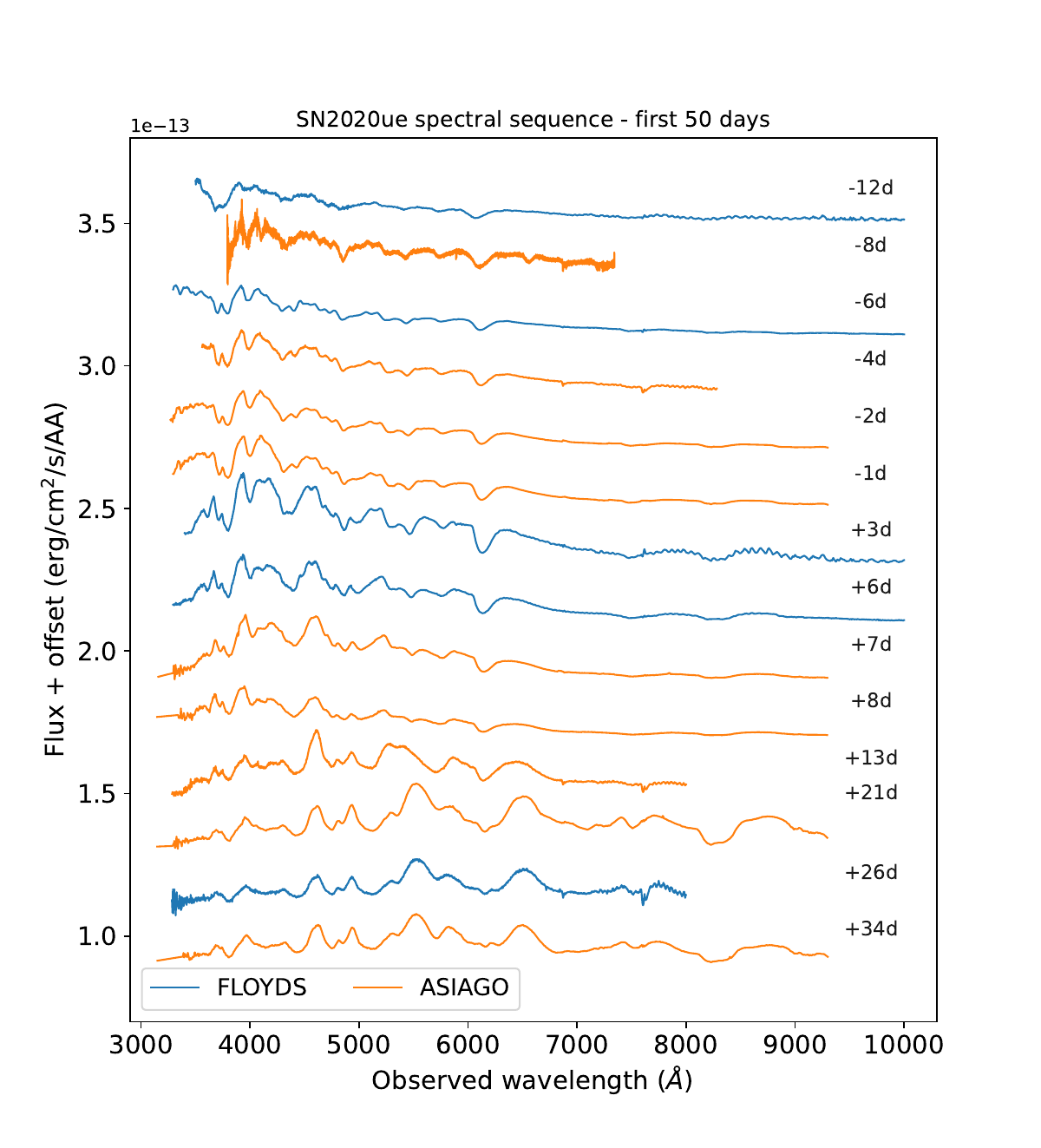}
    \includegraphics[width=0.49\linewidth]{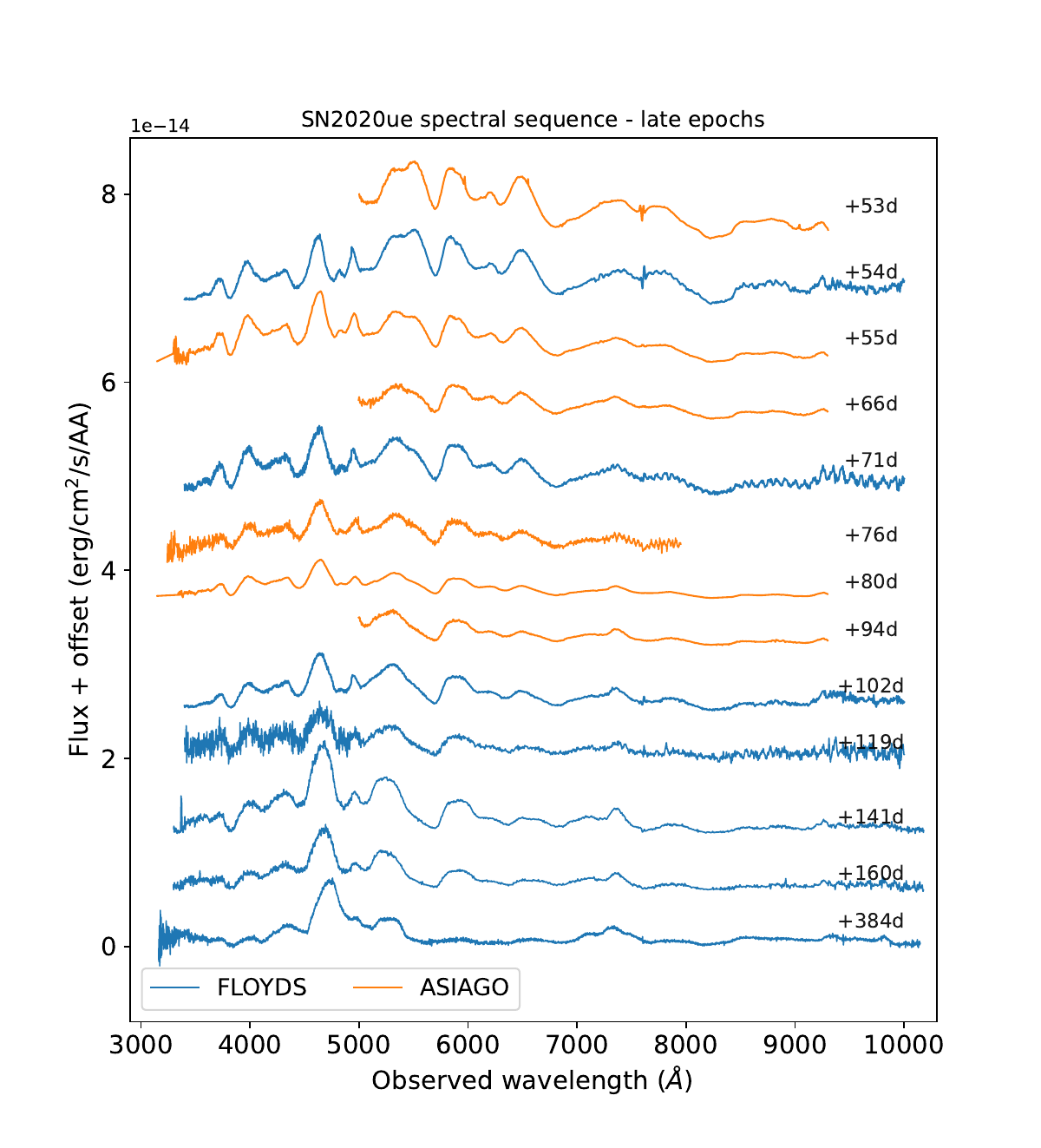}
    \caption{Optical spectral series of SN 2020ue. Spectra were obtained with the Copernico 1.82m and Galileo 1.22m in Asiago, and the FLOYDS telescopes during the first 50 days of the SN evolution (left panel) and up to 390 days from the $B$-band maximum (right panel).}
    \label{fig:2020uespec}
\end{figure*}

\begin{figure}
    \centering
    \includegraphics[width=0.99\linewidth]{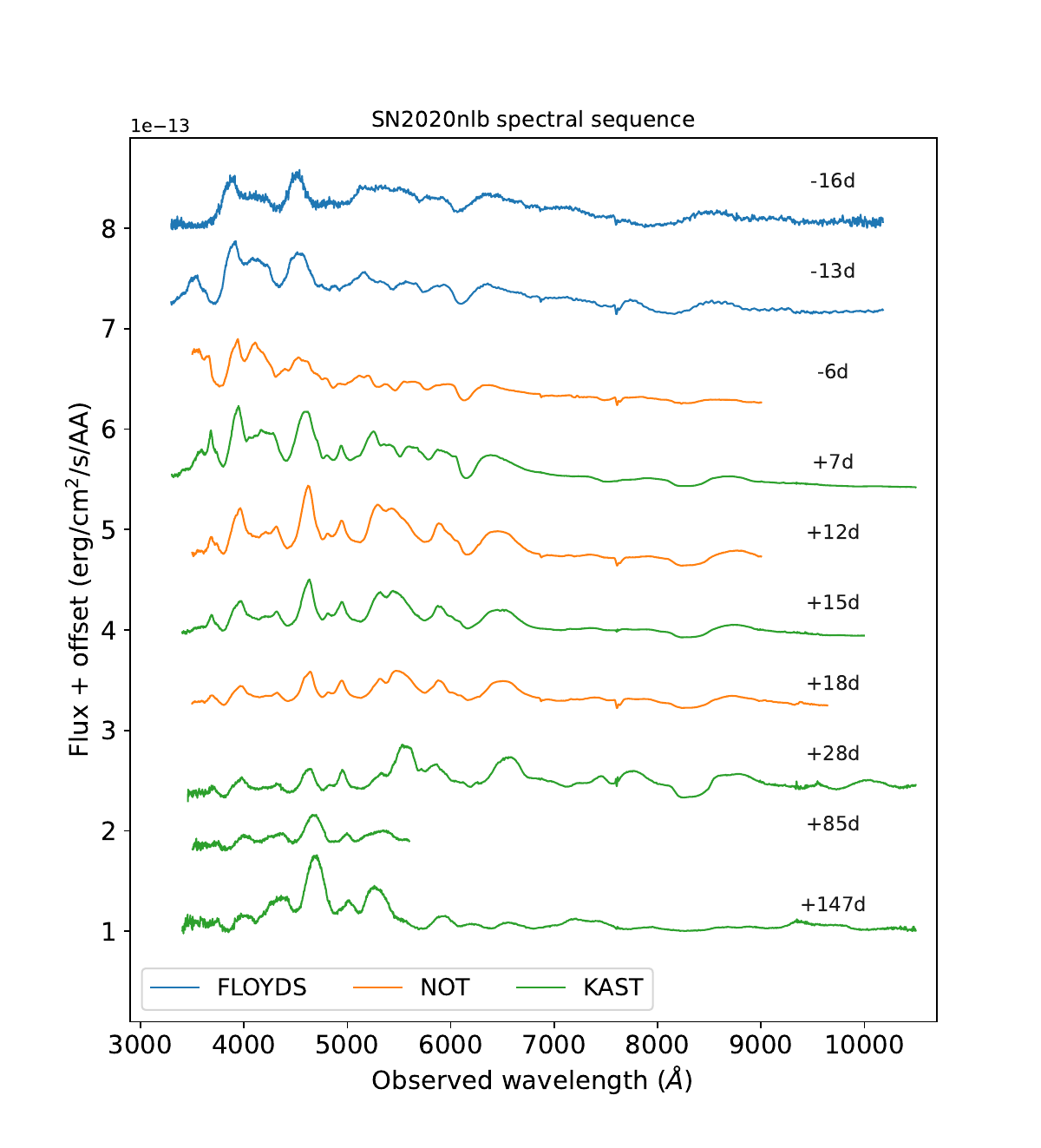}
    \caption{Optical spectral series of SN 2020nlb. Spectra were obtained with the NOT, the FLOYDS, and the KAST telescopes.}
    \label{fig:2020nlbspec}
\end{figure}

The entire spectral series, except for the high-resolution spectra obtained with the Lick/APF spectrograph, of both SNe discussed in this work are shown in Figs. \ref{fig:2020uespec} and \ref{fig:2020nlbspec}. We have obtained The spectral series covers a wide temporal range from $-$12 to 389 days for SN 2020ue, and from $-$16 to 273 days for SN 2020nlb. The phase is calculated with respect to the corresponding $B$-band maxima. The complete log of the spectroscopic observations is shown in Table \ref{tab:spexlog}. The initial flux calibration 
has been performed using spectro-photometric standard stars (Feige 34, Feige 56 and HR 3454, see also Appendix) observed each night. Thereafter, we have photometrically 'mangled' \citep{Hsiao2007} the spectra using available measured photometric data at each epoch. 

Spectroscopic observations of SN~2020ue and SN~2020nlb were obtained using a wide range of instruments and facilities covering optical to NIR wavelengths. For SN 2020nlb, five epochs were acquired with NOT/ALFOSC, reduced with the {\sc pyNot} pipeline. Both SNe were also observed with the FLOYDS spectrographs on the LCO network of telescopes. Extensive optical coverage of SN 2020ue was obtained at Asiago telescope with AFOSC and the echelle spectrograph. Additional spectra were collected with Low-Resolution Imaging Spectrograph (LRIS) at Keck telescope, Lick/Kast, and the Automated Planet Finder (APF) Levy spectrograph, providing both low- and high-resolution coverage. NIR spectra of SN 2020ue were secured with IRTF/SpeX, and integral-field observations for both SNe were obtained using WiFeS at Siding Spring Observatory. Finally, one ultraviolet spectrum of SN 2020ue was obtained with \textit{Swift}/UVOT. A detailed description of the data reduction and analysis of the spectral dataset obtained with each telescope - instrument pair is provided in the Appendix. 

\section{Photometric Analysis}
\label{s:photana}

\subsection{Peak brightness and intrinsic extinction}
\label{sec:3_1}

\begin{figure*}
    \centering
    \includegraphics[width=0.45\linewidth]{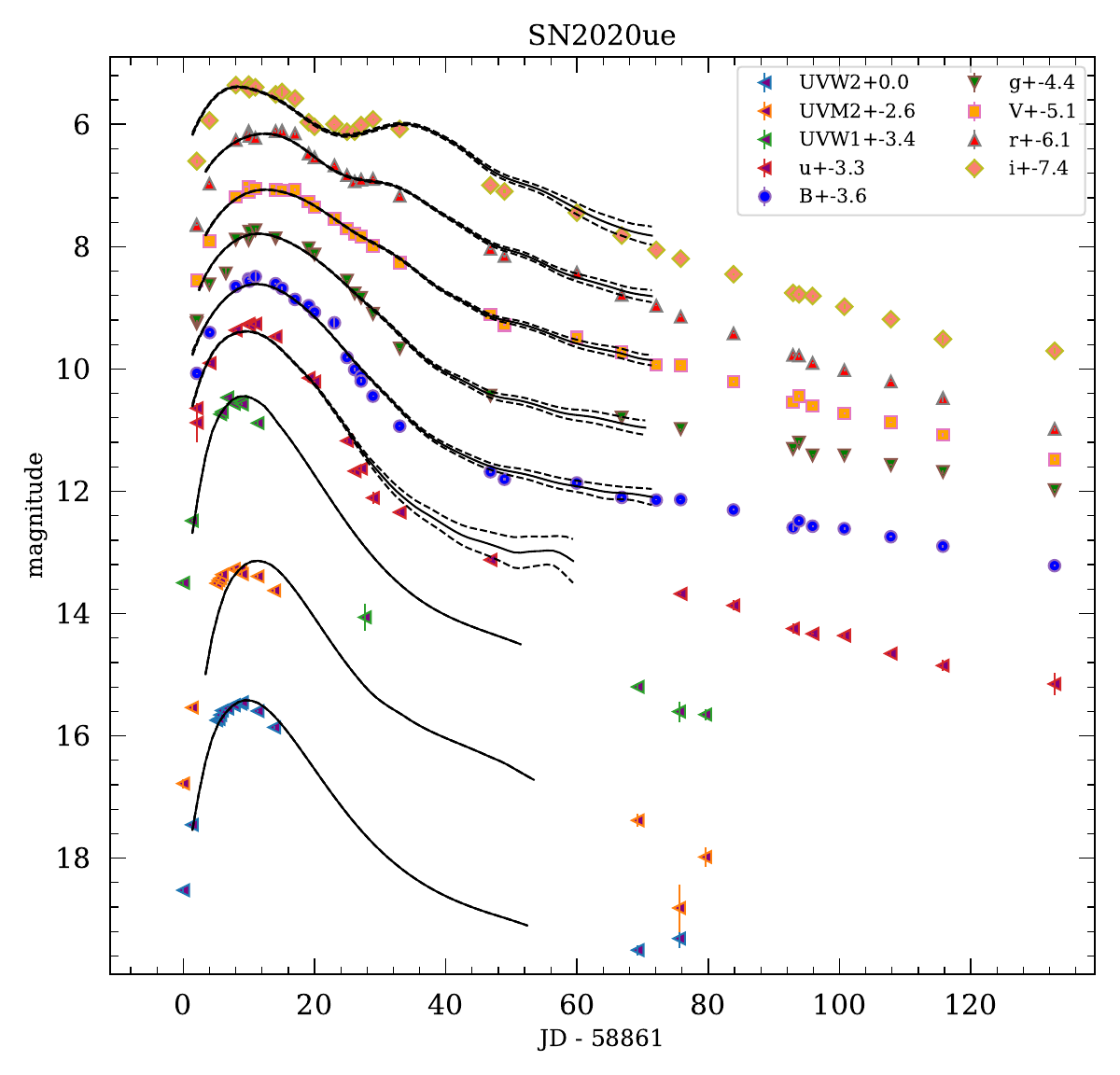}
    \includegraphics[width=0.46\linewidth]{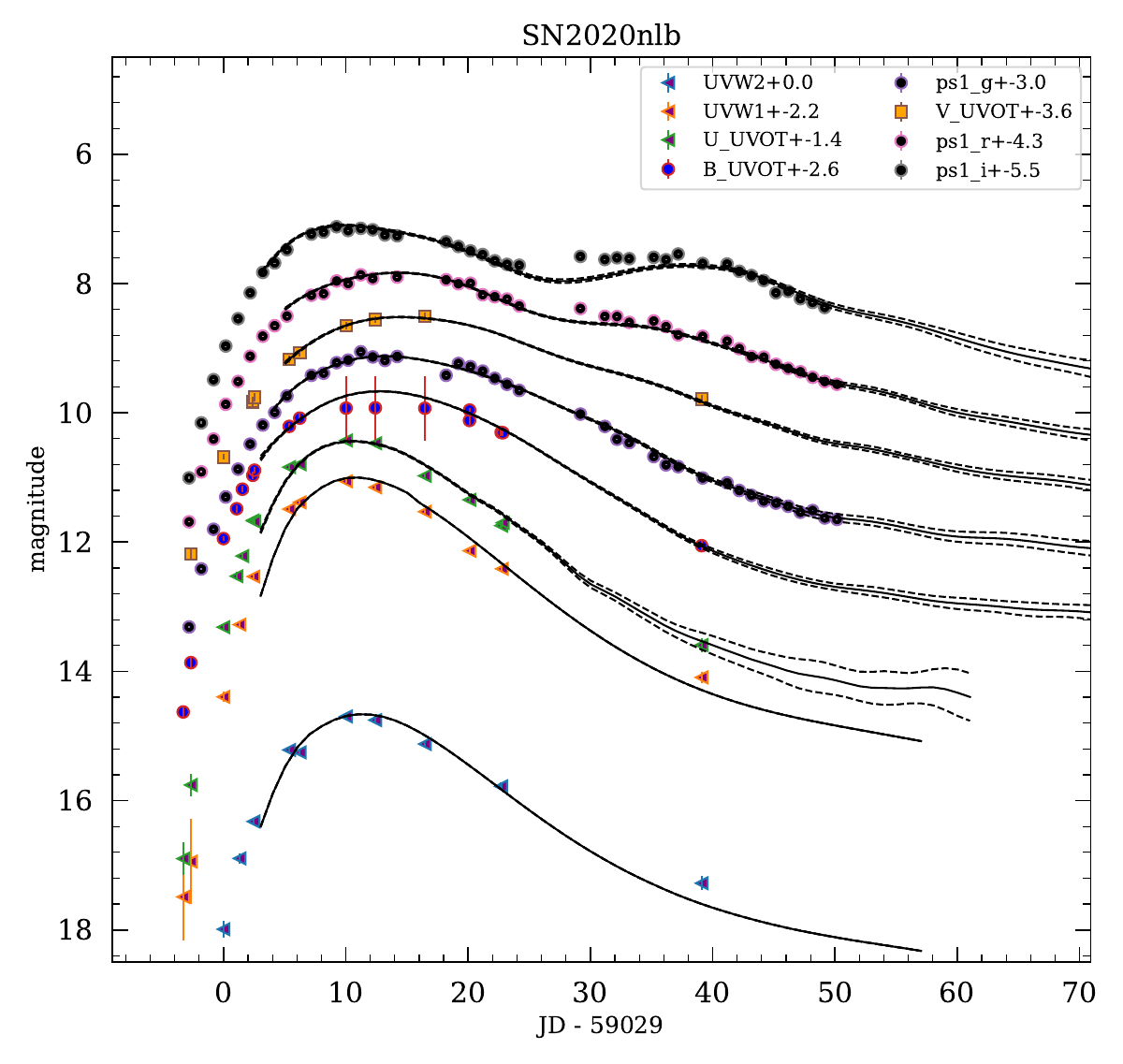}
    \caption{\textsc{SNooPy} fits (black curves) of {\it Swift}-UVOT and Pan-STARRS $gri$ data of SN 2020ue (left panel) and SN 2020nlb (right panel). The fits were obtained using the \texttt{max-model} light curve model function and the $s_{BV}$ stretch-color parameter. }
    \label{fig:SNooPyfit}
\end{figure*} 

We analyzed multi-filter light curves of SN 2020ue and SN 2020nlb using three methods: (1) the light-curve fitter \textsc{SNooPy} v2.5.3 \citep{Burns2011}, (2) the hierarchical Bayesian model \textsc{BayeSN} \citep{Mandel2022, Ward23}, and (3) a custom Gaussian process (GP) regressor algorithm \footnote{http://gaussianprocess.org/gpml/chapters/RW.pdf} implemented in {\sc scikit-learn} \citep{scikit-learn}. Figure \ref{fig:LC2020uenlb} shows multi-filter light curves of SN 2020ue and SN 2020nlb, both in days past $B$-band maximum, T$_{B,max}$, determined using \textsc{SNooPy}. This Python package analyzes and fits SN~Ia light curves in various photometric filters using templates and light curve behavior models. Prior to fitting, \textsc{SNooPy} corrects SN light curves for Galactic extinction and performs K-corrections based on \citet{Hsiao2007}. We estimated the $B$-band maximum using the \textit{max-model} and the color-stretch parameter $s_{BV}$, a dimensionless, normalized parameter representing the epoch past maximum light when the $B-V$ color curve peaks \citep{Burns2014}.  We also employed the $B$-band decay rate, $\Delta m_{15}(B)$, noting that our $B$-band maximum results are consistent with those from \citet{Sand2021} within uncertainties.

We used the entire {\it Swift}-UVOT dataset combined with Pan-STARRS $gri$ data for SN 2020nlb, and photometric data from NOT, Copernico 1.82m and Schmidt 67/92 Asiago telescopes for SN 2020ue to obtain light curve parameters (Table \ref{tab:1}). Using the {\it max-model}, we obtained the following results (with results for {\it EBV-model2} and $\Delta m_{15}$ decay rate parameter also reported in Table \ref{tab:1} for extinction estimation and SN bolometric light curve construction). For SN 2020ue, the $B$-band light curve peaks at $B_{max,20ue} = 11.982 \pm 0.011$ mag on $T_{B,max,20ue} = 58873.25$ MJD with color-stretch parameter $s_{BV} = 0.735 \pm 0.008$. For SN 2020nlb, the $B$-band light curve peaks at $B_{max,20nlb} = 12.164 \pm 0.094$ mag on $T_{B,max,20nlb} = 59042.12$ MJD with color-stretch parameter $s_{BV} = 0.923 \pm 0.009$.

We also employed a dedicated GP regressor algorithm to fit $B$ and $V$ band light curves separately, estimating $\Delta$m$_{15}$ without SN templates. Our GP kernel comprised three components: 1) a constant scaling kernel, 2) a rational quadratic kernel with length scale $l = 3$ and scale mixture parameter $\alpha = 3/2$, and 3) a white kernel representing white noise with variance $\sigma^2 = 1$. This method suits SN light curves with a temporal cadence near the length scale $l$. Similar to \textsc{SNooPy}, we corrected for Galactic extinction and applied K-corrections before fitting. Light curve parameters were measured using the predictor properties of the GP regressor function, including the stretch-color $s_{BV}$.

In Fig. \ref{fig:SNooPyfit}, we display the best-fit results for both SNe obtained using the GP method from \textsc{SNooPy}. In Appendix Fig. \ref{fig:LC2020uenlbGP}, we show the best-fit results using the GP regressor method developed in this work. We find the GP method produces a very good agreement with the \textsc{SNooPy} {\it max-model} fitting results that were obtained using the LCO light curve data for SN 2020ue and for SN 2020nlb using the OASDG and the Thacher telescope data (see Appendix Fig. \ref{fig:SNooPyfit_LCO_OASDG} and Table \ref{tab:app1}.

\begin{table*}
\centering
\caption{Results of the light curve fits made with \textsc{SNooPy}.}
\label{tab:1}
\begin{tabular}{lccccc}
\hline \hline
\multicolumn{6}{c}{SN 2020ue}\\
\hline
model parameter & \multicolumn{2}{c}{max-model} & \multicolumn{2}{c}{EBV-model2} & GP-model\\
\hline
$E(B-V)_{host}$ (mag) & - & - & $-0.105$ $\pm$ 0.013 & $-0.133$ $\pm$ 0.016 & $-0.002$ $\pm$ 0.001\\
$\mu$ (mag) & - & - & 31.139 $\pm$ 0.015 & 31.127 $\pm$ 0.015 & \\
$B_{max}$ (mag) & 11.982 $\pm$ 0.011 & 11.994 $\pm$ 0.010 & - & - & 12.079 $\pm$ 0.043\\
$V_{max}$ (mag) & 12.076 $\pm$ 0.009 & 12.067 $\pm$ 0.009 & - & - & 12.107 $\pm$ 0.060\\
$T_{B,max}$ (MJD) & 58873.25 $\pm$ 0.06 & 58873.21 $\pm$ 0.06 & 58873.27 $\pm$ 0.06 & 58873.23 $\pm$ 0.06 & 58872.11 $\pm$ 0.01\\
$s_{BV}$ & 0.735 $\pm$ 0.008 & -  & 0.731 $\pm$ 0.007 & - & 0.771 $\pm$ 0.067\\
$\Delta m_{15}(B)$ (mag) & - & 1.498 $\pm$ 0.012 & - & 1.509 $\pm$ 0.012 & 1.479 $\pm$ 0.060\\
    \hline
    \hline
\multicolumn{6}{c}{SN 2020nlb}\\
\hline
$E(B-V)_{host}$ (mag) & - & - & 0.113 $\pm$ 0.009 & 0.136 $\pm$ 0.010 & 0.102 $\pm$ 0.005  \\
$\mu$ (mag) & - & - & 30.961 $\pm$ 0.019 & 31.066 $\pm$ 0.021 & -\\
$B_{max}$ (mag) & 12.164 $\pm$ 0.094 & 12.181 $\pm$ 0.063 & - & - & 12.200 $\pm$ 0.042 \\
$V_{max}$ (mag) & 12.045 $\pm$ 0.118 & 12.112 $\pm$ 0.077 & - & - & 12.151 $\pm$ 0.027\\
$T_{B,max}$ (MJD) & 59042.12 $\pm$ 0.08 & 59041.96 $\pm$ 0.08 & 59042.35 $\pm$ 0.083& 59042.06 $\pm$ 0.09 & 59041.43 $\pm$ 0.01 \\
$s_{BV}$ & 0.923 $\pm$ 0.009 & - & 0.895 $\pm$ 0.010 & - & 0.829 $\pm$ 0.159\\
$\Delta m_{15}(B)$ (mag) & - & 1.089 $\pm$ 0.018 & - & 1.089 $\pm$ 0.021 & 1.245 $\pm$ 0.058\\
\hline
        \end{tabular}
\end{table*}

Our choice of the \textit{max-model} option in \textsc{SNooPy} is based on the estimation of the stretch-color parameter $s_{BV}$ and the cosmological application of the fit results. However, this model does not account for host galaxy extinction, focusing instead on fitting the maximum magnitude of each filter light curve \citep{Burns2014}. \textsc{SNooPy} offers an alternative approach: fitting the \textit{EBV-model2} before \textit{max-model}, which includes an additional free parameter for inferring host galaxy extinction, $E(B-V)_{host}$, and provides an estimate of the distance modulus $\mu$. The extinction curve parameter value for the host galaxy is fixed to $R_V = 3.1$. We obtained near-zero extinction for SN 2020ue and a significant extinction value of $E(B-V)_{20nlb}$ = 0.113 $\pm$ 0.009 mag for SN~2020nlb (with 0.060 mag systematic errors for both cases).

We used \textsc{BayeSN} for comparison, which models the continuous optical-to-NIR SED of SNe~Ia over time with intrinsic functional principal components scaled by dust extinction parameters. This latent reddened SN SED reconstruction enables quantifying dust extinction \citep{1999PASP..111...63F} (parameterized by total extinction in the $V$-band, $A_V$) and deriving the intrinsic SED's distance modulus. Fig.~\ref{fig:BAYESNfit} presents \textsc{BayeSN} model-fits for SN~2020nlb and SN~2020ue photometric datasets (as used with \textsc{SNooPy} in Fig.~\ref{fig:SNooPyfit}), including a corner plot of inferred parameters: distance modulus $\mu$, $V$-band absorption $A_V$, and light curve shape $\theta$ \citep{Mandel2022}.

Regardless of the fitting methods and models (\textsc{SNooPy} models, \textsc{BayeSN}), the inferred distance to SN 2020nlb is slightly lower than that of SN 2020ue. Individual distance estimates for both SNe, inferred using both methods, agree within statistical uncertainties. For SN 2020nlb, our findings align with \citet{Williams24}, who utilized \textsc{SNooPy}'s \textit{EBV-model2} for distance determination. Similar results emerge when using Pan-STARRS (PS1) and {\it Swift} data fitting results alone (Fig.~\ref{fig:SNooPyfit}). 

The extinction in the $V$-band parameters obtained with \textsc{BayeSN} suggests an environment for SN 2020nlb that is more attenuated ($A_V = 0.29 \pm 0.04$ mag) than for SN 2020ue ($A_V < 0.06$ mag, 95$\%$ cl.), although still in agreement with the results obtained using \textsc{SNooPy}. We have also used the empirical relation between the $B - V$ color at the SN maximum and the decay rate parameter $\Delta m_{15}(B)$ proposed by \citet{Phillips1999}. 
Adopting the $B-$ and $V-$band maximum magnitudes resulting from the \textsc{SNooPy} \textit{max-model} fits without prior extinction correction, we obtain $E(B-V)_{20ue} = -0.049 \pm$ 0.023 mag and $E(B-V)_{20nlb}$ = 0.140 $\pm$ 0.010 mag. This is consistent with the results from our other fitting methods and demonstrates that indeed, SN~2020ue is not affected by host galaxy dust extinction. All results of our \textsc{SNooPy} analysis are listed in Table \ref{tab:1}, where all the reported magnitudes have been corrected for Galactic extinction, $E(B-V)_{20ue,MW} = 0.024$ mag and $E(B-V)_{20nlb,MW} = 0.026$ mag.

\begin{table}
\centering
\caption{Summary of distance moduli (mag units) to the SN host galaxies. The systematic error from the \textsc{SNooPy} analysis has been included in parentheses for both supernovae.}
\label{tab:dist}
\begin{tabular}{lcc}
\hline \hline
method & NGC 4636 & NGC 4382 \\
 & (SN 2020ue) & (SN 2020nlb) \\
\hline
Cosmic-Flow 3 & 30.89$\pm$0.24 & 31.00$\pm$0.24 \\
FP      & 31.17$\pm$0.50 & 30.28$\pm$0.50 \\
SBF     & 30.80$\pm$0.13 & 31.22$\pm$0.07 \\
\hline
SNooPy & 31.13$\pm$0.02($\pm$0.12) & 30.96$\pm$0.02($\pm$0.12)\\
\textsc{BayeSN} & 31.02$\pm$0.10 & 30.86$\pm$0.11\\
    \hline
    \hline
        \end{tabular}
\end{table}

\subsection{Bolometric light curves}\label{sec:3_2}

Fig. \ref{fig:bolofits} displays estimated bolometric light curves for SN 2020ue and SN 2020nlb, constructed using \textsc{SNooPy}'s {\sc bolometric} function and {\sc direct} method, following \citet{Gall2018}. UV band light curves were built using GP spline functions within {\sc SNooPy}. A quasi-bolometric light curve was created through direct flux measurements from observed epochs, with spectral energy distributions interpolated via a `spline' fit method. Rayleigh-Jeans extrapolation based on UV and optical fluxes was used for the NIR range. We assumed zero host extinction for SN 2020ue (Sec. \ref{sec:3_1}), while $E(B-V) = 0.113$ mag was used for SN 2020nlb based on the EBV-model2 fit (Table \ref{tab:1}). The bolometric flux was placed on the absolute luminosity scale using distance moduli from \texttt{SNooPy} (Table \ref{tab:1}): $\mu = 31.139$ mag for SN 2020ue and $\mu = 30.961$ mag for SN 2020nlb. Light curve uncertainties were determined by iterating the process 50 times and averaging the results.

We employed the same GP regressor algorithm and kernel as in Section \ref{sec:3_1} to determine the peak bolometric luminosity, $L^{\mathrm{bol}}{\mathrm{peak}}$. For SN 2020nlb, the bolometric light curve peaked at $MJD_{peak}^{bol} = 59041.77 \pm 0.10$ days, occurring 0.41 $\pm$ 0.17 days before $T_{B,max}$ with a luminosity of $L_{peak}^{bol} = (9.96 \pm 0.21) \times 10^{42}$ erg s$^{-1}$. For SN 2020ue, the bolometric light curve peaked at $MJD_{peak}^{bol} = 58872.34 \pm 0.10$ days, 0.91 $\pm$ 0.12 days before $T_{B,max}$, with a luminosity of $L_{peak}^{bol} = (9.93 \pm 0.18) \times 10^{42}$ erg s$^{-1}$.

\subsection{SN explosion and rise-time to peak brightness}\label{sec:3_3}

For SN 2020nlb, the early high-cadence Pan-STARRS $g-$ and $r-$band observations, beginning 16.2 days before maximum light, together with stringent ATLAS upper limits obtained two days before discovery, provide a precise constraint on both the explosion time and the rise time. However, for SN 2020ue, our data coverage starts about 5 days later, at $-11.9$ days in $B-$ and $V-$ bands obtained from the Copernico 1.82m telescope at the Asiago Observatory. 
Thus, for SN 2020ue, we can only provide an estimate. 

We explored the applicability of an expanding fireball model for the very early emission of both SNe, where the SN luminosity evolution with time is proportional to the square of the explosion time, $L \propto t^{2}$ \citep{Riess1999}. This rise-time depends on factors such as the synthesized $^{56}$Ni amount \citep{Arnett1982,Shigeyama1992,Hoeflich1993} and its distribution within the ejecta. Progenitor WD metallicity, influencing the binding energy of the WD and ejecta expansion velocities \citep{Hoeflich2010}, is another factor affecting SNe~Ia rise-times. Distinct rise-times of $B$-band light curves have been determined for various SNe~Ia sub-types \citep{Riess1999, 2011MNRAS.416.2607G, GonzalezGaitan2012, 2015MNRAS.446.3895F}. Recently, a $t^2$ evolution was confirmed in a volume-limited sample of {\it ZTF} SNe~Ia \citep{Miller2020} but is not universally observed \citep{FausnaughTESS}.

\begin{figure}
    \centering
    \includegraphics[width=0.90\linewidth]{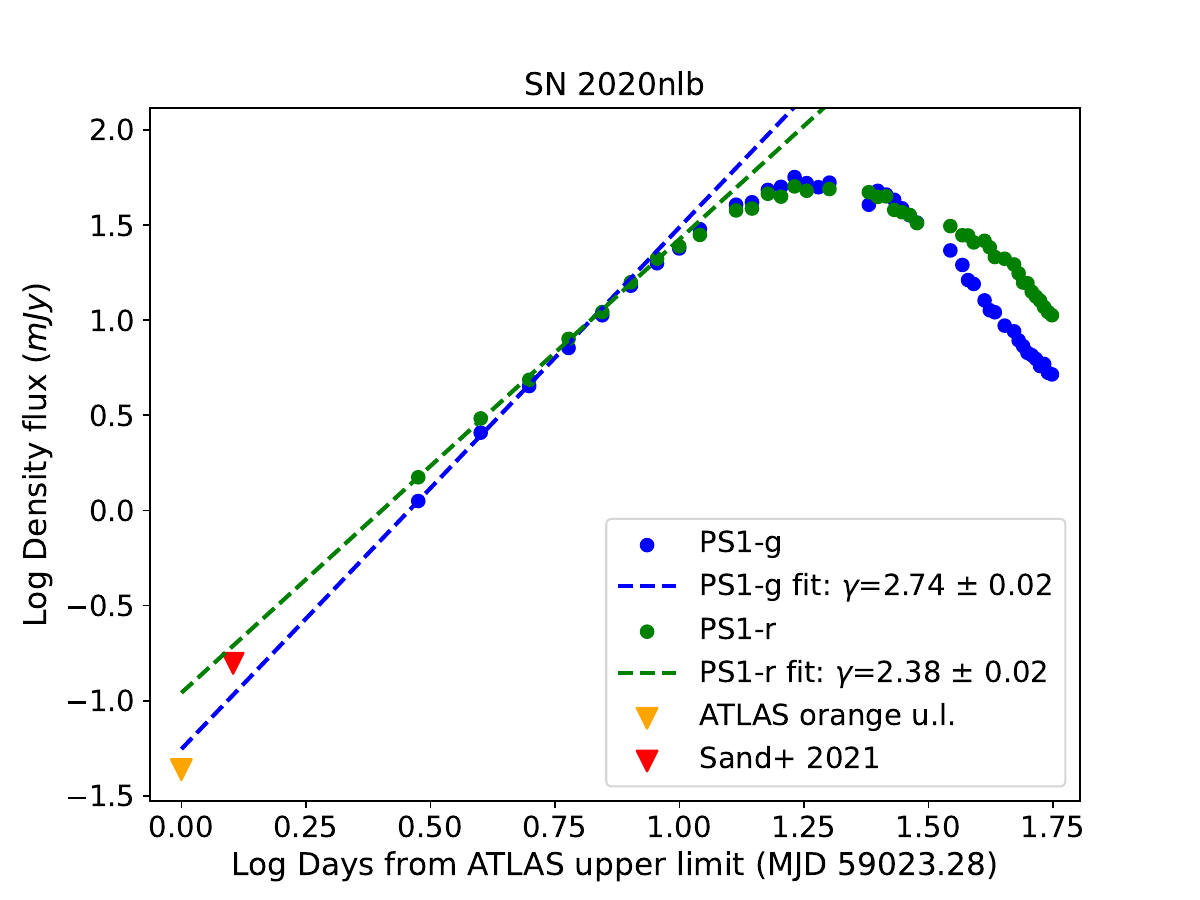}
    \includegraphics[width=0.90\linewidth]{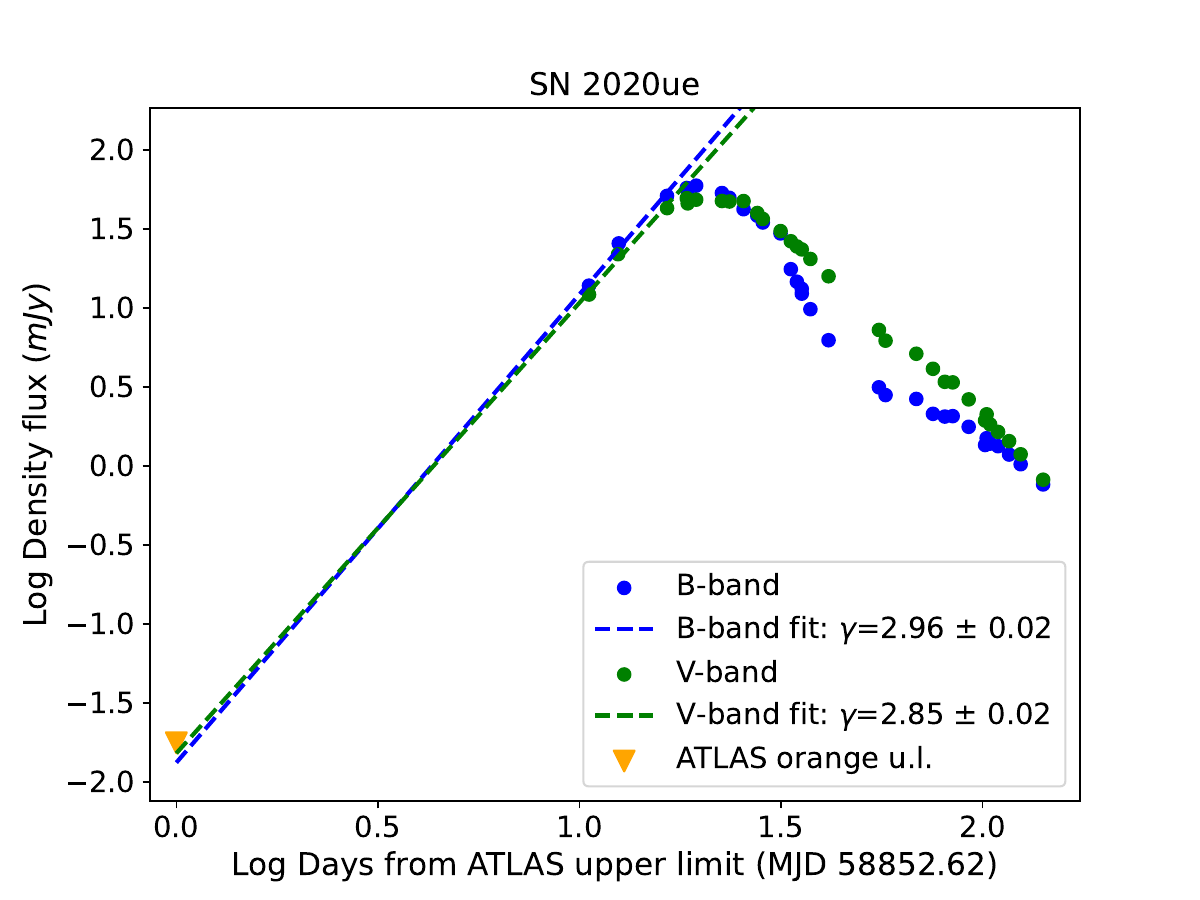}
    \caption{Pan-STARRS $g$- and $r$-band of SN 2020nlb (upper panel) and Copernico 1.82m B- and V-band SN~2020ue (lower panel) are shown, computed using the procedure outlined in Sec. \ref{sec:3_2}. The dashed lines represent the best fit considering a power-law model for the early light curve evolution \citep{GonzalezGaitan2012}. The most stringent ATLAS upper limit is indicated by an orange triangle, while an additional upper limit reported by \citet{Sand2021} 1.3 days after a red symbol marks the ATLAS observation. For SN 2020ue, the best-fit model extrapolation for the $r-$band aligns with these flux upper limits. }
    \label{fig:risefits}
\end{figure}

The fireball phase typically lasts about a week post-SN~Ia explosion, when the flux reaches 40--50\% of the SN~Ia peak brightness \citep{Miller2020}. For SN 2020nlb, we analyzed all $g-$ and $r-$band Pan-STARRS data between the last non-detection from ATLAS and up to 10 days after, and fit them using a power law model, $L \propto t^{\gamma}$. Our analysis shows that a power-law index of $\gamma_r = 2.38 \pm 0.02$ (statistical uncertainty only) effectively describes the initial rise to peak brightness. However, extrapolating the best-fit result to earlier times contradicts existing ATLAS upper limits (Fig. \ref{fig:risefits}), indicating an explosion time roughly 2 days before the last ATLAS non-detection, suggesting a discrepancy between inferred event onset and observational constraints. For the $g-$band, we obtained $\gamma_g = 2.74 \pm 0.02$ (stat. unc.), with the extrapolation consistent with non-detections at earlier times. Our findings are compatible with those of \citet{Williams24}, making SN 2020nlb a `single SN' Ia in the \citet{Hoogendam2024} rising light curve classification scheme.

An alternative approach to estimating the explosion time involves analyzing the evolution of the SN ejecta's photospheric velocity, well represented by the blue-shift minimum of the \ion{Si}{II} 6355 \AA\ absorption line. In a polytropic ejecta structure with an index $n=3$, the photospheric velocity evolution follows a power-law decay before peak brightness, with a power-law index of $\alpha = 0.22 \pm 0.02$ \citep{PiroNakar2014}. We measured the blue-shifted velocity of the \ion{Si}{II} 6355~\AA\ line for all available epochs around peak brightness for both SNe, as detailed in Section \ref{sec:4_2}. We then applied a Levenberg-Marquardt algorithm for non-linear fit minimization, as implemented in the {\it scipy} Python package, to fit the data series to a power-law model: $v(t) = v_0 \times (t - t_{exp})^{-0.22}$. Using this analysis, we determined the explosion time for both SNe via the \ion{Si}{II} 6355~\AA\ line's photospheric velocity evolution. For SN 2020nlb, our results indicate an explosion time of $MJD_{expl} = 59019.4 \pm 1.2$, occurring 2.6 days before the last epoch of non-detection by ATLAS. This estimate aligns with the results obtained before using the power-law fit of the rising light curve. For SN 2020ue, the explosion time is $MJD_{expl} = 58853.9 \pm 1.1$, consistent with non-detections and the early light curve analysis.

Deviations from the fireball value of $\gamma=2$ could have different explanations. Possible origins include enhanced circumstellar medium (CSM) density near the WD progenitor \citep{Moriya2023}, which can enhance and redden the early light curve, or an excess of $^{56}$Ni in the SN ejecta's outer region \citep{Magee2020,Ni2023a}. An early red color for SN 2020nlb is indeed evident in the $B-V$ evolution diagram of Fig. \ref{fig:Lira} (see also Sect. \ref{sec:3_3}). Early red excess has been observed in various SNe~Ia \citep{Dimitriadis+19,WangQ+21,WangQ+24,Ni2022}. In particular, SN~2020nlb shows strong similarities with early-red events such as SN~2021aefx \citep{Ni2023b} and KSN-2017iw \citep{Ni2025}, see Fig. \ref{fig:Lira}.  For SN 2020ue the paucity of data from the very early stages of the event prevents pinpointing the explosion time. 
Nevertheless, as shown in Figure \ref{fig:risefits}, the best-fit results for the power-law model reveal steep power-law indices for both $B$ and $V$ light curves. Extrapolations to earlier epochs are consistent with the constraints provided by our ATLAS non-detection.

\subsection{The ejected mass of $^{56}$Ni}\label{sec:3_4}

SNe~Ia emission is primarily driven by heating from radioactive $^{56}$Ni decay, synthesized in the thermonuclear explosion \citep{Pankey1962,Truran1967,Colgate1969}. Under simple initial assumptions, analytical models like that of Arnett \citep{Arnett1982} were developed to infer key physical properties, such as the ejected $^{56}$Ni mass, $M_{\mathrm{Ni}}$. Widely used in the literature, Arnett's model measures synthesized $^{56}$Ni in type I SNe from the time of the bolometric peak luminosity, assuming: 1) time-independent radiation energy density, partially limiting parameter estimate accuracy \citep{Pinto2000}, and 2) initial spherical symmetry. Recently, \citet{Khatami2019} presented an updated relation between bolometric peak luminosity and timing in SNe~Ia, considering the heating source's spatial distribution and non-constant opacity effects on the light curve. They introduced $\beta$, quantifying the degree of $^{56}$Ni mixing within the ejecta and how spatial distribution varies among SN types. For SNe~Ia, $\beta = 1.6$ was estimated. We applied both methods to infer the $^{56}$Ni mass of SN~2020nlb and SN~2020ue, using the updated formulation of \citet{Khatami2019} from \citet{Woosley2021}. 

Some SNe Ia discoveries occur shortly after the supernova explosion \citep{Nugent2011,Li2011b,Hosseinzadeh2017,Ashall2022}. For SN 2020nlb, both early light curve fitting and photospheric velocity methods suggest an explosion epoch about 2.6 days before the most constraining ATLAS upper limits. To validate these estimates, we used the analytical formulation for Type I supernova bolometric light curve evolution from \citet{Valenti2008}. This model relies on parameters such as total ejected mass, fraction of synthesized $^{56}$Ni, and ejecta kinetic energy, formulated in terms of photospheric velocity at peak brightness \citep{Valenti2008,Lyman2016}, provided as a fixed prior in our analysis. This method also enables explosion time estimation.
Additionally, we accounted for gamma-ray leakage correction, caused by gamma-ray photons from heavy element decay powering SN emission \citep{ClocchiattiWheeler}. Free parameters in this model include optical opacity $K$ and gamma-ray opacity $k_{\gamma}$. Gamma-ray leakage effects become noticeable $\sim$ a week after the SNe~I bolometric light curve peak. Our model only considered gamma-ray leakage from $^{56}$Co decay and positron annihilation, as high-energy photons from positron kinetic energy significantly impact SN light curves at $>$ 200 days post-explosion \citep{ClocchiattiWheeler,Tucker2022}. We performed a best-fit analysis of the bolometric light curve using the described analytic model, with final results shown in Table \ref{tab:Arnett} and Fig. \ref{fig:Arnett}.

\begin{figure}
    \centering
    \includegraphics[width=1.05\linewidth]{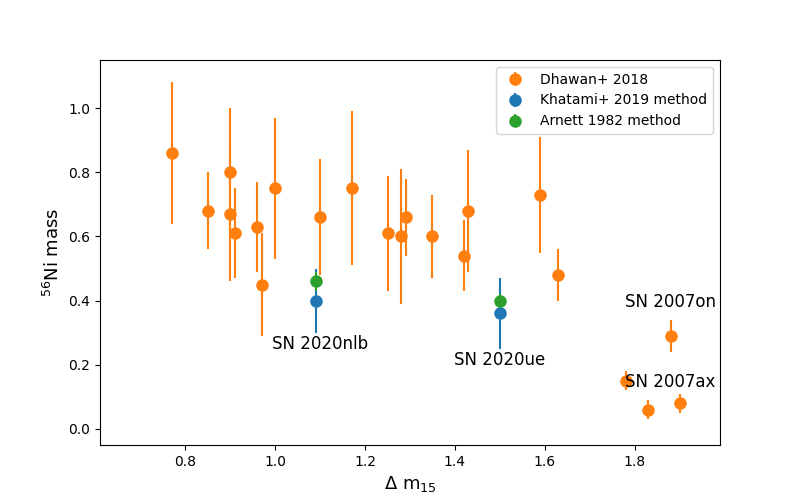}
    \caption{The distribution of $^{56}$Ni masses for SN 2020nlb and SN 2020ue as a function of $\Delta m_{15}$ was determined using both the \citet{Khatami2019} method (blue data) and the \citet{Arnett1982} analytical model (green data). For comparison, we included the Arnett model results (orange data) for the SNe~Ia sample from \citet{Dhawan2018}. The locations of transitional/fast-declining events SN 2007on \citep{Gall2018} and SN~2007ax \citep{Kasliwal2008} were highlighted.}
    \label{fig:Dhawan}
\end{figure}

The analytical Arnett model yields an SN 2020nlb explosion time consistent with the ATLAS upper limits, $MJD_{expl} = 59026.13 \pm 0.56$. Similarly, SN 2020ue's explosion epoch is $MJD_{expl} = 58858.77 \pm 0.58$, aligned with the ATLAS non-detection and previous estimates (Sec.~\ref{sec:3_3}). For $^{56}$Ni mass, we obtain $M{^{56}Ni} = 0.40 \pm 0.10$ M$_{\odot}$ for SN 2020nlb and $M{^{56}Ni} = 0.36 \pm 0.11$ M$_{\odot}$ for SN 2020ue. 
These $^{56}$Ni masses are consistent within uncertainties  and indicate that the self-similar energy density assumption by \citealp{Arnett1982} is suitable for modeling normal SNe~Ia \citep{Blondin2013}.
Our findings align with literature expectations \citep[e.g.,][]{Stritzinger2006, 2007ApJ...656..661K, Dhawan2018}, where SNe~Ia with lower $\Delta m_{15}$ values exhibit larger ejected $^{56}$Ni masses. As the bolometric peak luminosity of SNe~Ia is driven by the synthesized $^{56}$Ni amount, this relation offers an alternative interpretation of the Phillips relation \citep{Stritzinger2006}. Fig. \ref{fig:Dhawan} shows the distribution of $^{56}$Ni mass vs. $\Delta m_{15}$ for our studied SNe, alongside other SNe~Ia \citep{Dhawan2018}. Both SN 2020ue and SN 2020nlb occupy the region of normal SNe~Ia. The lower right corner hosts SNe~Ia with low $^{56}$Ni mass ($< 0.3$ M$_{\odot}$), including transitional events like SN 2007on \citep{Gall2018} and fast-decliner SNe such as SN~2005ke, SN 2007mr, and SN 2007ax \citep{Dhawan2018}. 

\begin{table}
\centering
\caption{Physical properties obtained from the best-fit analysis using the analytical model used for fitting the bolometric light curves of SN 2020nlb and SN 2020ue. $k$ and $k_{\gamma}$ are the optical and gamma-ray opacity \citep{Valenti2008}.}
\label{tab:Arnett}
\begin{tabular}{lcc}
\hline \hline
 & SN 2020ue & SN 2020nlb \\
\hline
$MJD_{expl}$ (days) & 58858.77$\pm$0.58 & 59026.12$\pm$0.56 \\
$MJD_{peak}^{bolo}$ (days) & 58872.34$\pm$0.10  & 59041.77$\pm$0.10\\
$L_{peak}^{bolo}$ (10$^{42}$ erg s$^{-1}$)    & 9.93$\pm$0.18 & 9.96$\pm$0.21 \\
$M_{ej}$ (M$_{\odot})$ & 1.40$\pm$0.06 & 1.70$\pm$0.12\\
$M_{^{56}Ni}$ (M$_{\odot})$ & 0.40$\pm$0.02 & 0.43$\pm$0.03 \\
$K.E.$ (10$^{51}$ erg) &1.20$\pm$0.05 & 1.24$\pm$0.04 \\
$k$ & 0.08$\pm$0.01& 0.08$\pm$0.01 \\
$k_{\gamma}$ & 0.02$\pm$0.01&0.03$\pm$0.01 \\
    \hline
    \hline
        \end{tabular}
\end{table}

\subsection{Color evolution}\label{sec:3_5}

\begin{figure}
    \centering
    \includegraphics[width=1.05\linewidth]{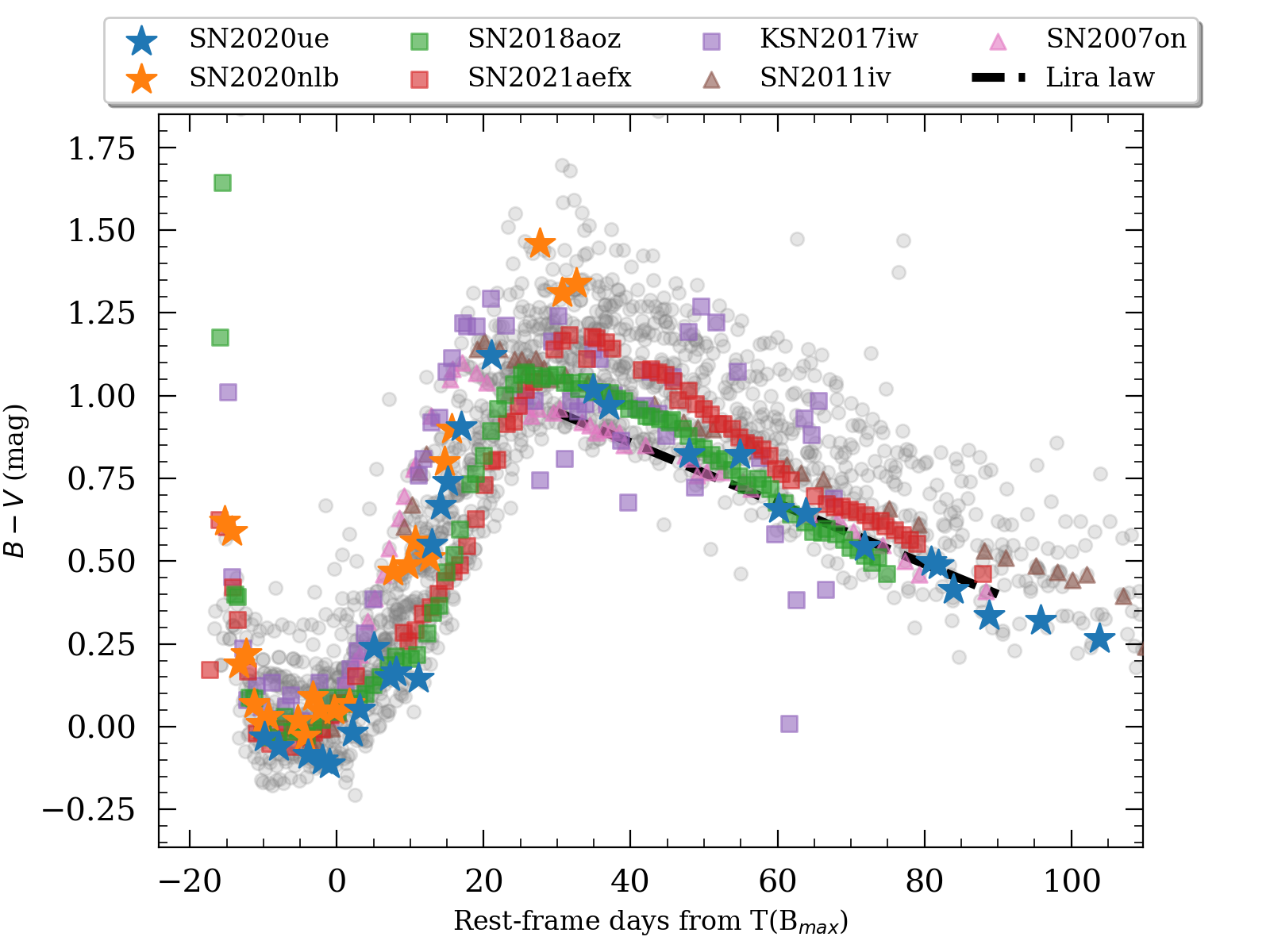}
    \caption{The temporal $(B-V)$ color evolution is presented, featuring SN 2020ue (blue stars) and SN 2020nlb (orange stars) alongside a sample of nearby SNe~Ia taken from \citet{Riess2016} and \citet{Khetan2021} (gray data), transitional SNe~Ia (SN 2007on - red stars, and SN 2011iv -green stars, \citet{Gall2018}), and  early-time red SNe~Ia (KSN 2017iw - violet stars, SN 2018aoz - green stars, and SN 2021aefx - red stars, \citet{Ni2023b,Ni2025}). The black dashed line indicates the Lira-law regime \citep{Phillips1999}.} 
    \label{fig:Lira}
\end{figure}

Fig. \ref{fig:Lira} presents the $B-V$ color evolution of SN 2020nlb and SN~2020ue compared to nearby calibrator SNe~Ia \citep{Riess2016,Khetan2021}. The $B-V$ evolution for SNe was computed using \texttt{SNooPy}'s dedicated built-in function, requiring a defined light-curve fit model and decay rate parameter. Fig.~\ref{fig:Lira} also includes the color evolution of transitional high-$\Delta m_{15}$ SNe~Ia SN 2007on and SN 2011iv \citep[][see Sec.~\ref{s:intro}]{Gall2018}.
SNe~Ia typically exhibit their bluest colors $\sim-10$ days before B-maximum, then abruptly turn redder, reaching the reddest colors $\sim$10 to $\sim$40 days post-maximum. Subsequent evolution features a linear decline towards bluer colors, with a similar decline rate for all SNe~Ia. Termed the `Lira-law' \citep{1996MsT..3L}, this behavior has been employed to estimate host galaxy extinction \citep{Phillips1999, 2010AJ....139..120F}, assuming SNe following the Lira-law (Fig. \ref{fig:Lira}'s dashed black line) experience low to zero extinction. Thus, color offsets from the Lira-law can be attributed to host galaxy extinction.

Around $B-$band maximum, SNe~Ia colors are confined within a small $B-V$ color range, while larger color variations emerge at later epochs (Fig.~\ref{fig:Lira}). Unfortunately, SN 2020nlb data beyond 40 days past $B-$band maximum are unavailable due to observational constraints. However, our data indicate that SN 2020nlb peaked at $B-V \sim 1.45$ mag around 25 days post-$B-$band maximum, confirming light curve fit results (Sec.~\ref{sec:3_1}, Table~\ref{tab:1}) that SN~2020nlb experiences significant host galaxy extinction. 

SN 2020ue's color evolution is comparable to transitional SNe~Ia SN 2007on and SN 20211iv, attaining $B-V$ peak magnitude $\sim$17-20 days post-$B-$band maximum. Intriguingly, SN 2020ue exhibits the bluest color at $B-$band maximum of all highlighted transitional and fast-declining SNe in Fig.\ref{fig:Lira}. However, its $B-V$ color evolution past $B-V$ peak in the Lira-law regime (30-90 days) aligns with SN 2007on. As discussed in Sec. \ref{s:intro}, $B-V$ color variations may be intrinsic, originating from physical differences in the progenitor WD system, such as distinct central densities of progenitor WDs \citep[e.g.,][]{Hoeflich2017, Gall2018} or explosion asymmetries \citep[e.g.,][]{Maeda2011, Foley2011}. 

\section{Spectral evolution}
\label{s:specana}\label{sec:4_1}

Both SNe display characteristic spectral features of SNe~Ia \citep[see, e.g.,][]{Morrell2024}. Our extensive spectral coverage, particularly near the SN explosion, enables analysis of the absorption line evolution from $\sim -$10 days to peak brightness. Key insights can be gleaned from line velocity and intensity evolutions, revealing information on the ejecta's stratified composition and unburned elements like carbon and oxygen (though oxygen is also a carbon-burning product). Although oxygen is commonly observed in optical SN~Ia spectra, detecting carbon remains challenging.
Carbon presence in optical SN~Ia spectra has been reported in early ($< 10$ days pre-peak brightness) spectra of numerous ($\sim$30-40\%) SNe~Ia \citep[see, e.g.,][and references therein]{Branch2003,Thomas2007,Thomas2011,Parrent2011,Folatelli2012,Silverman2012,Maguire2014,Dutta2021,Zeng2021}. Detecting carbon signifies incomplete burning, with crucial implications for interpreting the progenitor star's nature and the mechanisms driving element mixing in the ejecta \citep{Nomoto1984,Iwamoto1999,Folatelli2012}.

\begin{figure*}
    \centering
    \includegraphics[width=0.49\linewidth]{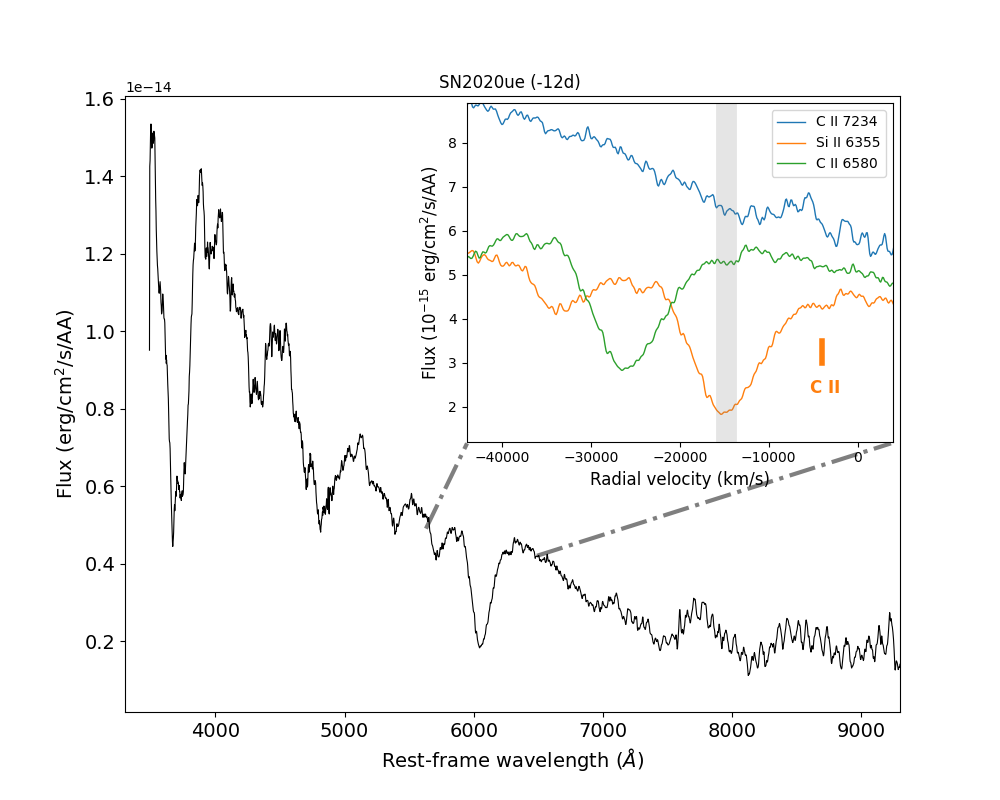}
    \includegraphics[width=0.49\linewidth]{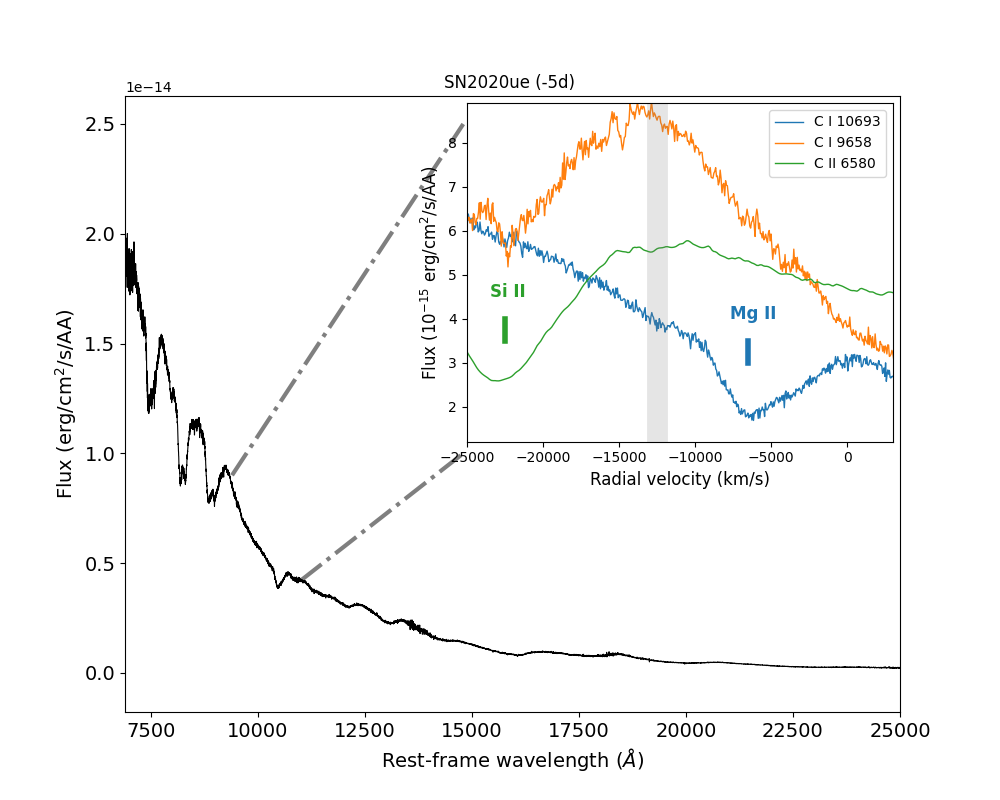}
    \caption{Spectrum of SN 2020ue at $-$12 days (left panel) and $-$5 days (right panel). The left panel upper inset shows the radial velocity spectrum centered on the \ion{Si}{II} 6355 \AA\, \ion{C}{II} 6580 \AA\ and \ion{C}{II} 7234~\AA, and the right panel upper inset shows carbon lines present in the optical and NIR spectra are shown. The gray strip marks the epoch expansion velocity of the \ion{Si}{II} 6355 \AA\ line as measured with \texttt{spextractor}. 
    }
    \label{fig:carbon2020ue}
\end{figure*}

The primary optical feature \ion{C}{II} 6580~\AA\, often blends with the prominent \ion{Si}{II} 6355~\AA\ line, making line identification challenging \citep{Folatelli2012}. When the \ion{Si}{II} line is intense, the upper level (3p-3d) \ion{C}{II} 7234~\AA\ transition can be detected as a faint line. Since un-burned elements only persist in the stratified ejecta's outermost layers, \ion{C}{II} 6580~\AA\ shows higher velocities than other absorption lines at similar epochs, like the nearby \ion{Si}{II} 6355~\AA\ line \citep{Fisher1997,Mazzali2001,Heringer2019}. However, carbon has also been detected at lower velocities in some instances \citep{Maguire2014}, implying that carbon often blends with the silicon line's red wing in many high- and normal-velocity SNe~Ia (see \citealt{Parrent2011,Folatelli2012} for discussion).
Based on this, we identify possible \ion{C}{II} 6580~\AA\ in the earliest spectrum (phase $-12$ d) of SN 2020ue, observed as a faint dip at the \ion{Si}{II} 6355~\AA\ line's red wing edge. This feature remains visible with reduced intensity until $-$4 days, displaying a similar expansion velocity ($\sim$15,000 km/s) to the \ion{Si}{II} line (Fig. \ref{fig:carbon2020ue}). 
Conversely, neither \ion{C}{II} 7234 \AA\ nor \ion{C}{II} 6580 \AA\ are visible in early SN 2020nlb spectra. This is unsurprising, as \ion{C}{I} detection is more likely in SNe~Ia with bluer peak colors, faster-declining light curves, and lower \ion{Si}{II} velocities \citep[e.g.,][]{Thomas2011, Silverman2012, Folatelli2012}. Strong \ion{C}{I} detection in pre-maximum NIR spectra of transitional SNe~Ia has been noted \citep[e.g.,][]{2002ApJ...568..791H,Hsiao2015,2021ApJ...914...57W}, and for more luminous, normal SNe~Ia, a \ion{C}{I} ``knee'' feature may be present \citep[e.g.,][]{Hsiao2013, Hoogendam2025}. Given this, the early \ion{C}{II} detection and absence of carbon features at day $-5$ in optical and near-IR wavelengths argue against a possible "transitional" nature for SN~2020ue.

\subsection{Pseudo-equivalent widths}
We quantified spectral properties by measuring the pseudo-equivalent width (pEW), Doppler velocity, and line depth of prominent absorption features using a modified version of the publicly available code \texttt{Spextractor} \citep{Papadogiannakis2019,Burrow2020}. It employs Gaussian process regression to fit the analyzed spectrum, then estimates pEWs for given absorption lines by identifying their flux minima (see \citealp{Burrow2020} for details).
We focused on absorption lines commonly observed in SN~Ia spectra, including \ion{Ca}{II} H\&K, \ion{Si}{II} 4130 \AA, \ion{Mg}{II} 4481 \AA, \ion{Fe}{II} 5062 \AA\ triplet, \ion{S}{II} 5468/5624 \AA\ doublet, \ion{Si}{II} 5972 \AA, \ion{Si}{II} 6355 \AA, and \ion{O}{I} 7775 \AA. Some lines blend with fainter transitions of other elements (e.g., \ion{Si}{II} 4130 \AA\ blends with \ion{Ti}{II} 4200 \AA\ around peak brightness, typically observed for fast-decliner events, \citealp{Taubenberger2017}).

We included an essential constraint in the prior distributions of each absorption line considered in the fit: all lines should have a similar offset for the low and high-velocity range defining each absorption line. This is particularly applicable to absorption lines not predominantly affected by blends up to a few days past peak brightness (e.g., \ion{Si}{II} 5972 \AA\ begins being affected by \ion{Na}{I} five days post-peak brightness), while for multiplets like \ion{Fe}{II}, we considered a wide range due to multiple components.
To establish absorption line boundaries, we considered the blue and red wavelength maxima of the absorption trough for \ion{Si}{II} 6355 \AA\, as it is least affected by blending with other absorption lines. This accounts for the weak \ion{C}{II} 6580 \AA\ presence in SN 2020ue, which minimally impacts the pseudo-continuum and pEW value.
Fig. \ref{fig:Spex_example} demonstrates the best-fit result from our \texttt{Spextractor} for SN 2020ue's $-$1 day spectrum.
Our analysis focuses on spectra from the early rising phase up to approximately 8 days post-$B$-band peak maximum. Table \ref{tab:spex} summarizes the measurements for SN 2020ue and SN 2020nlb.

\begin{figure}[ht!]
    \centering
    \vspace{-0.4 cm}
    \advance\leftskip-0.5cm
    \includegraphics[width=0.98\linewidth]{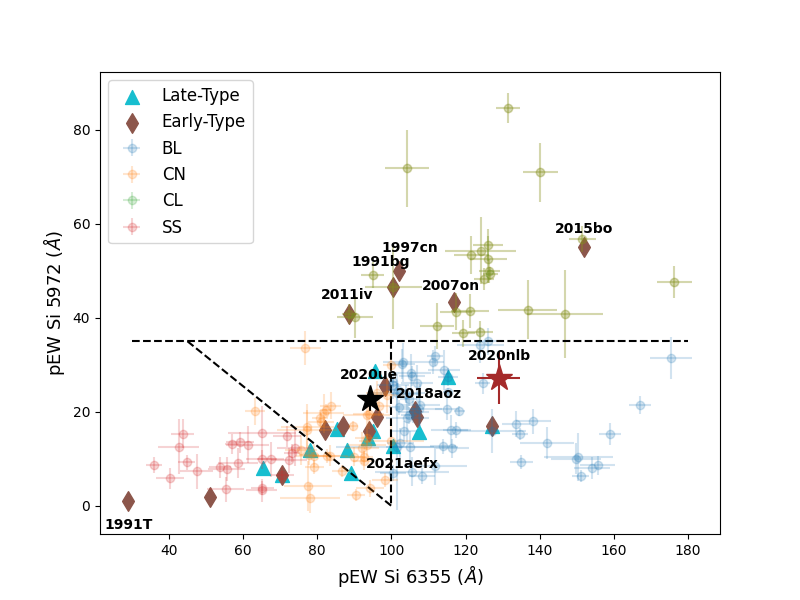}
     \vspace{-0.4 cm}
    \caption{The Branch spectroscopic classification scheme. SNe~Ia classes are shown in various colors: {\it shallow silicon} (SS, red), {\it broad line} (BL, blue), {\it cool} (CL, green), and {\it core normal} (CN, orange). Dashed lines separate these regions \citep{Phillips2022}. SNe~Ia in late-type \citep[Cepheid calibrators]{Riess2016} and early-type galaxies \citep[SBF calibrators]{Khetan2021} are shown as blue triangles and brown diamonds, respectively. SN 2020ue and SN 2020nlb are highlighted as black and red stars. Transitional and fast-declining SNe~Ia such as SN 1991bg, 1997cn, 2007on, 2011iv, and 2015bo, as well as early red events like 2018aoz and 2021aefx, are also included \citep{Gall2018,Hoogendam2022,Ni2023a,Ni2023b}.
    }
    \label{fig:Branch}
\end{figure}

Absorption line properties around peak brightness in SNe~Ia have been used to identify common patterns \citep{Nugent1995,Benetti2005,Bailey2009,Wang2009,Morrell2024}. We focus on the \citet{Branch2006} classification scheme (`Branch-classification'), which categorizes SNe~Ia into four spectroscopic sub-classes based on the pEWs of \ion{Si}{II} 5972/6355~\AA\ lines: {\it cool} (CL), {\it core normal} (CN), {\it shallow silicon} (SS), and {\it broad line} (BL) SNe.
To assess how SN 2020ue and SN 2020nlb fit within this classification, we utilize the \citet{Burrow2020} data set for comparison, as spectroscopic properties were measured using the same method as in our work. We consider pEW values for SN 2020ue from the $-$1 day spectrum, and for SN 2020nlb, the closest spectrum to $B$-band peak available at $-$6 days.
Fig. \ref{fig:Branch}'s top panel displays our two SNe's pEW measurements in the `Branch-plot'. Additional SNe~Ia observed in late-type and early-type galaxies are included as they were employed as Cepheid \citep{Riess2016} and/or SBF \citep{Khetan2021} calibrators of SN luminosity relations.

Firstly, we observe that SN 2020ue and SN 2020nlb are situated within the CN and BL regions, respectively. This positioning aligns with most calibrator SNe included in this analysis, distributed between these two classes. A few exceptions exist: SNe~Ia that belong to other spectral classes but were still included in calibrating samples. These include SN 1991T, and the transitional events SN 2007on and SN 2011iv.
SN 1991T, located at the extreme edge of the SS sub-class, is the prototype of a subclass of SNe Ia characterized by high-luminosity and slowly-evolving light curve behavior, accompanied by pre-maximum weak \ion{Si}{ii} and strong \ion{Fe}{ii} lines \citep{Filippenko1992,Phillips1992}. The two transitional events were considered SBF SN calibrators in \citet{Khetan2021} and represent the sole calibrating SNe~Ia within the CL sub-class, which also contains the fast-decliner prototype, SN 1991bg \citep{Leibundgut1993}. SNe~Ia in the CL-class are located in the region where we find type Ia SNe characterized by a very intense \ion{Si}{II} 5972~\AA\ line compared to those in other spectroscopic sub-classes. 
CL SNe photometrically differ from most SNe~Ia in other sub-classes, becoming evident when examining the distribution of \ion{Si}{II} 5972~\AA\ pEW and $R(Si) = pEW(\ion{Si}{II}~5972)/pEW(\ion{Si}{II}~6355)$ relative to $\Delta$m$_{15}$. Both the \ion{Si}{II} 5972 \AA\ pEW and \ion{Si}{II} pEW line ratio consistently demonstrate a linear relationship with SNe~Ia light-curve decay rates across various SN~Ia samples \citep{Hachinger2006,Taubenberger2008,Hachinger2008,Blondin2012,Zhao2021}. The tightest correlation emerges for \ion{Si}{II} 5972 \AA\ pEW (see Fig. \ref{fig:Branch}'s lower panel), causing a clear separation of CL SNe~Ia from BL, CN, and SS sub-classes in the upper-right region of Fig. \ref{fig:Branch} (middle and lower panels). This region is characterized by high \ion{Si}{II} 5972~\AA\ pEWs and elevated $\Delta$m$_{15}$ ($>$ 1.5) values. This distinction is noticeable for transitional SN~2007on and SN 2011iv, and fast-decliner SN~1991bg, markedly separated from other calibrator SNe. Intriguingly, despite SN 2020ue being a `border-line' event with $\Delta$m$_{15}$ values close to SN 2011iv and strikingly similar color evolution as SN 2007on, its spectroscopic behavior differs from transitional events. Conversely, SN 2020nlb exhibits typical spectroscopic evolution of normal SNe~Ia.

\begin{figure}
    \centering
    \includegraphics[width=0.99\linewidth]{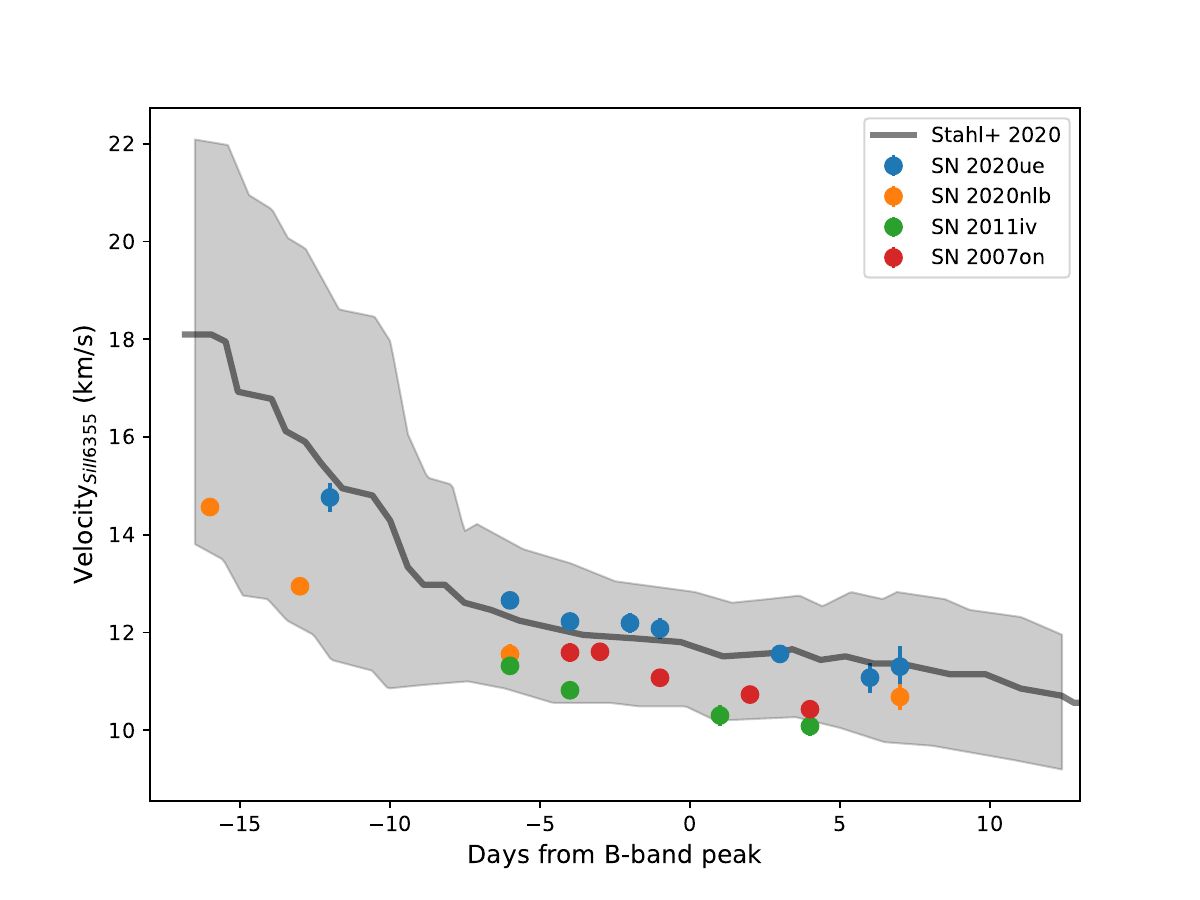}\\
    \vspace{-5mm}
    \includegraphics[width=0.99\linewidth]{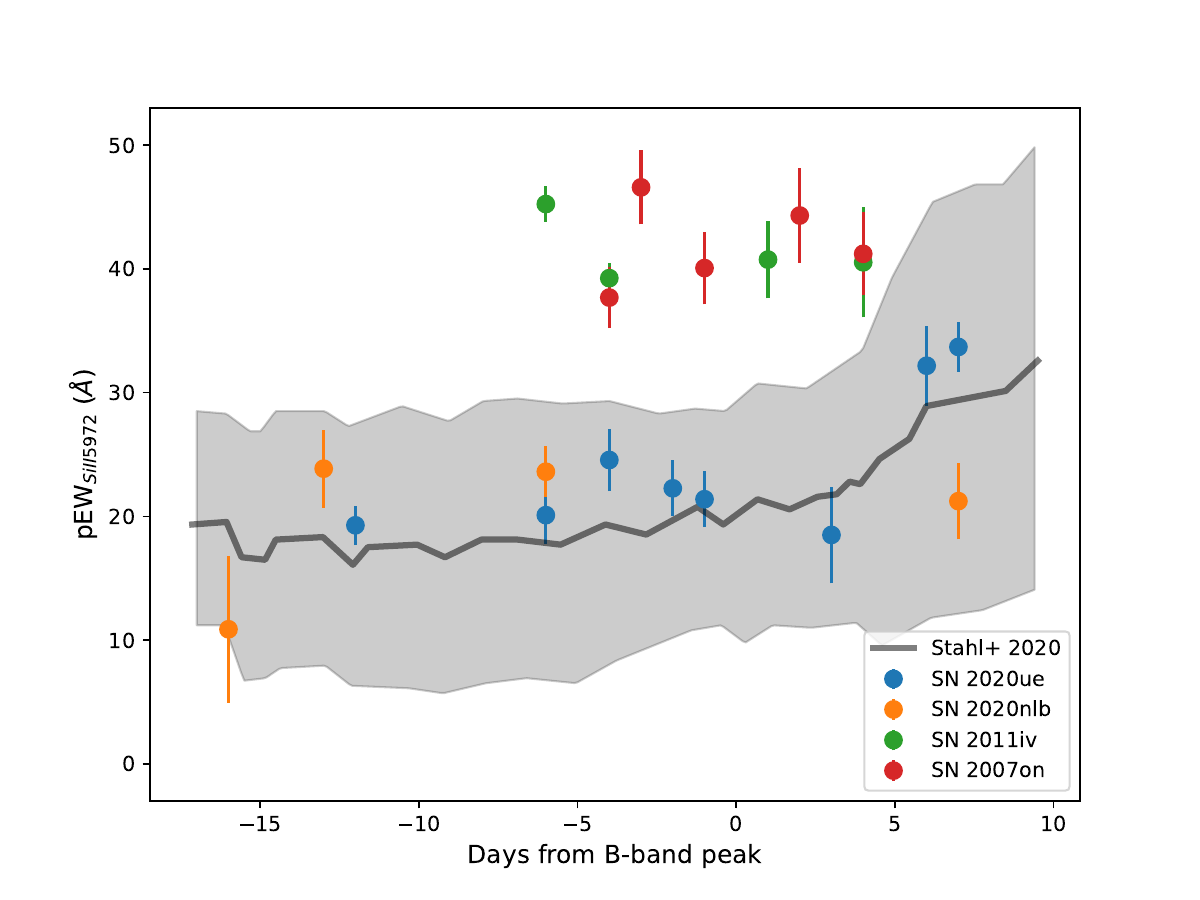}
    \caption{\textit{(Upper panel)} Blue-shifted velocity evolution inferred using \texttt{spextractor} on early spectra of SN 2020ue (blue), SN 2020nlb (orange), and transitional SNe SN 2011iv (green) and SN 2007on (red). The gray line and shaded region represent the average trend from a 247 SN sample in \citet{Stahl2020}, analyzed using the same method. \textit{Lower panel} Evolution of the pEW of the \ion{Si}{II} 5972 \AA\ line compared to the average trend reported in \citet{Stahl2020}.}
    \label{fig:lineprop}
\end{figure} 

\vspace{-5mm}
\subsection{Time-resolved analysis}\label{sec:4_2}

To further explore the temporal spectroscopic behavior of SN~2020ue and SN 2020nlb, we analyze selected spectroscopic features as a function of time. Specifically, we compute velocities and pEWs of the two \ion{Si}{II} transitions at $\lambda\lambda$5972 and 6355 \AA\ using \texttt{spextractor}.
We compare our results with a similar analysis of SNe 2007on and 2011iv, utilizing photometry and spectroscopy from \citet{Gall2018}. Our measurements align with the results reported by \citet{Burrow2020} for SN 2007on and SN 2011iv.
Figure \ref{fig:lineprop} places the results obtained from our analysis, the expanding velocity for the \ion{Si}{II} 6355 \AA\, line and the pEW of the \ion{Si}{II} 5972 \AA, in perspective to the average trend of the same quantities obtained by the analysis of 247 SNe~Ia from the Berkeley Supernova Ia Program \citep{Stahl2020}. 
Our decision to use the expansion velocity of \ion{Si}{II} 6355 \AA\ and the pEW value of \ion{Si}{II} 5972 \AA\ is motivated by the fact that \ion{Si}{II} 6355 \AA\ is easily saturated in SNe~Ia spectra \citep{Hachinger2008,Zhao2021}, while \ion{Si}{II} 5972 \AA\ is not.
Line blending is another concern; \ion{Si}{II} 5972 \AA\ is minimally affected, except for a small \ion{C}{II} contribution at pre-$B$-band maximum epochs (see Sect~\ref{s:specana}). However, \ion{Si}{II} 5972 \AA\ becomes strongly affected by other transitions like \ion{S}{II} at epochs $\sim$10 days after $B$-band maximum \citep{Stahl2020}. This is evident in Fig. \ref{fig:lineprop}, where pEW(\ion{Si}{II} 5972 \AA) values begin increasing at epochs $>5$ days past $B$-band maximum. We also notice that in post-peak epochs, the increasing intensity of the emission in P-Cygni profiles starts to affect the measurement of pEWs.

Our analysis reveals two key findings. First, no significant difference is observed between SNe velocity evolution analyzed here and the average trend reported by \citet{Stahl2020}. While SN 2007on matches SN 2020ue evolution, SN 2020nlb displays lower velocities earlier, becoming consistent with the general trend around the $B$-band maximum. Second, and most intriguingly, transitional SNe display significantly higher values (above \citet{Stahl2020} average) for \ion{Si}{II} 5972 \AA\ pEW, as expected for CL SNe. In contrast, SN 2020ue aligns with the average trend exhibited by normal SNe~Ia, indicating that SN 2020ue more closely resembles normal SNe Ia than extreme 1991bg-like events. Note that \ion{Si}{II} 5972 \AA\ pEW values in SNe~Ia stay relatively constant from $\sim$ $-$12 days to $\sim$ 5 days relative to $B$-band maximum. This finding implies that at epochs before peak brightness, SNe~Ia with pEW values deviating from the average trend in Fig. \ref{fig:lineprop} can be identified as part of the CL sub-class, and then classified as either transitional or fast-declining events, which constitute most events within this sub-class.

\subsection{Spectral synthesis analysis}

\begin{table}
\centering
\caption{Physical properties and element abundance obtained from the {\it TARDIS} simulation of day $-5$ spectra of SN 2020ue and day $-6$ of SN~2011iv.}
\label{tab:TARDIS-results}
\begin{tabular}{lcc}
\hline \hline
Phys. Property & SN 2020ue & SN 2011iv \\
\hline
v$_{\textsc{in}}$ (km/s) & 11873 & 11534\\
v$_{\textsc{out}}$ (km/s) & 15991 & 17725\\
T$_{\textsc{in}}$ (K) & 10425 & 11384\\
w$_{\textsc{in}}$ & 0.42 & 0.42\\
\hline
Element Ab. ($\%$) & SN 2020ue & SN 2011iv \\
\hline
O & 54.06 & 67.35\\
Mg & 7.20 & 2.30\\
Si & 33.07 & 18.25\\
S & 4.11 & 3.30\\
Ca & 0.15 & 0.31\\
Ti & 0.02 & 0.10\\
Cr & 0.05 & 0.22\\
Fe & 0.94 & 6.11\\
Co & 0.27 & 0.96\\
Ni & 0.13 & 1.10\\
\hline
\end{tabular}
\end{table}

\begin{figure}
    \centering
    \includegraphics[width=0.99\linewidth]{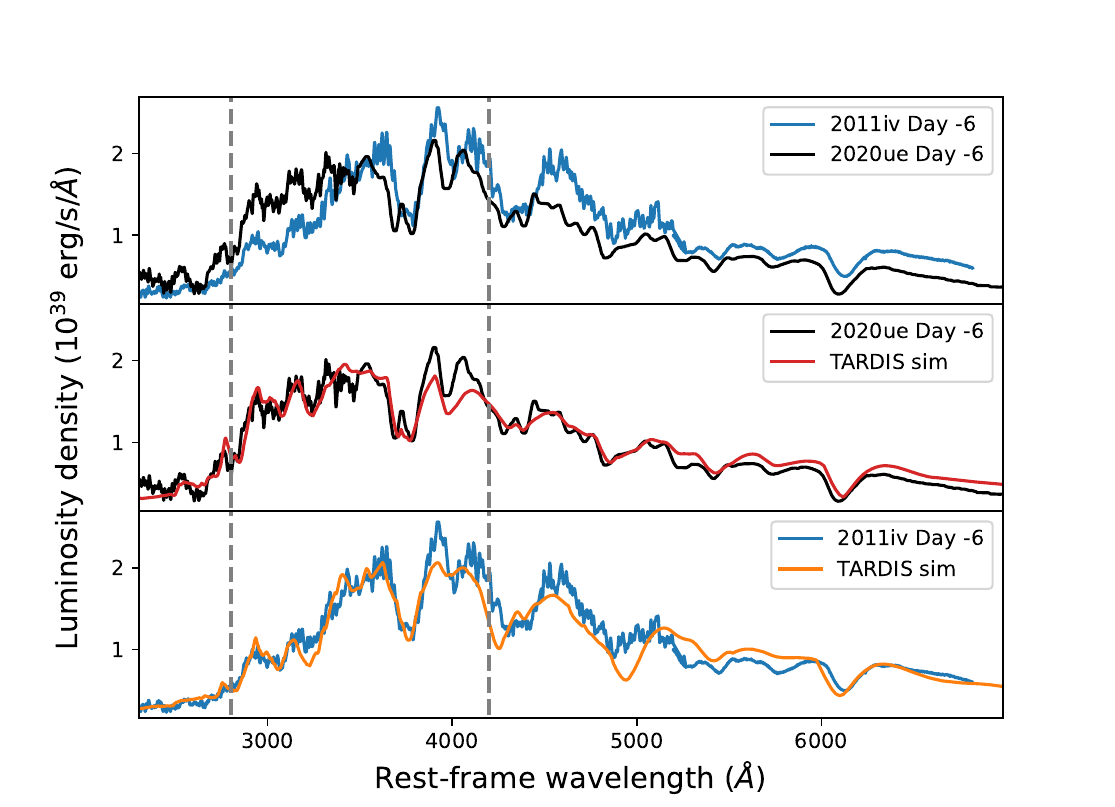}
    \caption{Spectral modeling and comparison between SN 2020ue and SN 2011iv. (Top panel) UV and optical spectra were observed at a similar epoch (about six days before peak brightness) for SN 2020ue (black) and SN 2011iv (blue, \citealp{Gall2018}). (Middle panel) Results of the {\it TARDIS} simulations (red curve) for the spectrum of SN 2020ue (black). (Lower panel) Results of the {\it TARDIS} simulations (orange curve) for the spectrum of SN 20211iv (black). The region between the two dashed gray lines corresponds to the wavelength range considered for the minimization of the root mean square value (see Sec.~\ref{s:specana}).}
    \label{fig:TARDIS}
\end{figure}

The differing temporal evolution of the \ion{Si}{II} 5972 \AA\ line in SN 2020ue and transitional SNe~Ia SN 2007on and SN 2011iv (Fig.\ref{fig:lineprop}) suggests inherent physical differences in the ejecta evolution. The three SNe exhibit similar absorption line evolution in the red wavelength range ($\lambda > 5000$ \AA). However, substantial differences emerge at blue wavelengths, where flux level variations are complex due to metallicity, line blanketing, and other physical effects like ionization \citep[e.g.,][]{2008MNRAS.391.1605S,2012ApJ...753L...5F,2014MNRAS.439.1959M,2015MNRAS.452.4307P}.
Near peak brightness, stronger absorption troughs are observed for transitional SNe than for SN 2020ue. Interestingly, the observed differences in the UV spectral range align with our results from the color evolution analysis (Sec.~\ref{s:photana} and Fig.~\ref{fig:Lira}), revealing bluer colors for SN~2020ue around $B-$band maximum than SN 2007on, SN 2011iv, and most other SNe in Fig. \ref{fig:Lira}. This suggests a physical origin for the observed spectral differences rather than a geometrical or viewing angle effect \citep[e.g.,][]{2009MNRAS.398.1809K}.

To further examine the identified spectral discrepancies among these SNe, we employed spectral synthesis analysis using the Monte-Carlo radiative transfer code \texttt{TARDIS} \citep{Kerzendorf2014,Kerzendorf2018}. \texttt{TARDIS} can simulate synthetic spectra directly compared to observed SN~Ia spectral emission based on specific assumptions: 1) ejecta density distribution, 2) abundance composition, and 3) bolometric SN luminosity. These quantities are analyzed at each epoch. We assume a homologous velocity field and spherically symmetric ejecta. The abundance profile is not fixed at the start of our analysis. Instead, it varies according to the best-fit synthetic spectrum simulating the observed emission. This approach is valid as we focus on a single spectral epoch (see below), assuming uniform element abundance within the SN shell boundaries for the line-forming region. With a continuous spectral series, one could reconstruct a comprehensive radial abundance pattern for the SN ejecta \citep[e.g.,][]{Stehle2005,Mazzali2008,2014MNRAS.439.1959M,Ashall2016,Ashall2018}. Here, we concentrate on the spectral analysis of SN 2020ue and SN~2011iv, for which we have UV and optical spectral coverage at a similar epoch (six days before B-band maximum), emphasizing spectral differences around their peak brightness spectra.

We directly estimate the photosphere's temperature from the best value obtained for the ejecta's inner radius at each spectral epoch and the corresponding luminosity used in the simulation. We adopted the \citet{Branch1985} W7 density model for the density distribution, along with the nebular ionization and {\it dilute-LTE} excitation mode for the plasma configuration, and the {\it downbranch} line interaction method. To select the best-fit result, we estimated the root mean square value from the SN luminosity and {\it TARDIS} simulation in the 2,800 \AA\ to 4,200 \AA\ wavelength range. This range was chosen for two reasons: 1) to maximize the match between the {\it TARDIS} simulation and observed data in the near-UV range, and 2) to avoid the blackbody approximation at longer wavelengths, which typically overestimates the spectral continuum at $\lambda > 5,500$ \AA. For more details on model parameters and the code, please refer to \citet{Kerzendorf2014}.

Fig. \ref{fig:TARDIS} showcases the analysis results. The {\it TARDIS} simulation accurately fits the entire spectral region of SN 2020ue, except for the bright emission features at 4,000 \AA, which are both underestimated in flux by about 10\%. Similar results are obtained for SN 2011iv, although the simulation does not reproduce the continuum at red wavelengths precisely. Notably, the \ion{Si}{II} 5972~\AA\ line is not accurately reproduced, likely due to the assumed uniform abundance distribution in the shell region. The main physical properties and abundance composition obtained from the best-fit model are presented in Table \ref{tab:TARDIS-results}. We observe that the photosphere is at similar radii (in the homologous expansion approximation, radial velocity serves as a scaled proxy of the distance from the center) in both SNe, while the outer boundary is at higher ($\sim$ 1,600 km/s) values in 2011iv. The temperature is similar in both SNe. The most significant difference in our analysis is that the SN ejecta is more enriched with Fe-peak elements in 2011iv, whereas intermediate-mass elements (IME) exhibit nearly identical abundance values to those seen in SN 2020ue.
SN 2011iv's spectral behavior was also analyzed by \citet{Ashall2018}. The availability of a spectral series covering a large time interval enabled \citet{Ashall2018} to perform an abundance tomography analysis, reconstructing the abundance pattern throughout the entire SN ejecta. The density distribution used in \citet{Ashall2018} differs from the W7 model employed here, leading to differing results from our {\it TARDIS} analysis. \citet{Ashall2018} found lower abundances for oxygen and iron and slightly higher values for Si, Mg, and Ni. However, \citet{Ashall2018} also concluded that SN 2011iv might have a high central density \citep{Ashall2016}. This finding suggests that transitional and fast-decliner SNe may originate from progenitors with distinct properties compared to normal type-Ia SNe.

\subsection{The nebular phase}\label{sec:4_4}

\begin{figure}
    \centering
    \includegraphics[width=1.05\linewidth]{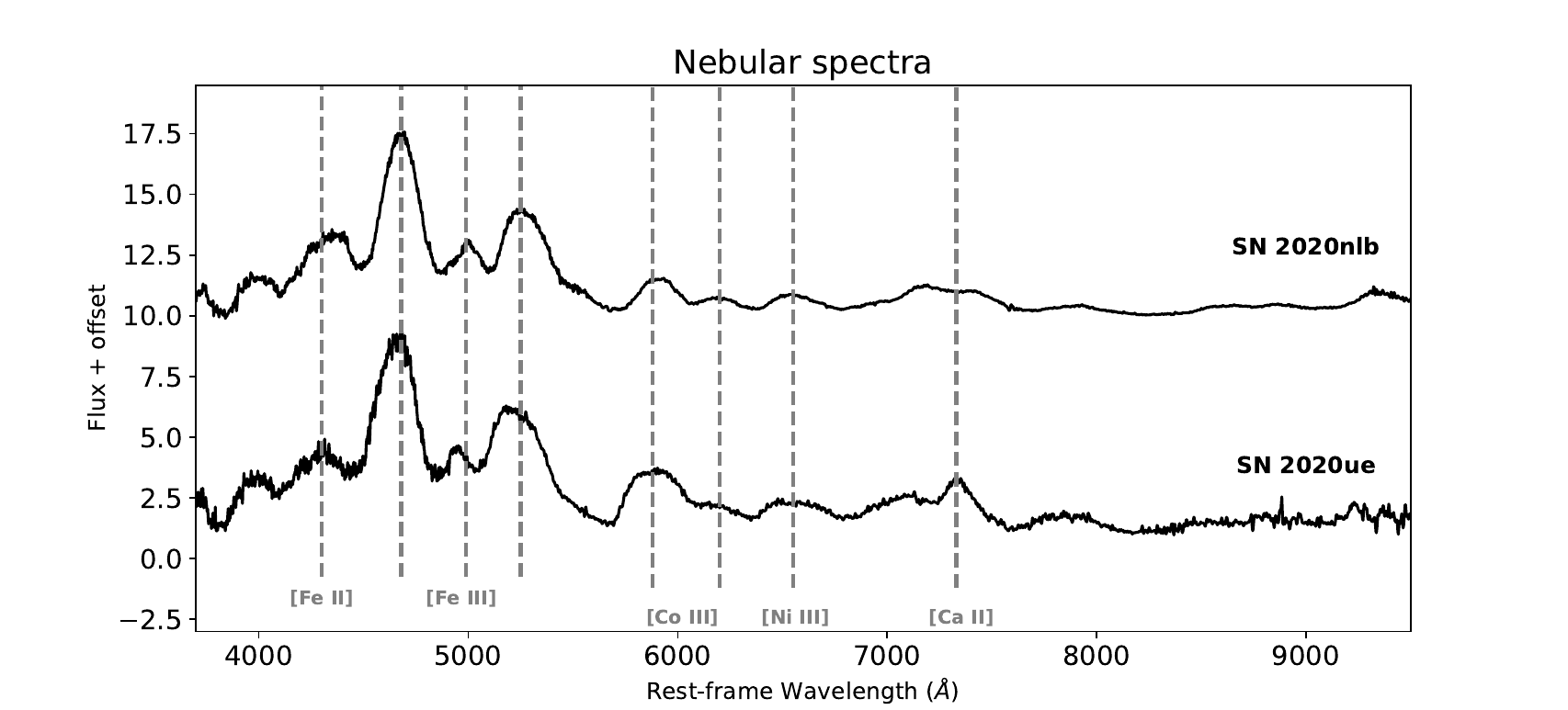}
    \caption{Nebular spectra of SN 2020ue and SN 2020nlb used for the analysis described in Sec. \ref{sec:4_4} with reported the identification of main line features.}
    \label{fig:nebular}
\end{figure}

Approximately 200 days after the SN explosion, the SN ejecta became optically thin, allowing scrutiny of the interior region. The spectrum during this nebular phase features emission lines from newly synthesized elements reflecting the ejecta's velocity distribution. Analyzing SN emission features in this phase can provide valuable insights into physical properties of the SN event, including $M_{Ni}$ synthesized in the thermonuclear explosion, total ejected mass, and clues about the (a)symmetry of the explosion. Spectral synthesis methods or emission line relations can yield quantitative measures \citep{Kuchner1994,Mazzali1998,Stritzinger2006,Mazzali2007,Graham2017,Mazzali2018}.
Here, we discuss the late ($\gtrsim 150$ days after the SN explosion) spectra of SN 2020ue and SN 2020nlb (see Table \ref{tab:spexlog} and Fig. \ref{fig:nebular}). Unfortunately, except for the WiFeS spectra (affected by underlying bright galaxy light), the available spectral resolution is insufficient to draw conclusions about the element distribution within the ejecta from emission line profiles. 

One method to infer the $^{56}$Ni mass involves reconstructing the luminosity evolution of the [\ion{Co}{III}] 5893 \AA\ line, enabling recovery of the line luminosity at specific epochs: 145 days for SN~2020ue and 147 days for SN 2020nlb after peak brightness. Faint [\ion{Co}{III}] 5893 \AA\ emission is also observed in the very late SN 2020ue spectra obtained with LRIS at 384 days post-peak brightness.
The [\ion{Co}{III}] line luminosity linearly correlates with the synthesized nickel mass, but only within a particular time range, e.g., during the nebular phase around $\sim$200 d from the explosion \citep{Childress2015}. Thus, we can estimate the $^{56}$Ni mass by correcting the measured luminosity at a given epoch for the corresponding $^{56}$Co decay lifetime shift, with the resulting uncertainties accounted for in the systematic error budget ($\sigma M_{Ni} = 0.2$ M$_{\odot}$).
We followed \citet{Childress2015}'s prescriptions for measuring the [\ion{Co}{III}] 5893 \AA\ line flux. For all spectra, we refined the flux calibration using available photometry (see Sec.~\ref{s:obs}). Additionally, we applied host galaxy extinction correction for SN 2020nlb using the value reported by \texttt{BAYESN} in Sec.~\ref{sec:3_1}. Results are reported in Table~\ref{tab:neb}. In summary, both SNe exhibit similar $^{56}$Ni masses: $M_{Ni} = 0.50 \pm 0.23$ M$_{\odot}$, consistent within uncertainties with estimates inferred in Sec. \ref{sec:3_4}. Note that systematic uncertainties contribute significantly to the $M_{Ni}$ uncertainty.

\begin{table}
\small
\centering

\caption{Measurements of the [\ion{Co}{III}] 5893\,\AA\ nebular line flux, and derived nickel mass, for both SNe. The reported epoch is in days from SN peak brightness.}
\label{tab:neb}
\begin{tabular}{lcccc}
\hline \hline
 & Epoch & Flux & Luminosity & $M_{Ni}$\\
  & (days) & (10$^{-13}$ erg/cm$^2$/s) & (10$^{39}$ erg s$^{-1}$) & ($M_{\odot}$)\\
\hline
2020ue & 160 & 1.71$\pm$0.01 & 4.64$\pm$0.02 & 0.50$\pm$0.23\\
2020nlb & 149 & 1.68$\pm$0.01 & 5.06$\pm$0.02 & 0.50$\pm$0.23\\
\hline
\end{tabular}
\end{table}

\subsection{Analysis of the environment}

Identifying interstellar medium (ISM) lines and diffuse interstellar bands (DIBs) in host galaxies from SN spectra allows for estimating host extinction along the line of sight \citep{Phillips2013}. Empirical relations can then provide extinction estimates from \ion{Na}{I} equivalent width (EW) measurements \citep[e.g.,][]{1997A&A...318..269M,2012MNRAS.426.1465P}. Furthermore, the presence and potential variations of these lines could indicate te existence of a circum-stellar medium around the SN progenitor star. To find out, we analyze high-resolution spectra obtained with Lick/APF at three epochs for SN 2020ue and two epochs for SN 2020nlb. 
Interestingly, neither the Lick/APF spectra nor the Asiago echelle spectrum of SN 2020nlb show Na ID lines (see Fig. \ref{fig:APF}). We simulated Na ID lines using a Gaussian function with $\sim 20$ km/s dispersion velocity and varying EW arbitrarily. We also considered a $\pm 100$ km/s velocity shift relative to each galaxy's center (gray regions in Fig. \ref{fig:APF}). However, no match with narrow ISM/CSM features was found in either SN's APF spectra, with an EW limit of $EW < 0.015$ \AA. This sets a limit of $E(B-V) < 0.02$ mag \citep{2012MNRAS.426.1465P}.

\section{Discussion}
\label{s:discu}

Photometric and spectroscopic analyses in Sec. \ref{s:photana} and \ref{s:specana} demonstrated that SN 2020nlb exhibits properties entirely consistent with normal SNe~Ia. However, SN 2020ue is intriguing, displaying very blue colors at peak brightness comparable to or even bluer than transitional events.
Additionally, SN 2020ue's light curve evolution resembles the fast-declining SNe~Ia sub-class, characterized by faster decline rates than normal SNe~Ia. Its $\Delta m_{15}$ value lies at the borderline between normal SNe~Ia and the fast-declining sub-class. The rise time computed from the bolometric light curve, $t_{r,20ue} = 14.48 \pm 0.20$ days, is similar to transitional and fast-declining events \citep{Taubenberger2008}. 

High-resolution spectra of both events show no detectable \ion{Na}{I} D absorption lines at the host's redshift. \ion{Na}{I} D EW measurements have been extensively used to estimate interstellar dust extinction based on empirical relations \citep[e.g.,][]{2012MNRAS.426.1465P}, although sodium is not a primary element of interstellar dust grains \citep{2003ARA&A..41..241D}. 
We therefore conclude that SN~2020nlb is intrinsically red due to SN ejecta temperature, rather than reddened by interstellar dust. However, as evident from its lightcurve characteristics, SN~2020nlb is a ''normal'' SN at the edge to the transitional region in the luminosity-decay rate relation. Thus, SN~2020nlb may be the first SN of a small population of SNe~Ia which can be identified only through detailed analysis of their local host galaxy environments. High-cadenced SN light curve coverage is also required. Furthermore, the distance estimate differences between SN 2020ue and SN 2020nlb align with their respective host galaxies' distance moduli obtained using different distance methods, considering the extent of the Virgo cluster $\Delta \mu$ = 0.32 $\pm$ 0.06 mag \citep{Mei2007}.
We emphasize that SBF- or FP-derived distances to the host galaxies of SN 2020ue and SN 2020nlb agree with those obtained via SNe~Ia light-curve fitting methods, but are characterized by larger uncertainties, see Table \ref{tab:dist}. 

Our spectral analysis revealed that SN 2020ue and SN~2020nlb belong to the CN and BL sub-classes, respectively, in the Branch classification scheme defined by \citet{Branch2006} (see Fig. \ref{fig:Branch}). This implies that both SNe share spectral properties with the bulk of normal SNe~Ia, unlike transitional and fast-declining sub-types, characterized by larger \ion{Si}{II} 5972 \AA\ pEWs and classified within the CL sub-class by definition.
This evidence suggests that spectroscopic diagnostics can clarify the true nature of SNe Ia explosions and provide crucial information about the underlying progenitor WD's physical properties. Numerous studies have linked faint SNe Ia light-curve properties to various WD progenitor properties \citep[e.g.,][]{Hoeflich2017,Ashall2018}. Within the single degenerate progenitor models, one possible explanation is that a higher central density of the WD can result in more stable $^{58}$Ni production at the expense of $^{56}$Ni \citep{Hoeflich2017,Ashall2018}. A lower synthesized $^{56}$Ni amount leads to less ejecta heating and consequently, faint peak luminosities, low ejecta temperatures, and enhanced low-ionization features like \ion{Ti}{II} in SN~Ia spectra. This contrasts with the large \ion{Si}{II} 5972 \AA\ pEWs observed in faint SNe Ia, given the higher excitation energy required for \ion{Si}{II} 5972 \AA\ compared to \ion{Si}{II} 6355 \AA. However, at lower temperatures, \ion{Si}{II} is more abundant than the higher ionization \ion{Si}{III} line, which is more prominent in luminous SNe~Ia spectra \citep[e.g.,][]{Ashall2016}. This indicates that high \ion{Si}{II} 5972 \AA\ intensity implies a low-ionization level for the entire SN ejecta, aligning with the lower excitation temperature observed for transitional events compared to normal SNe~Ia at similar epochs.

The detection of carbon in SN 2020ue and its non-detection in SN 2020nlb at very early epochs has significant implications for understanding their progenitor systems. Carbon features in SNe Ia spectra can provide crucial insights into the amount of unburned material, helping differentiate between scenarios involving the progenitor white dwarf's nature and explosion dynamics. Unburned material serves as a vital indicator of the explosion mechanism, suggesting that the progenitor white dwarf did not undergo complete thermonuclear burning before the explosion. Strong carbon absorption features are relatively rare \citep{Parrent2011}, typically observed during the pre-maximum phase of a small fraction of SNe Ia \citep{Folatelli2012,Maguire2014}. This could be attributed to low unburned material quantities or the blending of carbon lines with other spectral features like \ion{Si}{II} 6355 \AA. Conversely, the absence of carbon lines could indicate a more complete burning process, implying that the progenitor white dwarf might have been closer to the Chandrasekhar mass limit and experienced a more energetic explosion. \citealp{Thomas2011} present evidence of unburned carbon in multiple SNe Ia, supporting the notion that the lack of such features in other events might be connected to distinct progenitor properties, such as sub-Chandrasekhar double detonation models \citep{Sim2010,Polin2019,Pakmor+22,Zenati+23_carich} or explosion mechanisms \citep{Heringer2019}.

\section{Conclusions and Perspectives}
\label{s:concl}

In this work, we conducted a high-cadence photometric and spectroscopic follow-up campaign of two SNe~Ia, SN 2020ue and SN~2020nlb, in two elliptical galaxies, NGC 4636 and NGC 4382, respectively. Our combined analysis of multi-filter SN evolution using light-curve fitting codes such as {\sc SNooPy}, {\sc BAYESN}, and GP-based methods, along with pseudo-bolometric light curves and derived $^{56}$Ni masses, and the entire spectral series, has revealed the following:

\begin{itemize}
\item SN 2020nlb appears to be an intrinsically red yet otherwise normal SN~Ia, with its color arising from temperature rather than dust reddening. This provides evidence for intrinsic color variation in normal SNe~Ia, challenging standardization methods like the Tripp relation, and hinting at overlapping populations near the transitional regime;
\item The light curve and color evolution of SN 2020ue, the rise-time to peak derived from the bolometric light curve (see Sec. 3.3), and the decay rate parameterized by $s_{BV}$ and $\Delta m_{15}$ obtained from {\sc SNooPy} and {\sc BAYESN} collectively demonstrate that this SN closely resembles the transitional SN sub-class, including SN 2007on and SN 2011iv, whose accuracy as distance indicators remains debated \citep{Gall2018,Hoogendam2022};
\item The spectroscopic evolution shows that SN 2020ue is typical of the core-normal (CN) spectral class \citep{Branch2006}, whereas transitional and fast-decliner events display properties of the cool (CL) spectral class;
\item A spectral synthesis analysis of the UV + optical peak brightness spectrum of SN 2020ue has been compared to a similar analysis of the UV + optical spectrum at the same epoch of transitional SN 2011iv. The spectra show differences, particularly in the near-UV region due to line blanketing, so we conclude that the peak spectrum of transitional SN 2011iv is richer in Fe-peak elements than SN 2020ue ejecta, evidence that can be attributed to mixing during the deflagration phase in SN 2011iv \citep{Ashall2018};
\item The distance of these two SNe inferred from light-curve fits and the luminosity-decay relation \citep{Phillips1993} aligns with their host distances estimated as a weighted average between the SBF and FP distance methods \citep{Kourkchi2020}, but display smaller uncertainties on the inferred distance moduli, see also Sec. \ref{sec:2_1} and Sec. \ref{s:discu};
\item The host extinction inferred from light-curve fitting methods agrees with the absence of \ion{Na}{I} D lines in SN 2020ue spectra but not SN 2020nlb, for which considerable color excess ($E(B-V) \sim 0.1-0.2$ mag) has been estimated using different methods. This situation parallels the fast-decliner SN~2015bo \citep{Hoogendam2022}, suggesting caution when employing absorption line empirical correlations to estimate host extinction in SNe.
\end{itemize}

Our results emphasize the importance of incorporating spectroscopic data to differentiate normal and peculiar events beyond traditional light curve fitting methods. Dedicated spectroscopic surveys are essential for constructing cosmological samples of SNe free from photometric biases, which is critical for their role as cosmological probes of the Universe's expansion history and advancing our knowledge of the progenitor systems behind these phenomena.

\section*{Acknowledgements}
\begin{small}
We warmly thank the referee, Nicholas Suntzeff, for his careful reading and insightful comments, which have greatly improved both the clarity and quality of our manuscript.
We thank G.\ Benetti for the realization of trichomes in Figure~\ref{fig:images},
P.\ Butler and A.\ Kuehnel for their assistance with the reduction of the Lick/APF spectra,
and C.\ Onken for his support in observing the environments of both SNe with WiFES at the 2.3-meter telescope of the Australian National University. This work is supported by a VILLUM FONDEN Young Investigator Grant (project number 25501, PI C.\ Gall), Villum Experiment grant (VIL69896) and by research grants (VIL16599, VIL54489) from VILLUM FONDEN.
This material is based upon work supported by the National Science Foundation Graduate Research Fellowship Program under Grant Nos.\ 1842402 and 2236415. Any opinions, findings, conclusions, or recommendations expressed in this material are those of the authors and do not necessarily reflect the views of the National Science Foundation.
L.I.\ acknowledges financial support from the INAF Data Grant Program ``YES'' (PI: Izzo) {\it Multi-wavelength and multi messenger analysis of relativistic supernovae}.
S.M.W.\ was supported by the UK Science and Technology Facilities Council.
S.T.\ was supported by funding from the European Research Council (ERC) under the European Union's Horizon 2020 research and innovation programmes (grant agreement no.\ 101018897 CosmicExplorer).
Supernova and astrostatistics research at Cambridge University is supported by the European Union’s Horizon 2020 research and innovation programme under European Research Council Grant Agreement No 101002652 (PI K.\ Mandel) and Marie Skłodowska-Curie Grant Agreement No 873089.
The UCSC team is supported in part by NASA grants 80NSSC23K0301 and 80NSSC24K1411; NSF grants AST--1815935 and AST--1720756; and a fellowship from the David and Lucile Packard Foundation to R.J.F.
S.D.\ acknowledges support from  UK Research and Innovation (UKRI) under the UK government’s Horizon Europe funding Guarantee EP/Z000475/1  and a Junior Research Fellowship at Lucy Cavendish College.
D.O.J.\ acknowledges support from NSF grants AST--2407632 and AST--2429450, NASA grant 80NSSC24M0023, and HST/JWST grants HST-GO-17128.028, HST-GO-16269.012, and JWST-GO-05324.031, awarded by the Space Telescope Science Institute (STScI), which is operated by the Association of Universities for Research in Astronomy, Inc., for NASA, under contract NAS5--26555.
C.D.K.\ gratefully acknowledges support from the NSF through AST--2432037, the HST Guest Observer Program through HST-SNAP-17070 and HST-GO-17706, and from JWST Archival Research through JWST-AR-6241 and JWST-AR-5441.
G.N.\ is funded by NSF CAREER grant AST--2239364, supported in-part by funding from Charles Simonyi. GN also gratefully acknowledges NSF support from AST--2206195, OAC--2311355, AST--2432428, as well as AST--2421845 and funding from the Simons Foundation for the NSF-Simons SkAI Institute. G.N.\ is also supported by the DOE through the Department of Physics at the University of Illinois, Urbana-Champaign (\# 13771275), and support from the HST Guest Observer Program through HST-GO-16764 and HST-GO-17128 (PI: R.\ Foley).
A.R.\ acknowledges financial support from the GRAWITA Large Program Grant (PI P.\ D’Avanzo) and the PRIN-INAF 2022 \textit{``Shedding light on the nature of gap transients: from the observations to the models''}.
M.R.S.\ is supported by the STScI Postdoctoral Fellowship.
S.J.S., S.S., and K.W.S.\ acknowledge funding from STFC Grants ST/Y001605/1, a Royal Society Research Professorship and the Hintze Charitable Foundation. The Young Supernova Experiment (YSE) and its research infrastructure is supported by the European Research Council under the European Union's Horizon 2020 research and innovation programme (ERC Grant Agreement 101002652, PI K.\ Mandel), the Heising-Simons Foundation (2018-0913, PI R.\ Foley; 2018-0911, PI R.\ Margutti), NASA (NNG17PX03C, PI R.\ Foley), NSF (AST--1720756, AST--1815935, PI R.\ Foley; AST--1909796, AST--1944985, PI R.\ Margutti), the David \& Lucille Packard Foundation (PI R.\ Foley), VILLUM FONDEN (project 16599, PI J.\ Hjorth), and the Center for AstroPhysical Surveys (CAPS) at the National Center for Supercomputing Applications (NCSA) and the University of Illinois Urbana-Champaign.
Pan-STARRS is a project of the Institute for Astronomy of the University of Hawaii, and is supported by the NASA SSO Near Earth Observation Program under grants 80NSSC18K0971, NNX14AM74G, NNX12AR65G, NNX13AQ47G, NNX08AR22G, 80NSSC21K1572, and by the State of Hawaii.  The Pan-STARRS1 Surveys (PS1) and the PS1 public science archive have been made possible through contributions by the Institute for Astronomy, the University of Hawaii, the Pan-STARRS Project Office, the Max-Planck Society and its participating institutes, the Max Planck Institute for Astronomy, Heidelberg and the Max Planck Institute for Extraterrestrial Physics, Garching, The Johns Hopkins University, Durham University, the University of Edinburgh, the Queen's University Belfast, the Harvard-Smithsonian Center for Astrophysics, the Las Cumbres Observatory Global Telescope Network Incorporated, the National Central University of Taiwan, STScI, NASA under grant NNX08AR22G issued through the Planetary Science Division of the NASA Science Mission Directorate, NSF grant AST--1238877, the University of Maryland, Eotvos Lorand University (ELTE), the Los Alamos National Laboratory, and the Gordon and Betty Moore Foundation. Parts of this research are based on observations made with the Nordic Optical Telescope (programme 60-408, PI: Izzo; programme 61-022, PI Angus) owned in collaboration by the University of Turku and Aarhus University, and operated jointly by Aarhus University, the University of Turku and the University of Oslo, representing Denmark, Finland and Norway, the University of Iceland and Stockholm University at the Observatorio del Roque de los Muchachos, La Palma, Spain, of the Instituto de Astrofisica de Canarias.  Data presented here were obtained in part with ALFOSC, which is provided by the Instituto de Astrofisica de Andalucia (IAA) under a joint agreement with the University of Copenhagen and NOT.
Some of the data presented herin were obtained at the Infrared Telescope Facility, which is operated by the University of Hawaii under contract 80HQTR24DA010 with the National Aeronautics and Space Administration.
Some of the data presented herein were obtained at Keck Observatory, which is a private 501(c)3 non-profit organization operated as a scientific partnership among the California Institute of Technology, the University of California, and the National Aeronautics and Space Administration. The Observatory was made possible by the generous financial support of the W.\ M.\ Keck Foundation.
The authors wish to recognize and acknowledge the very significant cultural role and reverence that the summit of Maunakea has always had within the indigenous Hawaiian community.  We are most fortunate to have the opportunity to conduct observations from this mountain.
A major upgrade of the Kast spectrograph on the Shane 3~m telescope at Lick Observatory was made possible through generous gifts from the Heising-Simons Foundation as well as William and Marina Kast. Research at Lick Observatory is partially supported by a generous gift from Google.
Partly based on observations collected at the Copernico 1.82-m Telescope and the robotic Schmidt 67/92 telescope operated by INAF - Osservatorio Astronomico di Padova at Asiago, Italy, and the Galileo 1.22m telescope operated by University of Padova (Dept.\ of Physics and Astronomy, DFA).
This work makes use of data from the Las Cumbres Observatory global telescope network.
We acknowledge the use of public data from the Swift data archive.
YSE-PZ was developed by the UC Santa Cruz Transients Team, supported in part by NASA grants NNG17PX03C, 80NSSC19K1386, and 80NSSC20K0953; NSF grants AST--1518052, AST--1815935, and AST--1911206; the Gordon \& Betty Moore Foundation; the Heising-Simons Foundation; a fellowship from the David and Lucile Packard Foundation to R.\ J.\ Foley; a Gordon and Betty Moore Foundation postdoctoral fellowship and a NASA Einstein fellowship, as administered through the NASA Hubble Fellowship program and grant HST-HF2-51462.001, to D.\ O.\ Jones; and a National Science Foundation Graduate Research Fellowship, administered through grant No.\ DGE-1339067, to D.\ A.\ Coulter.

\end{small}



\bibliographystyle{aa}

\newpage

\begin{appendix}

\section{Data reduction}

\subsubsection*{Photometry}
In addition to Pan-STARRS data, we have also obtained photometric coverage for SN~2020ue and SN~2020nlb with other facilities. Regarding SN~2020ue, a photometric coverage of the peak brightness until late epochs was obtained with the Alhambra Faint Object Spectrograph and Camera (ALFOSC) mounted at the Nordic Optical Telescope (NOT) located in La Palma, Canary Islands, Spain, using $uBVgriz$ filters, and both SNe 2020ue and 2020nlb were observed with the Asiago Faint Object Spectrograph and Camera (AFOSC) mounted at the 1.82m telescope and with the robotic Schmidt 67/92 telescope (both located at Cima Ekar, Asiago, Italy), using the same filter setup. 
The calibrated photometry was derived with the \textsc{ecSNooPy} pipeline\footnote{\textsc{ecSNooPy} is a package for SN photometry through PSF fitting and template subtraction developed by E. Cappellaro, 2014. See also http://sngroup.oapd.inaf.it/SNooPy.html}, using the PSF-fitting technique on the science images. 
The instrumental magnitudes were then calibrated to the Sloan AB system, using Sloan stars in the field of the host galaxy. 
 Given the smoothness of the underlying stellar background, PSF photometry has been performed by fitting the underlying background up to three times the FWHM of the target, and finally checking the smoothness of the residual image, after subtracting the PSF fit of the target. 
Error estimates for the SN magnitudes are obtained through artificial star experiments in which fake stars with similar magnitudes as the SN are placed in the fit residual image in a position close to, but not coincident with, the SN location. The simulated images were processed using the same procedure, and the standard deviation of the measured magnitudes of the fake stars was taken as an estimate of the instrumental magnitude error. In practical terms, this error mainly reflects uncertainty in the background subtraction.

Both SNe were also observed by the Neil Gehrels {\it Swift} observatory \citep{Gehrels2004}. We reduced {\it Swift}/UVOT data using {\sc uvotsource} using the \textsc{HEAsoft} v6.26 software package following the standard procedures as described in \citet{Brown14}.  We used a 3\arcsec\ aperture size with a 30\arcsec\ region to estimate the background contribution around each source.  The final {\it Swift}/UVOT light curve for both sources is shown in Fig.~\ref{fig:LC2020uenlb}.

We observed SN\,2020ue, and SN\,2020nlb in $uBVgri$ bands with the Direct 4K$\times$4K imager on the Swope 1\,m telescope at Las Campanas Observatory, Chile.  Following procedures described in \citet{Kilpatrick18}, all imaging data were reduced using {\sc photpipe} \citep{Rest05}, including corrections for bias, flat-fielding, amplifier crosstalk corrections, astrometric distortion corrections and regridding of each image using {\sc SWarp} \citep{swarp}, photometry using {\sc DoPhot} \citep{Schechter93}, and photometric calibration using the Pan-STARRS DR2 catalog \citep{Flewelling16} transformed to the Swope in-band natural magnitudes using Supercal \citep{Scolnic15}.  Final Swope photometry was obtained at the site of both sources using forced {\sc DoPhot} photometry in the final images.

We observed SN\,2020nlb in $griz$ bands with the Thacher 0.7\,m telescope at the Thacher School Observatory in Ojai, CA \citep{Swift21}.  Following the same procedures described above for Swope, we obtained the final photometry of the SN using {\sc photpipe}.

We observed SN\,2020ue and SN\,2020nlb in $uBVgri$ bands with the Sinistro imagers on the Las Cumbres Observatory (LCO) 1\,m telescopes.  Taking the {\sc BANZAI} processed images from the LCO pipeline \citep{BANZAI}, we corrected the reduced images for geometric distortion using {\sc SWarp} and performed photometry and flux calibration using the same {\sc photpipe} procedures described above.  The final photometry for both sources is forced {\sc DoPhot} photometry in the final reduced frames.

Observations of SN 2020nlb were conducted using the 0.5-meter telescope at the Osservatorio Astronomico "Salvatore Di Giacomo" (OASDG) in Agerola, Italy \footnote{https://osservatorio.astrocampania.it/}. BVR images were acquired during the initial 60 days of SN emission, with data reduction and calibration performed via a dedicated pipeline based on the \texttt{Astropy} package \citep{Astropy2022}. Magnitudes were calculated using an \texttt{astropy}-based aperture photometry routine on single-epoch stacked images and converted to BVR magnitudes using transformation formulas from \citep{Kostov}. The resulting OASDG light curve is depicted in Fig. \ref{fig:SNooPyfit_LCO_OASDG} alongside the complete dataset. 

\subsubsection*{Spectroscopy}

Five spectroscopic epochs of SN 2020nlb were observed using grism \#4 on the NOT with ALFOSC, covering the 3,800-9,400 \AA\ wavelength range and a resolving power of $R = 360$. Data reduction was performed with {\sc pyNot}\footnote{\url{https://github.com/jkrogager/PyNOT}}, a Python-based GUI tailored for ALFOSC. The software enables standard calibrations, background subtraction using Chebyshev polynomial approximation, and flux calibration using a standard star.

We observed SN\,2020ue and SN\,2020nlb with the FLOYDS spectrographs \citep{floyds} on the Faulkes 2\,m North and South telescopes at Haleakala, Hawaii and Siding Spring Observatory, Australia, respectively.  All FLOYDS data were processed using the LCO FLOYDS pipeline as described in \citet{BANZAI}.

Twenty spectroscopic epochs of SN 2020ue were obtained in Asiago, Italy, using both the Copernico 1.82m telescope, equipped with AFOSC or with an Echelle high-resolution spectrograph, and with the Galileo 1.22m telescope equipped with a Boller \& Chivens spectrograph. The slit was set at the parallactic angle. The spectra were reduced using standard \texttt{IRAF} tasks for the data reduction. After bias and flat-field correction, the SN spectrum was extracted and calibrated in wavelength with reference to standard arc lamp spectra. For the flux calibration, nightly sensitivity functions were derived from observations of spectrophotometric standard stars such as Feige 34, Feige 56 and HR 3154, and also used to derive the corrections for the telluric absorption bands. The flux calibration of each spectrum was verified against coeval broadband photometry. 

We obtained additional optical follow-up spectroscopy with the Low-Resolution Imaging Spectrograph \citep[LRIS;][]{LRIS} on the 10-m Keck~I telescope and the Kast dual-beam spectrograph \citep{KAST} on the Lick Shane 3-m telescope. We obtained two epochs of SN\,2020ue, and one epoch of SN\,2020nlb with LRIS, and we obtained five epochs of SN\,2020ue, and SN\,2020nlb, respectively, with Kast. Observations with Kast were taken with the slit at the parallactic angle \citep{Filippenko82} to reduce flux losses associated with aligning the slit at non-parallactic angles. Observations with LRIS were aligned to the nucleus of the host galaxy since LRIS is equipped with an atmospheric dispersion corrector. We reduced the Kast and LRIS spectral data in a standard manner using our custom data-reduction {\sc UCSC Spectral Pipeline}\footnote{\url{https://github.com/msiebert1/UCSC\_spectral\_pipeline}} \citep{Siebert2019}. The two-dimensional (2D) spectra were bias-corrected, flat-field corrected, adjusted for varying gains across different chips and amplifiers, and trimmed. We combined multiple frames using a sigma-clipping cosmic algorithm to remove cosmic rays.  One-dimensional spectra were then extracted using the optimal algorithm \citep{Horne1986}. The spectra were wavelength-calibrated using internal arc lamps spectra. The wavelength solution was refined by applying linear shifts determined by cross-correlating the observed night-sky lines to a master night-sky spectrum, and a heliocentric correction was applied. Flux calibration was performed relative to standard stars using the same setup and, where possible, a similar airmass to that of the science exposures with ``blue'' (hot subdwarfs; i.e., sdO) and ``red'' (low-metallicity G/F) standard stars corrected for atmospheric extinction. To remove atmospheric absorption features in the spectra, a telluric correction was determined by fitting the continuum of the flux-calibrated red standard stars to create a normalized spectrum only including telluric features. The relative strength of the absorption was determined by calculating the relative airmass between the standard star and the science image. The telluric bands of the red standard star were cross-correlated with the supernova spectrum at the B-band and the A-band to calculate small wavelength shifts, which were averaged and applied. We then combined the red and blue sides by rescaling one spectrum to match the flux of the other in the overlap region between the blue and red sides. 

We acquired three epochs of SN\,2020ue and two epochs of SN\,2020nlb using the Automated Planet Finder \citep[APF;][]{Vogt+2014}, a robotic 2.4-m telescope located at Lick Observatory. The APF is equipped with the Levy Spectrograph, a high-resolution (R $\simeq$ 100,000) echelle spectrograph, with a wavelength coverage of 3740-9700 \AA. These data were reduced using a custom raw reduction package developed for Iodine-based precision velocity spectrometers by the APF team. The reduction package performed standard flat fielding, scattered light subtraction, order tracing, and cosmic ray removal. The spectra were wavelength calibrated against the NIST Fourier Transform Spectrometer spectral atlas of the APF Iodine cell, with a resolution of 1,000,000 and a S/N $\sim$1,000. Telluric lines were identified by comparing the spectrum of a rapidly rotating B star to a synthetic telluric atlas. With the exception of telluric lines, the spectrum of the rapidly rotating B star is nearly featureless. Pixels with telluric lines were given zero weight for the wavelength and velocity determination. The final spectra are presented in vacuum wavelengths.

SN 2020ue was observed with the SpeX medium-resolution spectrograph \citep{Rayner03} mounted on the NASA Infrared Telescope Facility (IRTF), located on Mauna Kea (Hawaii, USA). We present two NIR spectra obtained on Days $-$5 and +56 (Program ID 2019B087, PI M. Shahbandeh). 
Data were reduced using the \texttt{Spextool} package with standard procedures and telluric corrections performed with an A0 star \citep{Cushing04}

Both SN 2020ue and SN 2020nlb were observed using the Wide Field Spectrograph (WiFeS) on the 2.3 m telescope at Siding Spring Observatory \citep{2007Ap&SS.310..255D,2010Ap&SS.327..245D} on the 6th of March 2021 and the 12th of May 2021, respectively. Each SN was observed using an RT-560 beam splitter and a B3000 and R3000 diffraction gratings that cover the $3200-5900$ \AA\, and $5300-9800$ \AA\  wavelength ranges. All observations have a 1x1 arcsec spaxel. The average seeing on each night was 1.7 arcsecs. Calibration data and standard star spectra were taken on each night, while each WiFeS observation was reduced using PyWiFeS \citep{2014Ap&SS.349..617C}, resulting in a 3D cube file having bad pixels and cosmic rays removed.

Finally, a single spectrum of SN 2020ue was obtained with the UltraViolet and Optical Telescope (UVOT) on board the Neil Gehrels {\it Swift} spacecraft \citep{Gehrels2004}. The spectrum was obtained on January 19, 2020 (MJD 58867.47), using the slitless grism mode. We used the same recipes for data reduction and spectra extraction that are reported in \citet{Pan2018}.

\section{Additional figures and tables}

This section includes additional Figures and Tables that support the results presented in the main text. 

Table \ref{tab:spexlog} shows the diary of the spectroscopic observations presented in this work.

Table \ref{tab:spex} shows the measurements of the pEWs measured in SN~2020ue and SN~2020nlb using  the {\sc Spextractor} code.

\begin{figure*}
    \centering
    \includegraphics[width=0.48\linewidth]{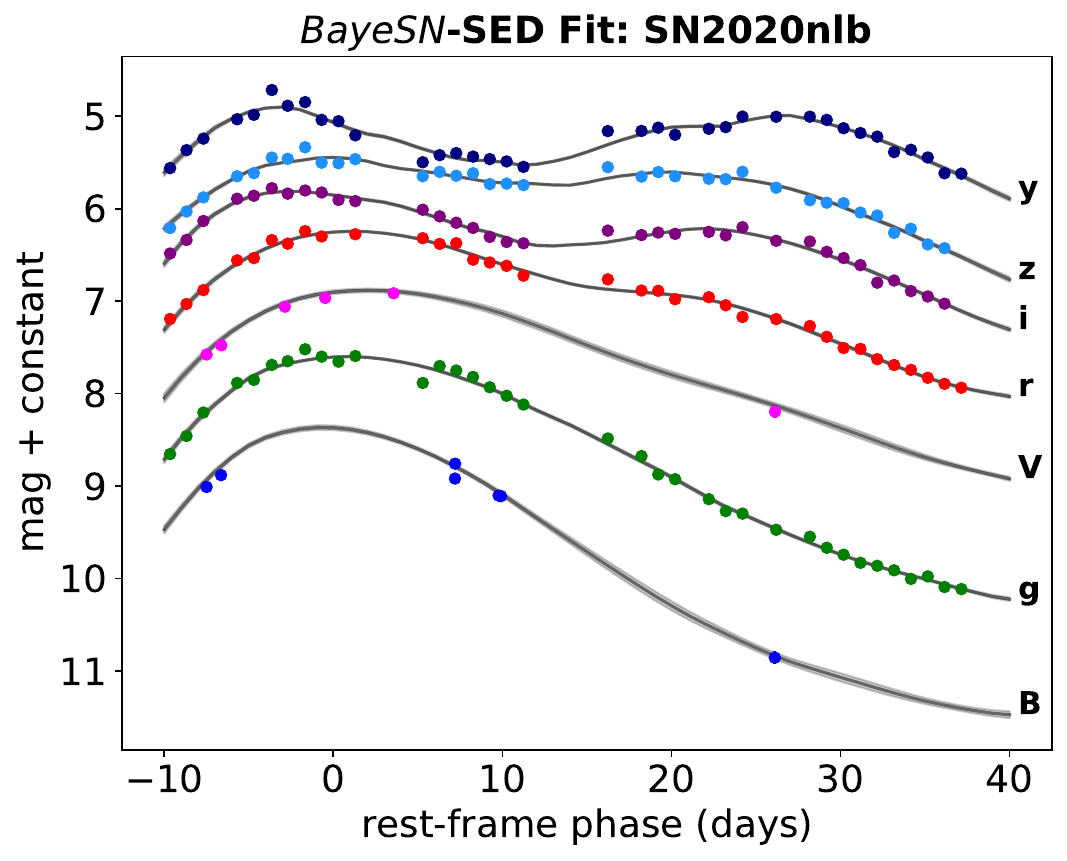}
    \includegraphics[width=0.48\linewidth]{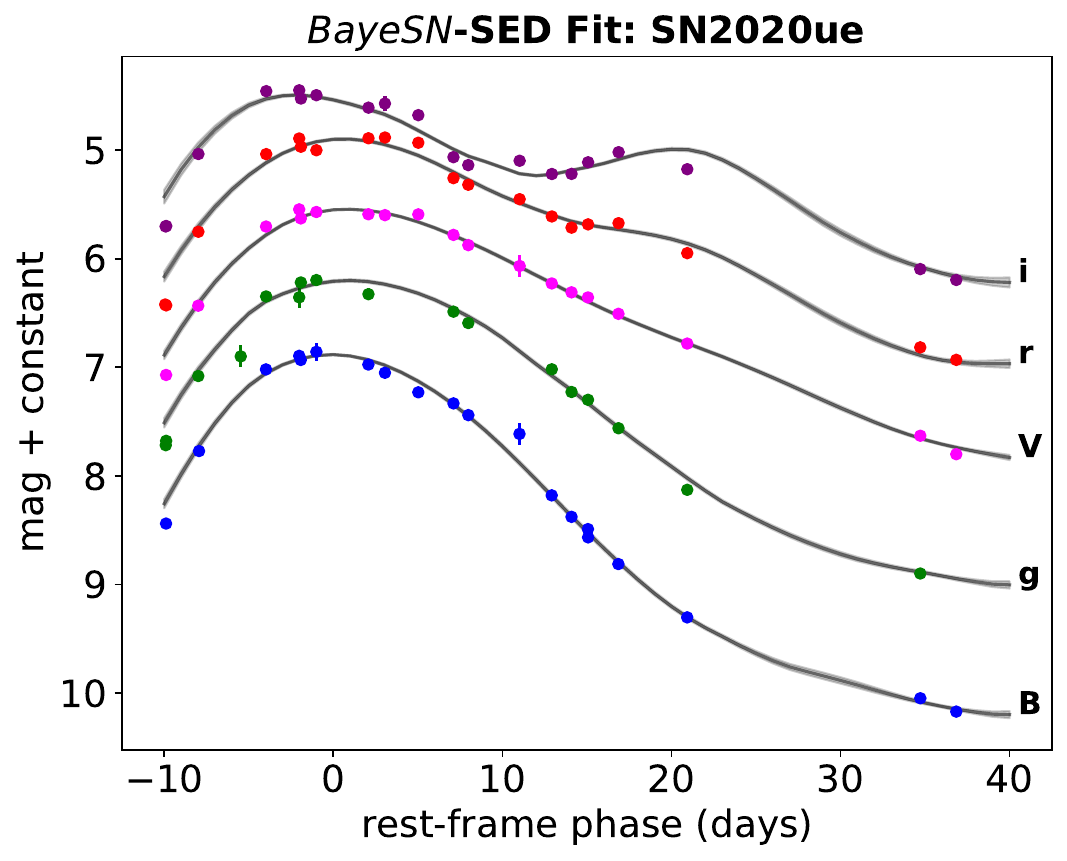}\\
    \includegraphics[width=0.48\linewidth]{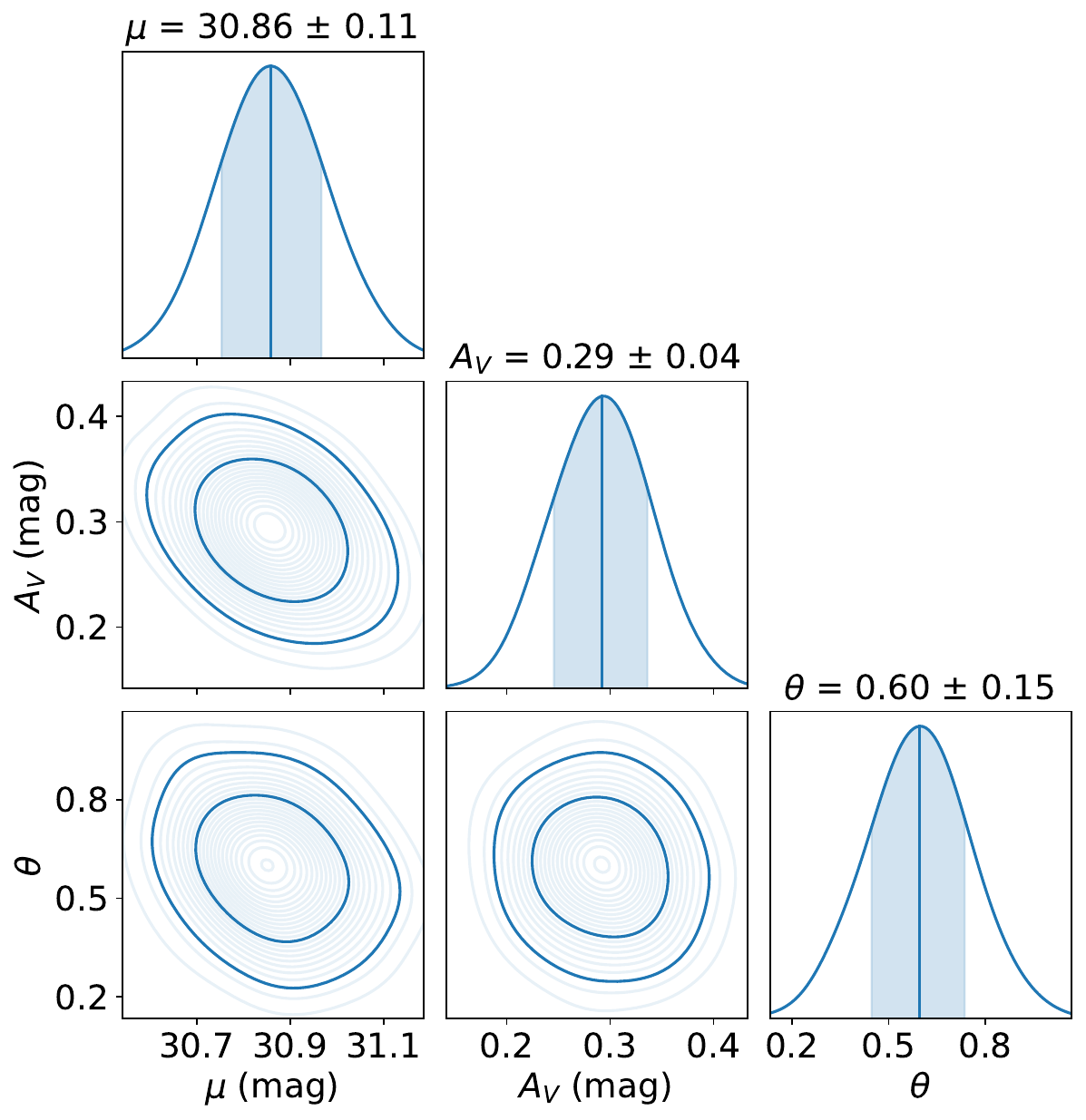}
    \includegraphics[width=0.48\linewidth]{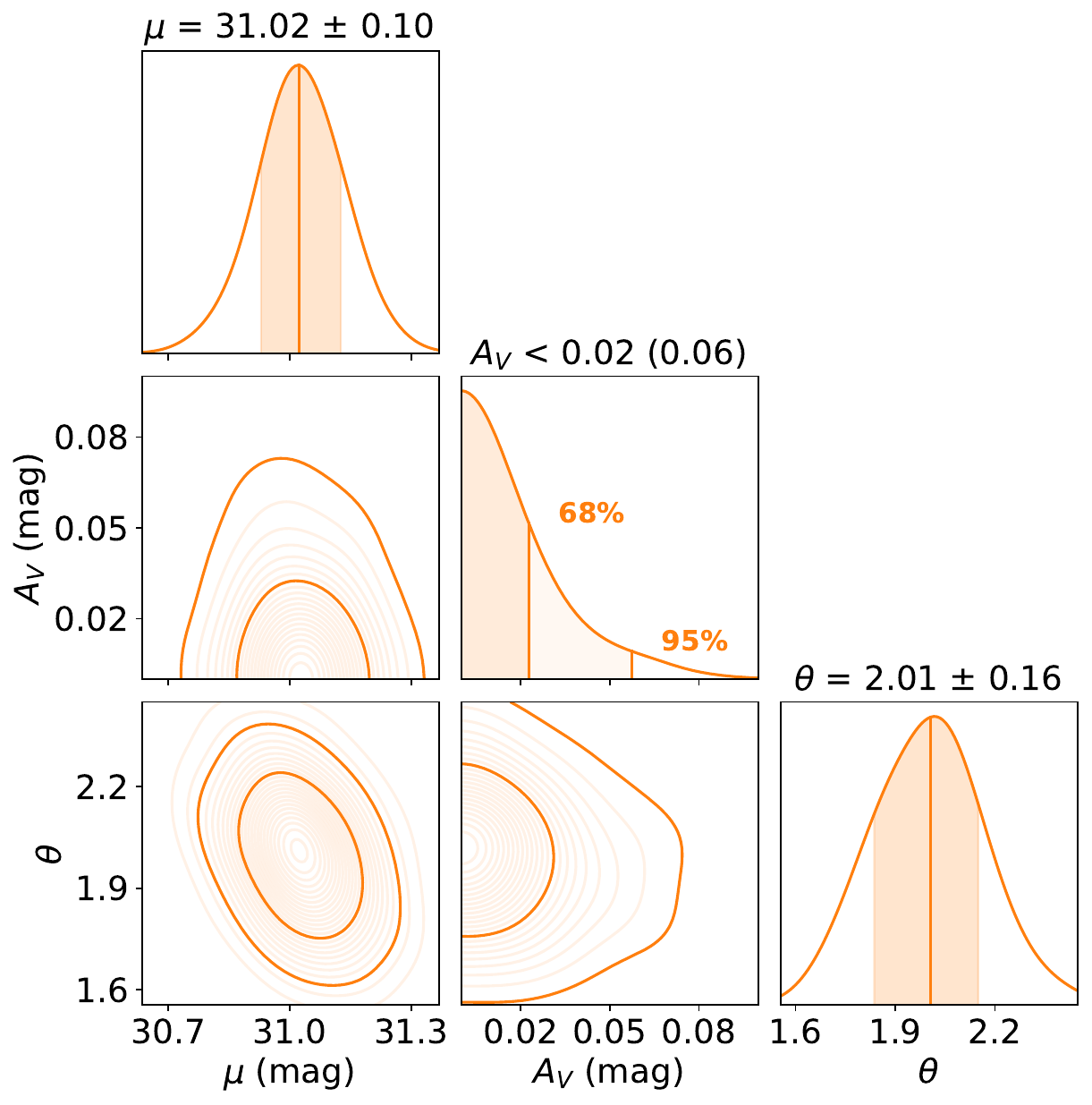}
    \caption{(Upper panels) Fit of the PS1 light curve of SN 2020nlb (left) and SN 2020ue (right). (Lower panels) The results from the MC analysis with \textsc{BayeSN} of the PS1 light curves of SN 2020nlb (left) and SN 2020ue (right).}
    \label{fig:BAYESNfit}
\end{figure*}

Figure \ref{fig:BAYESNfit} shows the best-fit results and parameters posterior distribution obtained applying {\sc BAYESN} to the multi-filter light curves of SN~2020ue and SN~2020nlb.

\begin{figure*}
    \centering
    \includegraphics[width=0.48\linewidth]{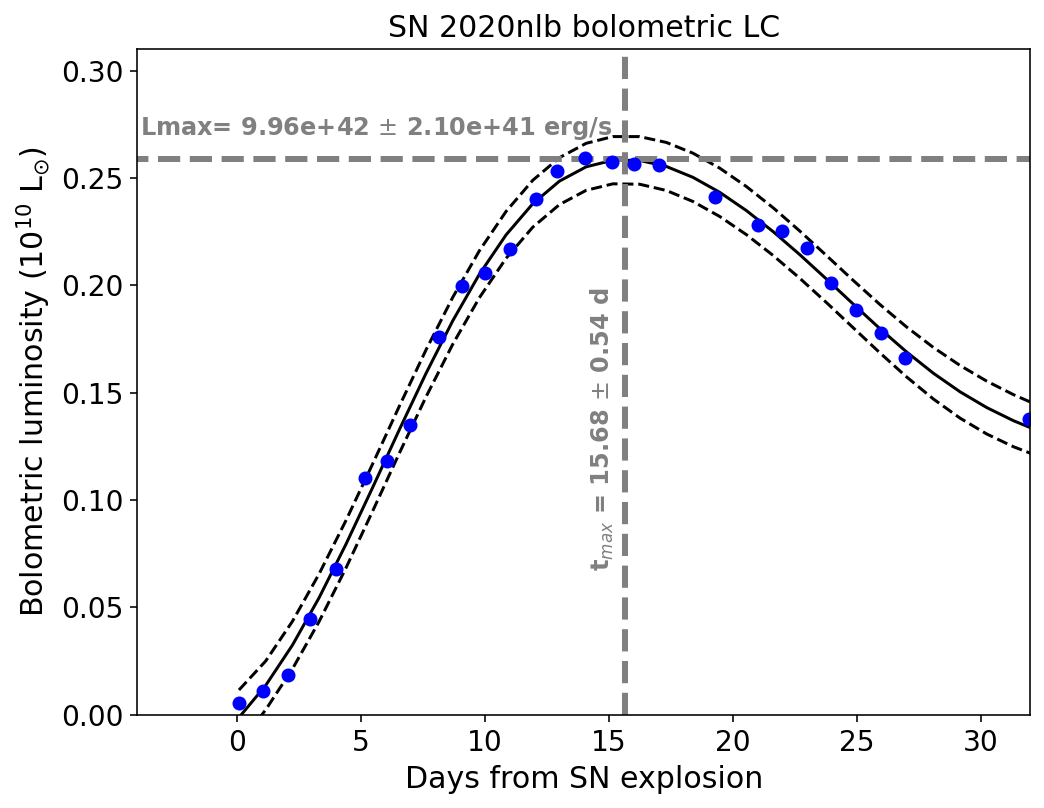}
    \includegraphics[width=0.48\linewidth]{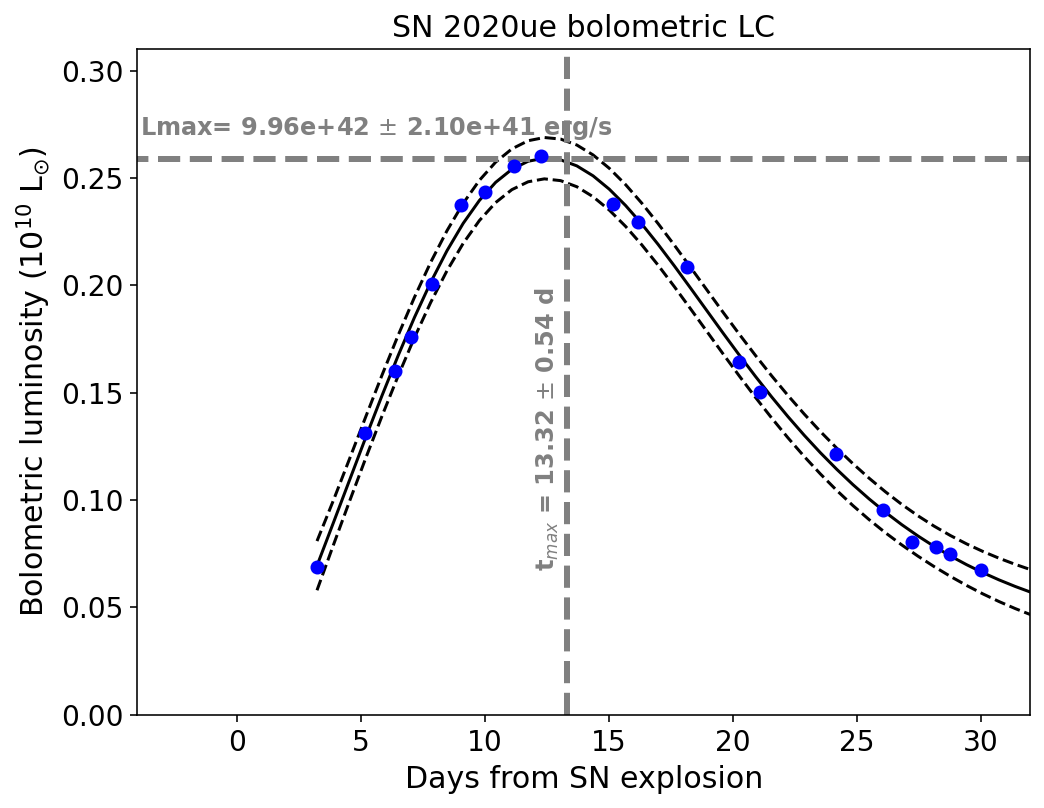}
    \caption{Bolometric light curves of SN 2020nlb (left panel) and SN 2020ue (right panel). The black curves represent the Gaussian Process (GP) model employed to estimate the peak luminosity and epoch of each light curve. The determined epochs for the SN explosions correspond to those obtained using the photospheric velocity method proposed by \citet{PiroNakar2014}}.
    \label{fig:bolofits}
\end{figure*}

\begin{figure*}
    \centering
    \includegraphics[width=0.45\linewidth]{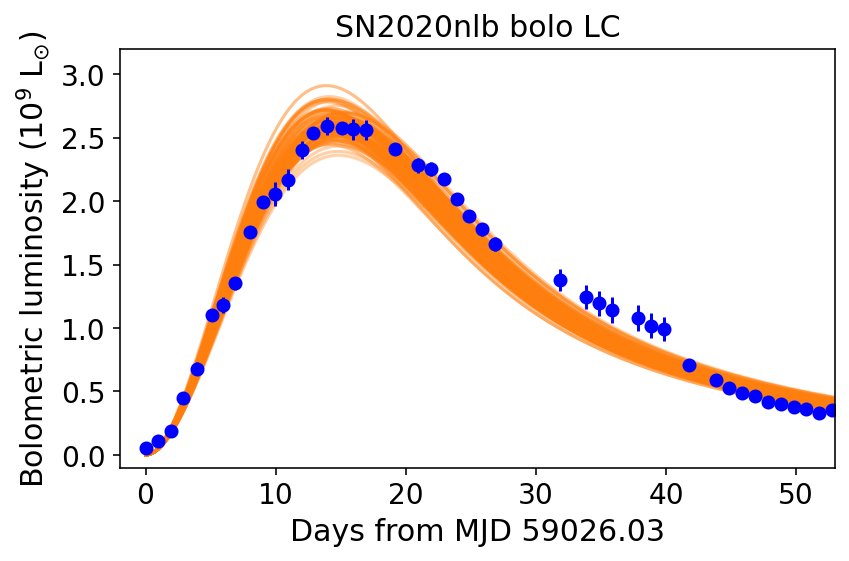}
    \includegraphics[width=0.45\linewidth]{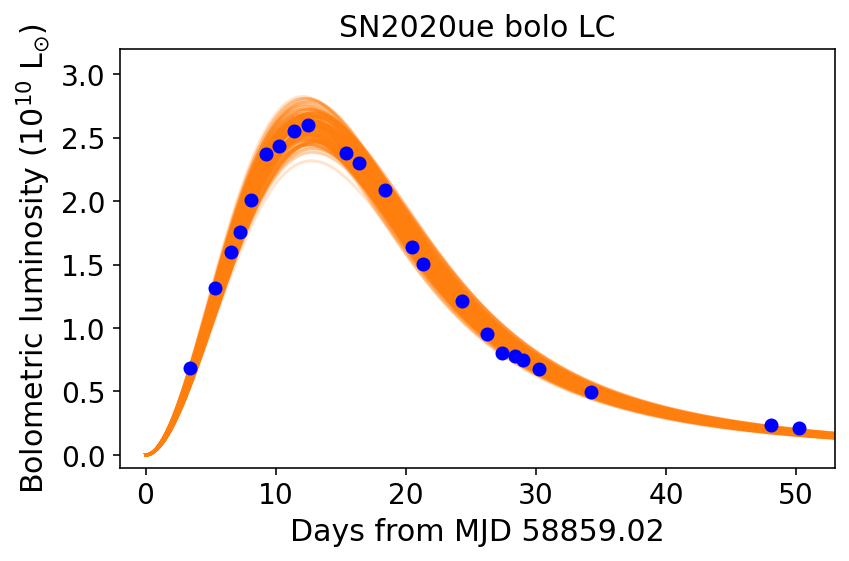}
    \caption{The fit of the bolometric light curves of SN 2020nlb (left panel) and of SN 2020ue (right panel). The orange curves correspond to 200 samples from the posterior distributions of the best-fit results shown in Table \ref{tab:Arnett}.  }
    \label{fig:Arnett}
\end{figure*}

Figures \ref{fig:bolofits}, \ref{fig:Arnett} refer to the analysis of the bolometric luminosity of SN~2020ue and SN~2020nlb discussed in Secs. \ref{sec:3_2}-\ref{sec:3_4}.

\begin{figure*}
    \centering
    \includegraphics[width=0.49\linewidth]{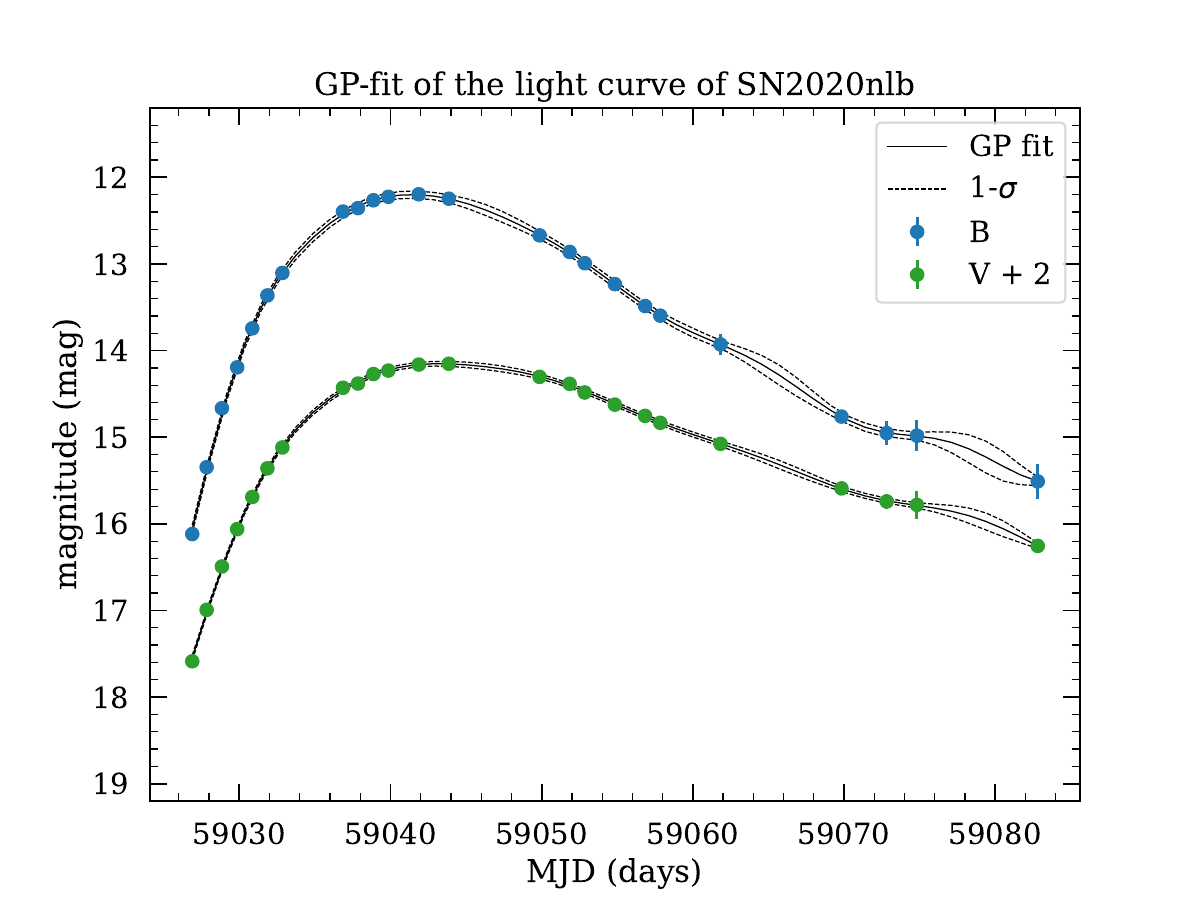}
    \includegraphics[width=0.49\linewidth]{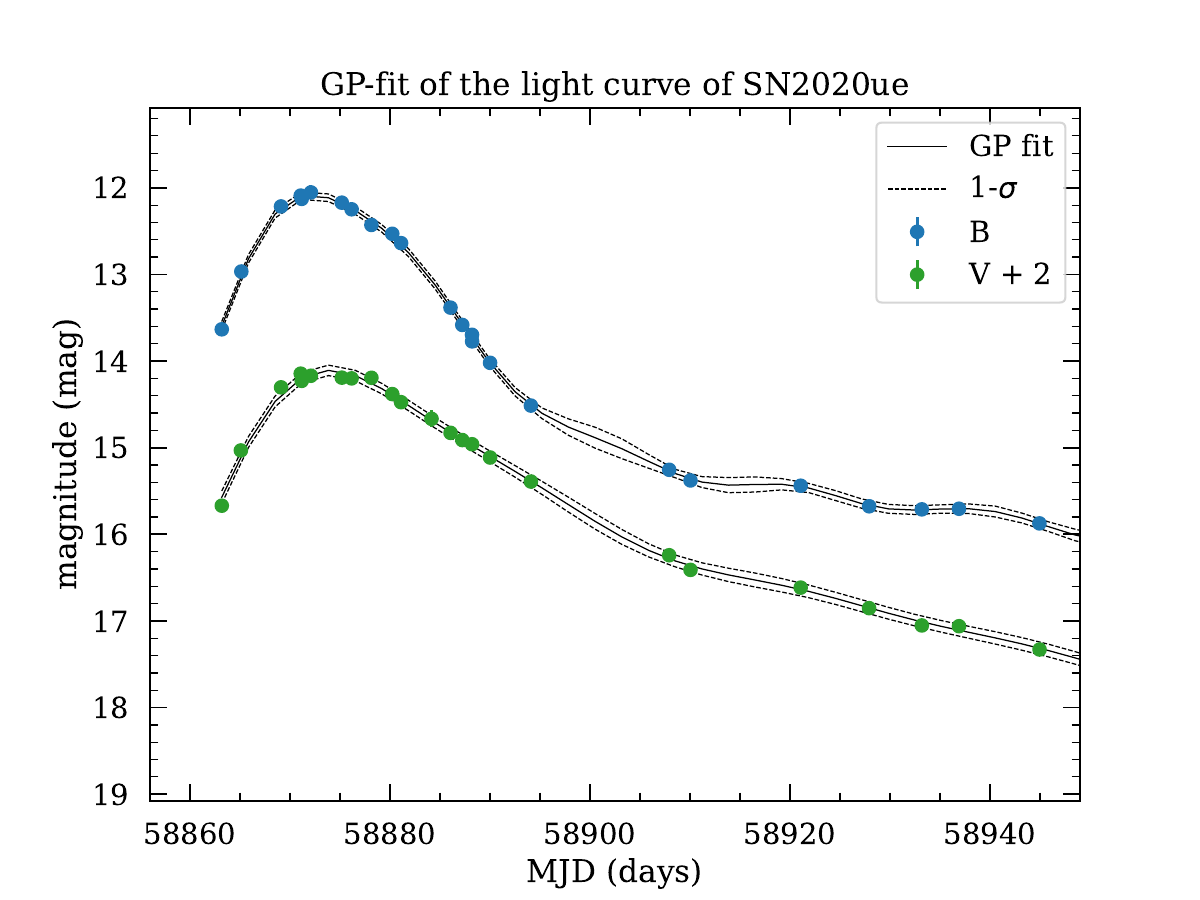}
    \caption{The light curve evolution of SN 2020nlb (left panel) and SN 2020ue (right panel) obtained using the GP regressor method.}
    \label{fig:LC2020uenlbGP}
\end{figure*}

\begin{figure*}
    \centering
    \includegraphics[width=0.32\linewidth]{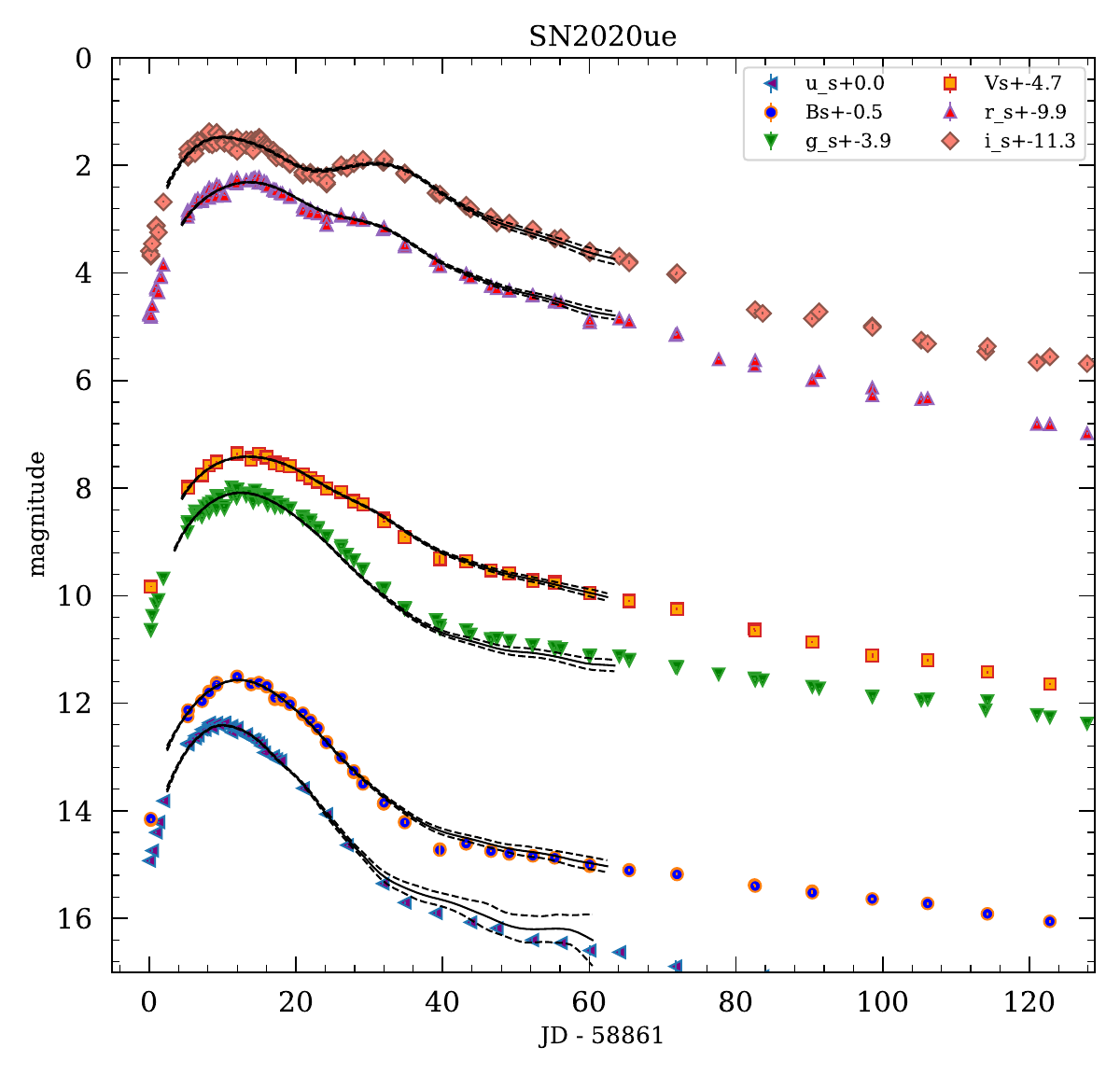}
    \includegraphics[width=0.32\linewidth]{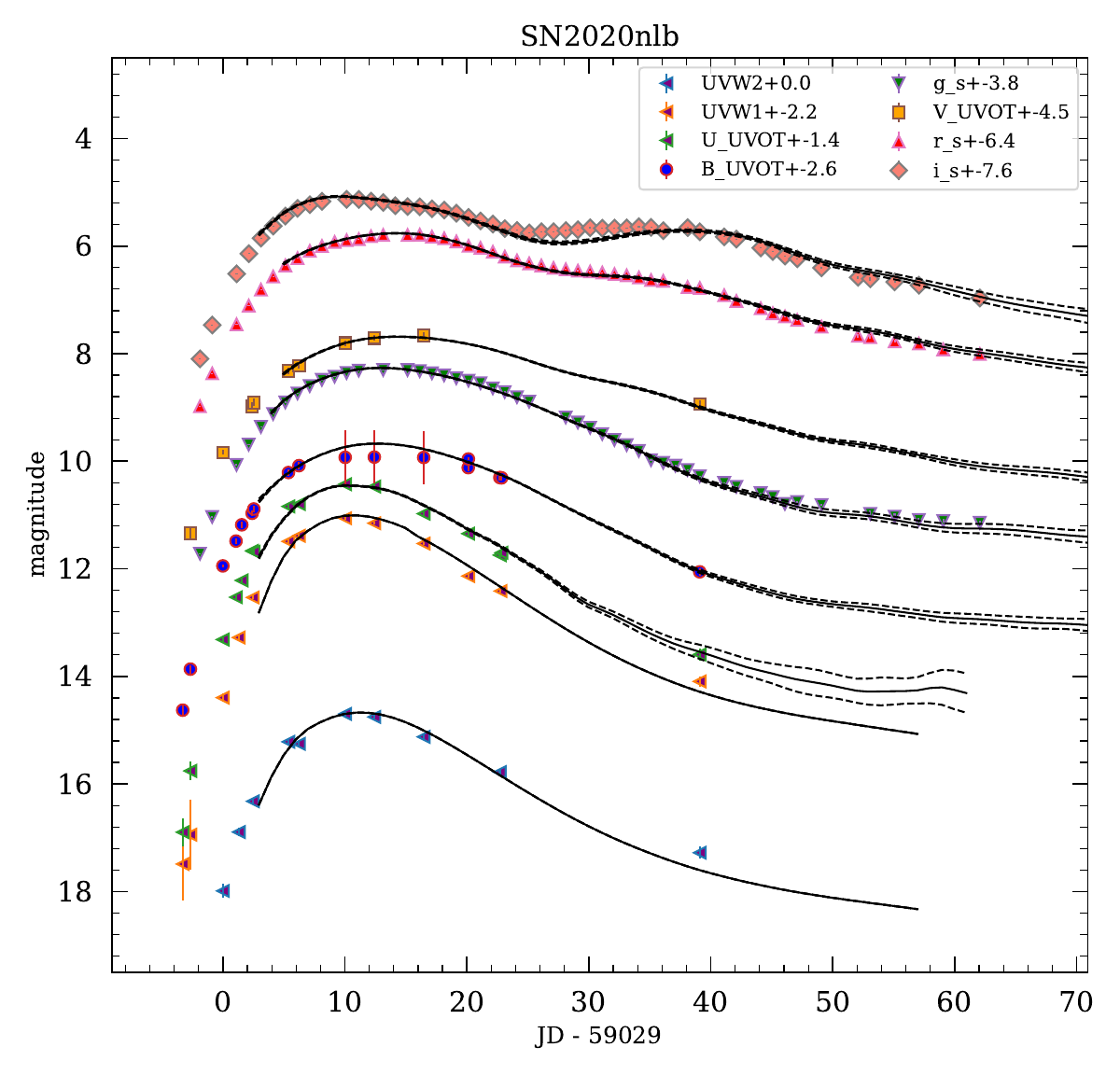}
    \includegraphics[width=0.32\linewidth]{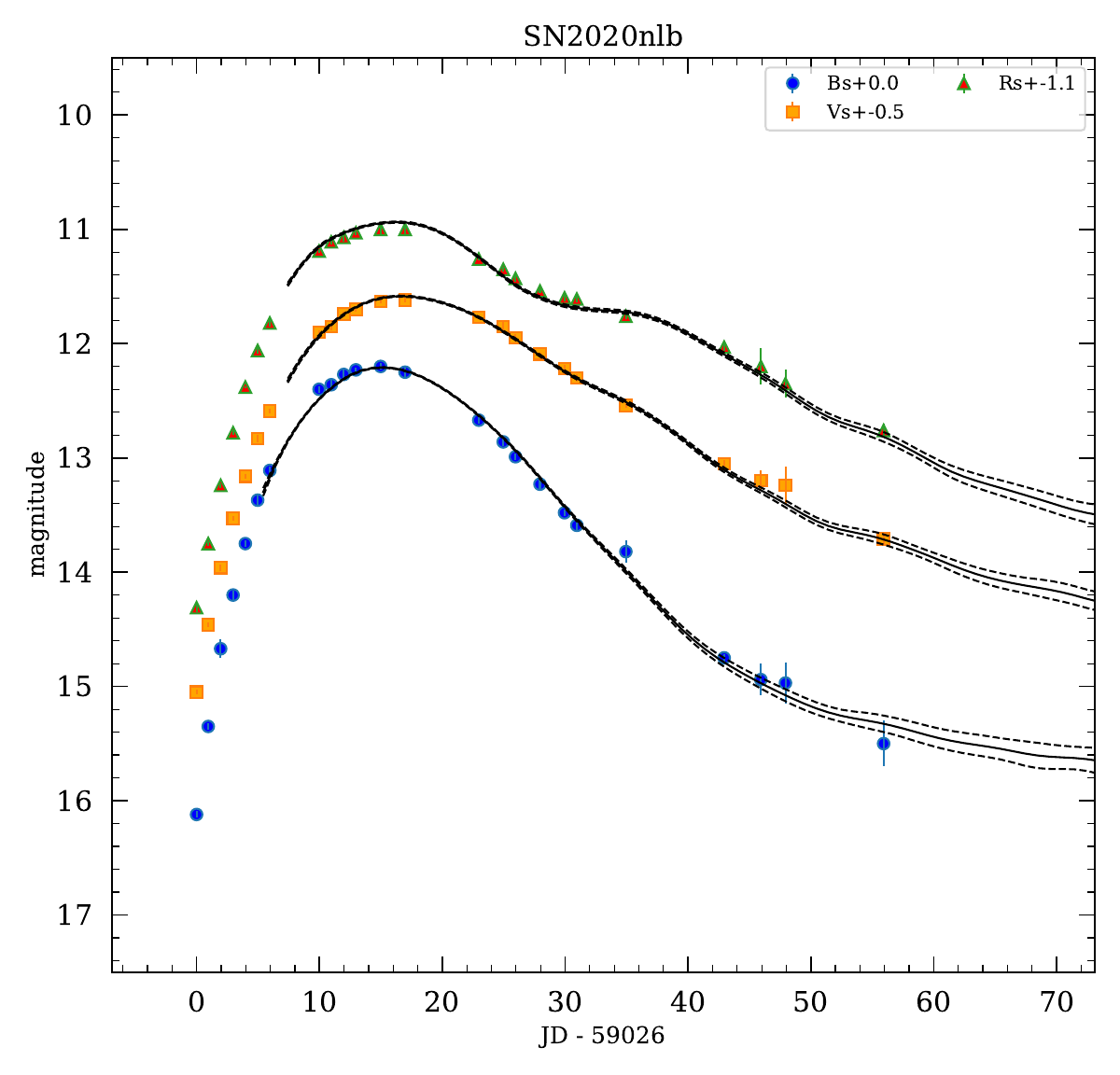}
    \caption{The fit obtained with \textsc{SNooPy} (black curves) using LCO data ($uBVgri$) for SN 2020ue (left panel), and using Thacher ($gri$) + UVOT data for SN 2020nlb (middle panel), and OASDG data ($BVR$, right panel). The plot shows the results obtained using the \texttt{max-model} light curve model function and the $s_{BV}$ stretch-color parameter. }
    \label{fig:SNooPyfit_LCO_OASDG}
\end{figure*}

Figure \ref{fig:LC2020uenlbGP} shows the best fit results of the OASDG $B$ and $V$ light curves of SN~2020ue and SN~2020nlb obtained using the GP process described in Sec. \ref{sec:3_1}, with the results reported in Table \ref{tab:1}. Similarly, Figures \ref{fig:SNooPyfit_LCO_OASDG} shows the {\sc SNooPy} best fit results of the multi-filter light curves of SN~2020ue and SN~2020nlb obtained with LCO, Thatcher and OASDG, with results reported in Table \ref{tab:app1}.

\begin{figure}
    \centering
    \advance\leftskip-0.6cm
    \includegraphics[width=0.95\linewidth]{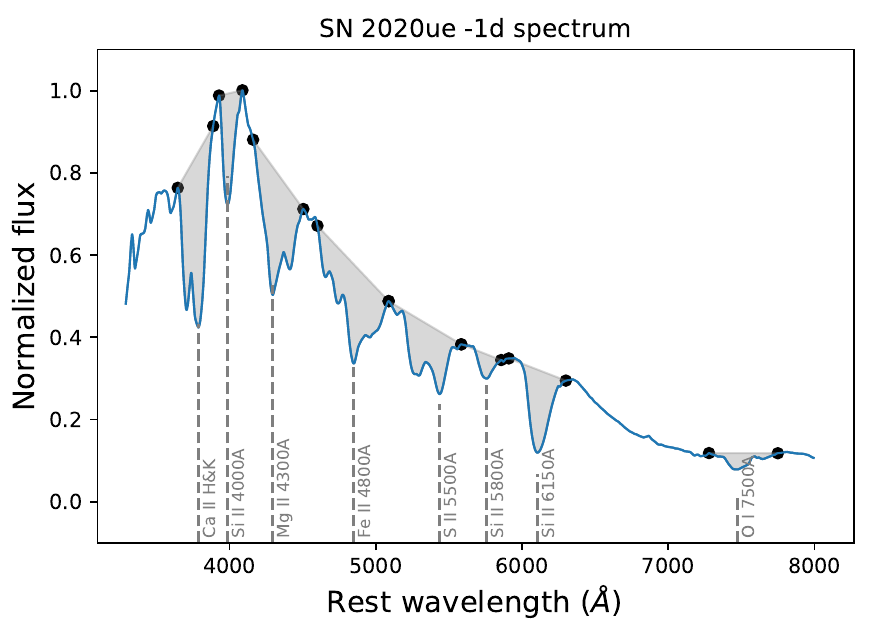}
    \caption{Measurement of the pEWs for SN 2020ue at $-$1 day using \texttt{Spextractor}. Shaded regions correspond to the area used to estimate the pEW, with black circles marking the boundaries of each absorption feature. Black circles mark the position of the low and high boundary wavelength regions found by the Gaussian process algorithm to identify each absorption line \citep{Papadogiannakis2019}.} 
    \label{fig:Spex_example}
\end{figure}

Figure \ref{fig:Spex_example} shows the best-fit model obtained for the peak spectrum of SN~2020ue using the {\sc Spextractor} code, with the identification of the main absorption features highlighted with gray regions.

\begin{figure}
    \centering
    \vspace{-0.4 cm}
    \advance\leftskip-0.5cm
    \includegraphics[width=0.98\linewidth]{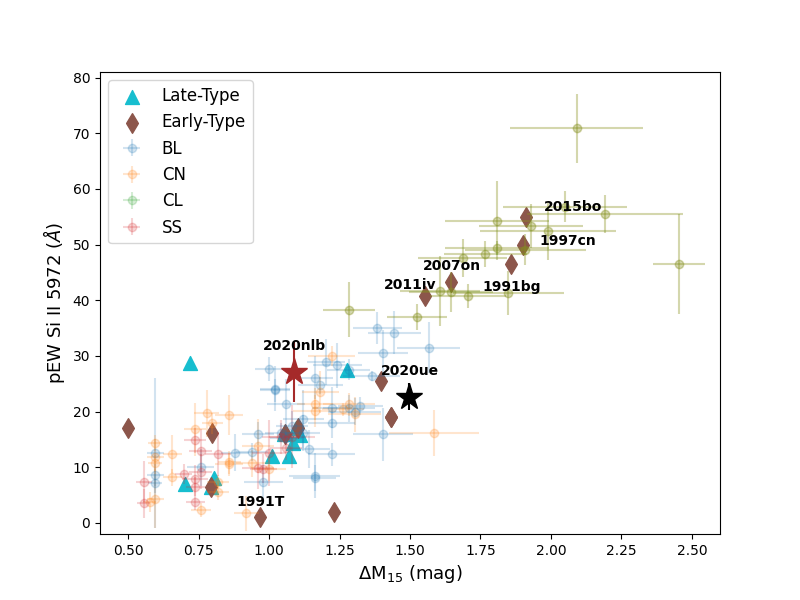}
    \includegraphics[width=0.98\linewidth]{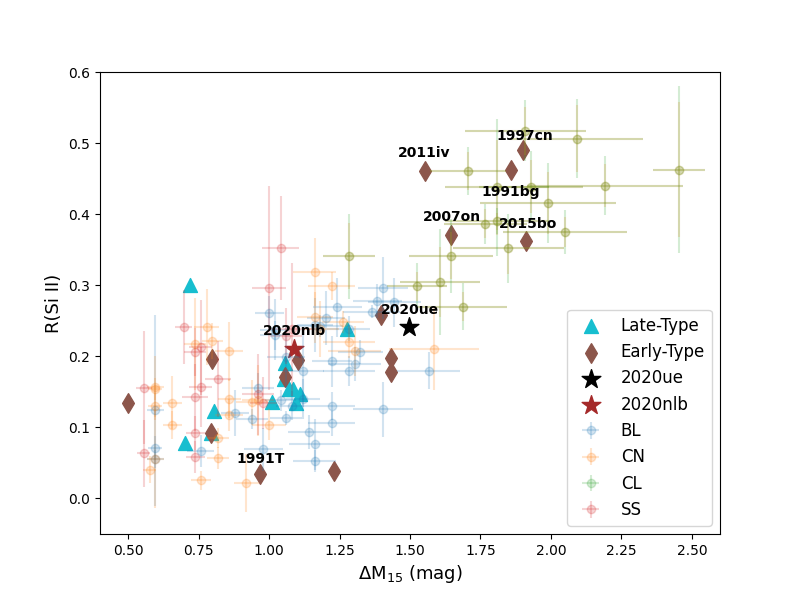}
     \vspace{-0.4 cm}
    \caption{Same as Fig. \ref{fig:Branch}, but here we show the distribution of the $\Delta m_{15}$ parameter against the pEW of the \ion{Si}{ii} 5972 \AA\, line (upper panel) and against the \ion{Si}{ii} ratio $R(Si)$ (lower panel. Data shown in red, blue, green, and orange symbols are taken from \citet{Burrow2020} and color-coded according to the Branch spectroscopic sub-classes {\it shallow silicon} (SS), {\it broad line} (BL), {\it cool} (CL), and {\it core normal} (CN), respectively. SN 2020ue and SN 2020nlb are highlighted as black and red stars, respectively. 
    }
    \label{fig:Branch2}
\end{figure}

Figure \ref{fig:Branch2} shows the Branch plot distribution of the pEW(\ion{Si}{ii} 5972) against the $\Delta m_{15}$ value, and the ratio of pEW of both \ion{Si}{ii} lines, R(\ion{Si}{ii}) versus $\Delta m_{15}$ value, for a set of SNe~Ia for which the pEW has been measured using the same methodology adapted in this work, e.g. with {\sc Spextractor} \citep{Burrow2020}.

\begin{figure}
    \centering
    \includegraphics[width=0.98\linewidth]{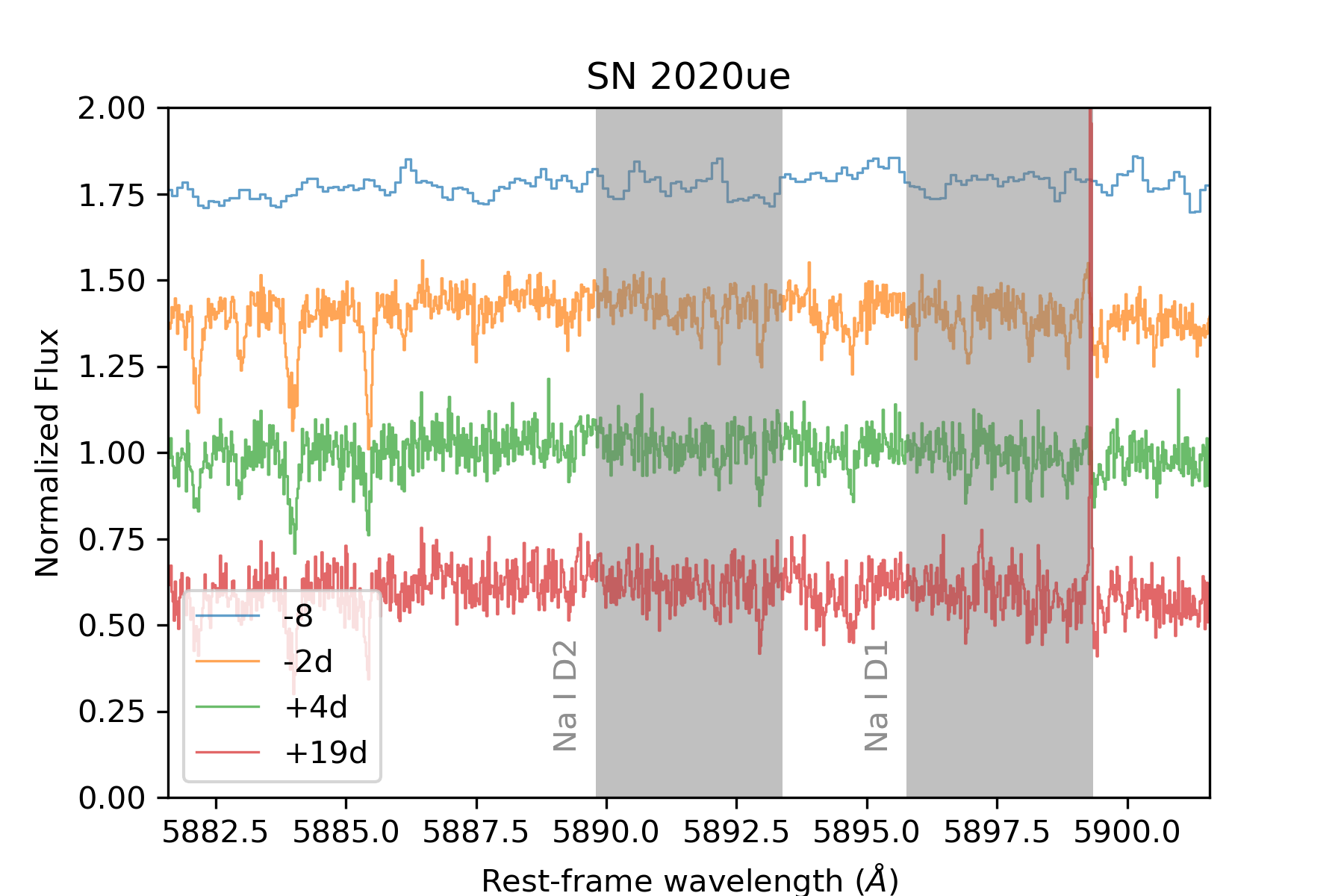}
    \includegraphics[width=0.98\linewidth]{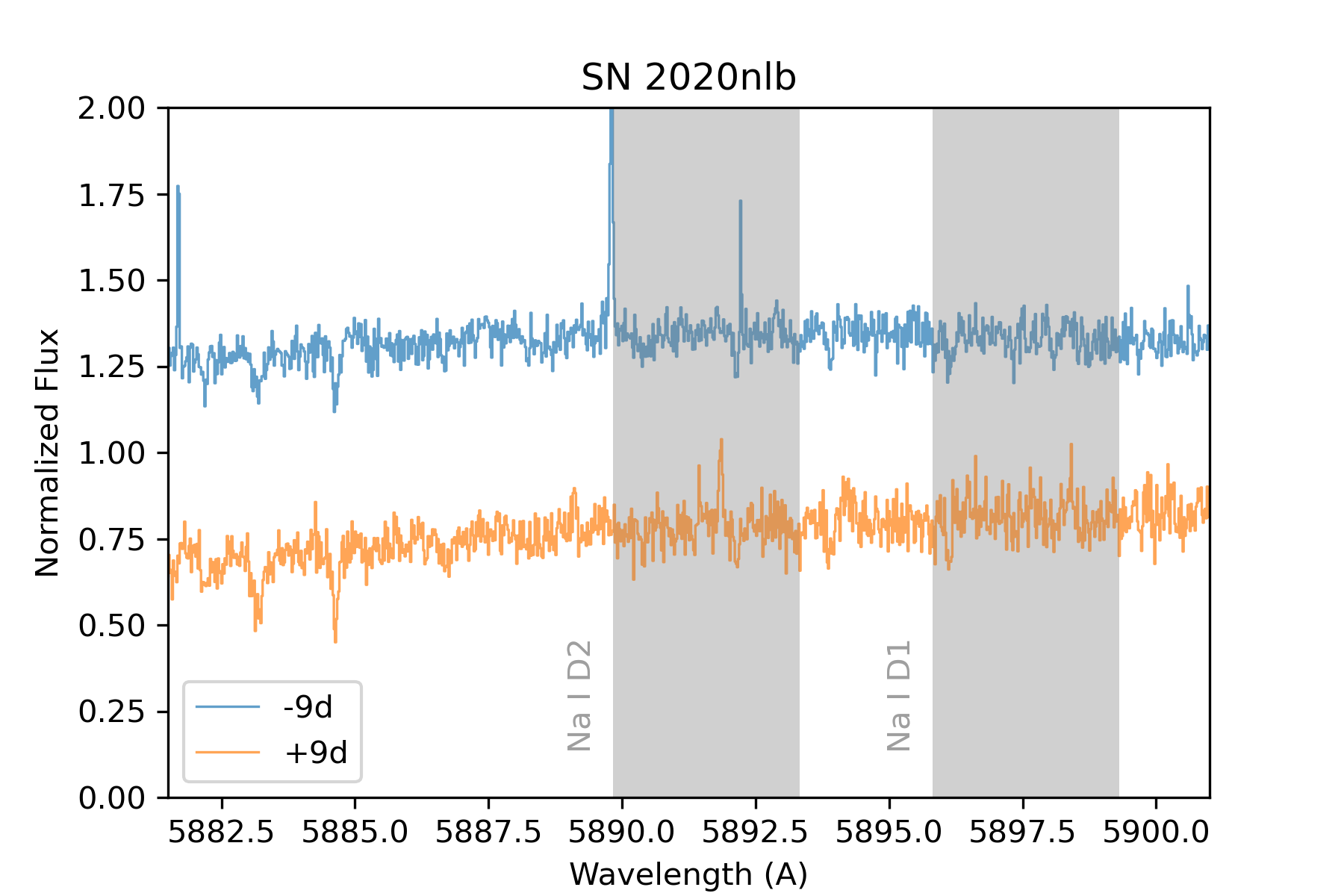}
    \caption{Asiago/Echelle spectrum ($-$8 d, blue curve) and Lick/APF spectra of SN 2020ue (top panel), and Lick/APF spectra of SN 2020nlb (lower panel) obtained around the peak brightness of both SNe. The gray regions mark the position of the expected Na ID lines at the redshift of each SN, including a shift of $\pm 100$ km/s due to the possible rotation velocity at each SN position. 
    }
    \label{fig:APF}
\end{figure}

Figure \ref{fig:APF} shows the high-resolution spectra of SN~2020ue and SN~2020nlb obtained with the Asiago/Echelle instrument and the Lick/APF instrument in the rest-frame region around the \ion{Na}{i} D line, showing absence of this feature in both SN spectra.

\begin{table}[h!]
\small
\centering
\caption{Log of spectroscopic observations.}
\label{tab:spexlog}
\begin{tabular}{lcccc}
\hline \hline
\multicolumn{5}{c}{SN 2020ue}\\
\hline
MJD & Phase & Spectrograph & Range & Resolution \\
(days) & (days) &  & (\AA) & (\AA) \\
\hline
58861.720 & -12 & FLOYDS & 3500-10000 & 3.5 \\
58865.650 & -8 & 1.82m+Echelle & 3792-7342 & 0.7 \\
58867.550 & -6 & FLOYDS & 3500-10000 & 3.5 \\
58869.605 & -4 & 1.22m+B\&C  & 3561-8282 & 9\\
58869.624 & -4 & 1.82m+AFOSC & 3218 - 9296 & 14 \\
58871.562 & -2 & 1.82m+AFOSC & 3270-9302 & 14 \\
58871.950 & -2 & APF+Levy & 3740-9700 & 100,000 \\
58872.583 & -1 & 1.82m+AFOSC & 3294-9302 & 14\\
58876.730 & +3 & FLOYDS & 3500-10000 & 3.5 \\
58877.950 & +4 & APF+Levy & 3740-9700 & 100,000 \\
58879.580 & +6 & FLOYDS & 3500-10000 & 3.5 \\
58880.730 & +7 & 1.82m+AFOSC & 3158-9299 & 14\\
58881.594 & +8 & 1.82m+AFOSC & 3147-9298 & 14\\
58886.588 & +13 & 1.22m+B\&C  & 3285-8000 & 9\\
58891.503 & +18 & 1.22m+B\&C & 3312-7893 & 9\\
58892.950 & +19 & APF+Levy & 3740-9700 & 100,000 \\
58893.750 & +20 & FLOYDS & 3500-10000 & 3.5 \\
58894.580 & +21 & 1.82m+AFOSC & 3148-9299 &14 \\
58899.643 & +26 & 1.22m+B\&C & 3284-7999 & 9\\
58899.690 & +26 & FLOYDS & 3500-10000 & 3.5 \\
58907.560 & +34 & 1.22m+B\&C & 3274-7988 & 9\\
58908.427 & +35 & 1.82m+AFOSC & 3151-9302 & 14\\
58926.476 & +53 & 1.22m+B\&C & 3088-7800 & 9\\
58927.520 & +54 & FLOYDS & 3500-10000 & 3.5 \\
58928.426 & +55 & 1.82m+AFOSC & 5002-9306 & 14\\
58939.354 & +66 & 1.82m+AFOSC & 3149-9299 & 14\\
58944.510 & +71 & FLOYDS & 3500-10000 & 3.5 \\
58945.463 & +72 & 1.82m+B\&C & 4996-9301 & 14\\
58949.452 & +76 & 1.22m+B\&C & 3240-7951 & 9\\
58953.384 & +80 & 1.82m+AFOSC & 3149-9300 & 14\\
58960.500 & +87 & FLOYDS & 3500-10000 & 3.5 \\
58970.397 & +97 & 1.82m+AFOSC & 4998-9302 & 14 \\
58977.490 & +104 & FLOYDS & 3500-10000 & 3.5 \\
58999.430 & +126 & FLOYDS & 3500-10000 & 3.5 \\
59018.350 & +145 & FLOYDS & 3500-10000 & 3.5 \\
59258.066 & +384 & Keck+LRIS & 3162-10148 & 6.9 \\

    \hline
    \hline
\multicolumn{5}{c}{SN 2020nlb}\\
\hline
MJD & Phase & Spectrograph & Range & Resolution \\
(days) & (days) &  & (\AA) & (\AA) \\
\hline

59026.290 & -16 & FLOYDS & 3500-10000 & 3.5 \\
59029.250 & -13 & FLOYDS & 3500-10000 & 3.5 \\
59033.650 & -9 & APF+Levy & 3740-9700 & 100,000 \\
59036.900 & -5 & NOT+AlFOSC & 3800-9000 & 16.2 \\
59049.184 & +7 & Shane+KAST & 3500-10000 & 16.2 \\
59051.650 & +9 & APF+Levy & 3740-9700 & 100,000 \\
59054.884 & +12 & NOT+AlFOSC & 3800-9000 & 16.2 \\
59057.198 & +15 & Shane+KAST & 3500-10000 & 16.2 \\
59060.883 & +18 & NOT+AlFOSC & 3800-9000 & 16.2 \\
59070.190 & +28 & Shane+KAST & 3500-10000 & 16.2 \\
59127.750 & +85 & Shane+KAST & 3500-5600 & 16.2 \\
59189.529 & +147 & Shane+KAST & 3500-5600 & 16.2 \\
    \hline
    \hline
        \end{tabular}
\end{table}

\begin{table}
\tiny
\centering
\caption{The pEWs (in units of m\AA) values obtained for selected absorption lines using {\sc spextractor} to the spectra of SN 2020ue and SN 2020nlb. Line transitions here refer to the average position of their absorption wavelength in the observed spectra.}
\label{tab:spex}
\begin{tabular}{lcccc}
\hline \hline
\multicolumn{5}{c}{SN 2020ue}\\
\hline
line & -12d & -6d & -4d & -2d \\
\hline
\ion{Ca}{II} H\&K & 84.1$\pm$ 9.1 & 64.5$\pm$17.0 & 62.3$\pm$13.0 & 61.5$\pm$12.9\\
\ion{Si}{II} 4000 & 7.9$\pm$5.9 & 19.3$\pm$16.5 & 20.8$\pm$14.6 & 3.6$\pm$3.3\\
\ion{Mg}{II} 4300 & 35.2$\pm$7.2 & 60.8$\pm$11.9 & 54.0$\pm$11.3 & 64.1$\pm$11.9\\
\ion{Fe}{II} 4800 & 89.8$\pm$4.2 & 71.2$\pm$6.4 & 82.6$\pm$6.8 & 90.9$\pm$6.9\\
\ion{S}{II} 5500 & 40.6$\pm$2.7 & 61.4$\pm$3.4 & 63.7$\pm$3.5 & 75.3$\pm$3.6\\
\ion{Si}{II} 5800 & 19.3$\pm$1.6 & 20.1$\pm$2.3 & 24.5$\pm$2.5 & 22.3$\pm$2.2\\
\ion{Si}{II} 6150 & 110.1$\pm$1.5 & 91.6$\pm$2.2 & 87.6$\pm$2.3 & 93.5$\pm$1.9\\
\hline
line & -1d & +3d & +6d & +7d \\
\hline
\ion{Ca}{II} H\&K & 67.3$\pm$13.3 & 67.2$\pm$12.3 & 39.1$\pm$7.1 & 52.9$\pm$15.2\\
\ion{Si}{II} 4000 & 22.4$\pm$16.7 & 25.5$\pm$16.9 & 21.9$\pm$13.2 & 27.5$\pm$13.8\\
\ion{Mg}{II} 4300 & 67.9$\pm$11.1 & 73.9$\pm$14.7 & 79.0$\pm$13.8 & 84.2$\pm$9.9\\
\ion{Fe}{II} 4800 & 67.8$\pm$5.0 & 139.9$\pm$11.6 & 158.1$\pm$13.2 & 102.2$\pm$5.8\\
\ion{S}{II} 5500 & 84.8$\pm$3.5 & 88.1$\pm$5.9 & 51.2$\pm$5.7 & 39.7$\pm$3.5\\
\ion{Si}{II} 5800 & 18.2$\pm$2.0 & 18.5$\pm$3.9 & 32.2$\pm$3.2 & 12.5$\pm$1.6\\
\ion{Si}{II} 6150 & 98.3$\pm$1.9 & 100.6$\pm$3.2 & 95.3$\pm$1.9 & 114.3$\pm$1.6\\
\ion{O}{I} 7500 & 50.9$\pm$0.3 & 58.7$\pm$0.7 & 63.8$\pm$0.1 & 82.3$\pm$0.1\\    
    \hline
    \hline
\multicolumn{5}{c}{SN 2020nlb}\\
\hline
line & -16d & -13d & -6d & +7d\\
\hline
\ion{Ca}{II} H\&K & 198.7$\pm$4.7 & 65.6$\pm$6.3 & 77.6$\pm$10.1 & 81.1$\pm$11.0\\
\ion{Si}{II} 4000 & 14.2$\pm$8.4 & 17.2$\pm$12.2 & 22.9$\pm$15.3 & 40.9$\pm$12.7\\
\ion{Mg}{II} 4300 & 103.9$\pm$11.7 & 25.4$\pm$7.4 & 16.3$\pm$6.8 & 71.9$\pm$8.4\\
\ion{Fe}{II} 4800 & 233.2$\pm$15.9 & 70.7$\pm$5.4 & 79.9$\pm$4.6 & 208.2$\pm$11.6\\
\ion{S}{II} 5500 & 8.5$\pm$10.2 & 53.4$\pm$5.2 & 73.6$\pm$3.4 & 49.5$\pm$6.0\\
\ion{Si}{II} 5800 & 10.9$\pm$5.9 & 23.8$\pm$3.1 & 23.6$\pm$2.0 & 21.2$\pm$3.1\\
\ion{Si}{II} 6150 & 99.9$\pm$6.8 & 108.0$\pm$3.7 & 104.3$\pm$2.1 & 108.1$\pm$2.6\\
\ion{O}{I} 7500 & 25.3$\pm$1.4 & 85.6$\pm$1.1 & 54.4$\pm$0.5 & 75.9$\pm$0.3\\
    \hline
        \end{tabular}
\end{table}

\begin{table*}
\centering
\caption{The results of the light curve fits made with \textsc{SNooPy} using LCO data for SN 2020ue and OASDG data for SN 2020nlb.}
\label{tab:app1}
\begin{tabular}{lccccc}
\hline \hline
\multicolumn{6}{c}{SN 2020ue - LCO}\\
\hline
model parameter & \multicolumn{2}{c}{max-model} & \multicolumn{2}{c}{EBV-model2} & GP-model\\
\hline
$E(B-V)_{host}$ (mag) & - & - & -0.155 $\pm$ 0.008 & -0.196 $\pm$ 0.009 & -0.198 $\pm$ 0.009\\
$\mu$ (mag) & - & - & 31.219 $\pm$ 0.010 & 31.203 $\pm$ 0.009 & \\
$B_{max}$ (mag) & 11.960 $\pm$ 0.010 & 11.964 $\pm$ 0.008 & - & - & 11.861 $\pm$ 0.107\\
$V_{max}$ (mag) & 12.031 $\pm$ 0.009 & 12.009 $\pm$ 0.007 & - & - & 11.972 $\pm$ 0.024\\
$T_{B,max}$ (MJD) & 58873.53 $\pm$ 0.04 & 58873.48 $\pm$ 0.04 & 58873.59 $\pm$ 0.05 & 58873.51 $\pm$ 0.04 & 58873.51 $\pm$ 0.04\\
$s_{BV}$ & 0.735 $\pm$ 0.005 & -  & 0.725 $\pm$ 0.006 & - & 0.726 $\pm$ 0.006\\
$\Delta m_{15}(B)$ (mag) & - & 1.506 $\pm$ 0.007 & - & 1.522 $\pm$ 0.008 & 1.521 $\pm$ 0.008\\
    \hline
    \hline
\multicolumn{6}{c}{SN 2020nlb - OASDG}\\
\hline
 & \multicolumn{2}{c}{max-model} & \multicolumn{2}{c}{EBV-model2} & GP-model\\
\hline
$E(B-V)_{host}$ (mag) & - & - & 0.206 $\pm$ 0.019 & 0.206 $\pm$ 0.012 & 0.102 $\pm$ 0.005  \\
$\mu$ (mag) & - & - & 30.839 $\pm$ 0.026 & 30.941 $\pm$ 0.021 & -\\
$B_{max}$ (mag) & 12.129 $\pm$ 0.014 & 12.144 $\pm$ 0.009 & - & - & 12.200 $\pm$ 0.042 \\
$V_{max}$ (mag) & 12.056 $\pm$ 0.008 & 12.084 $\pm$ 0.006 & - & - & 12.151 $\pm$ 0.027\\
$T_{B,max}$ (MJD) & 59042.31 $\pm$ 0.08 & 59042.28 $\pm$ 0.06 & 59042.29 $\pm$ 0.122 & 59042.29 $\pm$ 0.02 & 59041.45 $\pm$ 0.01 \\
$s_{BV}$ & 0.879 $\pm$ 0.010 & - & 0.890 $\pm$ 0.016 & - & 0.829 $\pm$ 0.159\\
$\Delta m_{15}(B)$ (mag) & - & 1.188 $\pm$ 0.006 & - & 1.089 $\pm$ 0.009 & 1.249 $\pm$ 0.072\\
\hline
\hline
\multicolumn{6}{c}{SN 2020nlb - Thacher \& UVOT}\\
\hline
 & \multicolumn{2}{c}{max-model} & \multicolumn{2}{c}{EBV-model2} & GP-model\\
\hline
$E(B-V)_{host}$ (mag) & - & - & 0.076 $\pm$ 0.009 & 0.094 $\pm$ 0.007 & -  \\
$\mu$ (mag) & - & - & 31.040 $\pm$ 0.019 & 31.102 $\pm$ 0.015 & -\\
$B_{max}$ (mag) & 12.203 $\pm$ 0.053 & 12.201 $\pm$ 0.066 & - & - & - \\
$V_{max}$ (mag) & 12.093 $\pm$ 0.066 & 12.152 $\pm$ 0.081 & - & - & - \\
$T_{B,max}$ (MJD) & 59042.04 $\pm$ 0.05 & 59041.90 $\pm$ 0.09 & 59042.37 $\pm$ 0.09 & 59042.20 $\pm$ 0.06 & - \\
$s_{BV}$ & 0.928 $\pm$ 0.006 & - & 0.893 $\pm$ 0.008 & - & -\\
$\Delta m_{15}(B)$ (mag) & - & 1.089 $\pm$ 0.017 & - & 1.179 $\pm$ 0.018 & -\\
\hline
        \end{tabular}
\end{table*}

\end{appendix}

\end{document}